\documentclass{ar-astro2e}
\usepackage{ARAstroBib}
\usepackage{amsfonts}
\usepackage{amssymb}
\usepackage[]{color,graphicx}
\usepackage{deluxetable}
\def\ltsima{$\; \buildrel < \over \sim \;$}
\def\lsim{\lower.5ex\hbox{\ltsima}}
\def\gtsima{$\; \buildrel > \over \sim \;$}
\def\gsim{\lower.5ex\hbox{\gtsima}}

\newcommand{\be}{\begin{equation}}
\newcommand{\ee}{\end{equation}}

\newcommand\arcsec{\mbox{$^{\prime\prime}$}}%
\newcommand\fdg{\mbox{$.\!\!^\circ$}}%

\begin{document}

\input psfig.sty

\jname{Annu.\ Rev.\ Astron.\ Astrophys.}
\jyear{2015}
\jvol{}
\ARinfo{ }

\title{The Occurrence and Architecture\\of Exoplanetary Systems}

\author{
  Joshua N.\ Winn\affiliation{
    Department of Physics, Massachusetts Institute of Technology,\\
    Cambridge, Massachusetts, 02139; email:~{\tt jwinn@mit.edu}}
  Daniel C.\ Fabrycky\affiliation{
    Department of Astronomy and Astrophysics, University of Chicago,\\
    Chicago, IL, 60637; email:~{\tt fabrycky@uchicago.edu}}
}


\begin{abstract}

  The basic geometry of the Solar System---the shapes, spacings, and
  orientations of the planetary orbits---has long been a subject of
  fascination as well as inspiration for planet-formation
  theories. For exoplanetary systems, those same properties have only
  recently come into focus. Here we review our current knowledge of
  the occurrence of planets around other stars, their orbital
  distances and eccentricities, the orbital spacings and mutual
  inclinations in multiplanet systems, the orientation of the host
  star's rotation axis, and the properties of planets in binary-star
  systems.

\end{abstract}

\maketitle


\section{INTRODUCTION}
\label{sec:intro}

Over the centuries, astronomers gradually became aware of the
following properties of the Solar System:
\begin{itemize}
\item The Sun has eight planets, with the four smaller
  planets ($R_{\rm p} =0.4$--1.0~$R_\oplus$) interior to the four larger planets (3.9--11.2~$R_\oplus$).
\item The orbits are all nearly circular, with a mean eccentricity
  of 0.06 and individual eccentricities ranging from 0.0068--0.21.
\item The orbits are nearly aligned, with a
  root-mean-squared inclination of $1\fdg9$ relative to the plane
  defined by the total angular momentum of the Solar System (the
  ``invariable plane''), and individual inclinations ranging
  from $0\fdg33$--$6\fdg3$.
\item The Sun's rotational angular momentum is much smaller than the
  orbital angular momentum of the planets ($L_\odot/L_{\rm orb}
  \approx 0.5\%$).
\item The Sun's equator is tilted by $6\fdg 0$ relative to the
  invariable plane.
\item The sizes of neighboring orbits have ratios in the range of 1.4--3.4.
\end{itemize}
These regularities have long been recognized as important clues about
the formation of the Solar System. Isaac Newton realized that the
Solar System is more orderly than required by the laws of motion and
took this as evidence for God's hand in creation. Pierre-Simon Laplace
was inspired by the same facts to devise a mechanistic theory for the
formation of the Solar System. Since then, it became traditional to
begin any article on the formation of the Solar System with a list of
observations similar to the one provided above.

Many authors of such articles have expressed a wish to have the
same type of information for other planetary systems. In a
representative example, \cite{WilliamsCremin1968} wrote, ``It is
difficult to decide whether all the major properties of the Solar
System have been included above, or indeed whether all the above
properties are essential properties of any system formed under the
same conditions as the Solar System. This is because only one Solar
System is known to exist and so it is impossible to distinguish
between phenomena that must come about as a direct consequence of some
established law and phenomena that come about as a result of unlikely
accidents.''

In the past few decades these wishes have begun to be fulfilled. In
this review, we have attempted to construct an exoplanetary version of
the traditional list of elementary properties of the Solar System. Our
motivation is mainly empirical. The measurement of exoplanetary
parameters is taken to be an end unto itself. Theories to explain the
values of those parameters are discussed but not comprehensively.
For brevity, we have also resisted the temptation to review the history
of exoplanetary science or the techniques of exoplanet
detection.\footnote{For history, see the collection of articles edited
  by \cite{Lissauer2012}. For detection techniques, see
  \cite{Seager2011} or \cite{WrightGaudi2013}. For theory, see
  \cite{Ford2014}.} Our main concern is the geometry of exoplanetary
systems---the parameters that Newton and Laplace would have
immediately appreciated---as opposed to atmospheres, interior
compositions, and other aspects of post-Apollo planetary
science. However, we have added a parameter to the list that may
have surprised Newton and Laplace: the number of host stars, which
need not be one.

This article is organized as follows. Section~\ref{sec:occurrence}
answers the first question our predecessors would have asked: How
common, or rare, are planets around other stars? Section~\ref{sec:pe}
discusses the sizes and shapes of individual planetary orbits. This is
followed in section~\ref{sec:multis} by a discussion of multiplanet
systems, particularly their orbital spacings and mutual
inclinations. Section~\ref{sec:rotation} considers the rotation of the
host star, and section~\ref{sec:binaries} considers planets in
binary-star systems. Finally, Section~\ref{sec:discussion} summarizes
the exoplanetary situation, comments on the implications for
planet-formation theory, and discusses the future prospects for
improving our understanding of exoplanetary architecture.


\section{OCCURRENCE RATE}
\label{sec:occurrence}

No simple physical principle tells us whether planets should be rare
or common. Before the discovery of exoplanets, there was ample room
for speculation. Monistic planet-formation theories, in which
planet and star formation are closely related, predicted that planets
should be ubiquitous; an example was Laplace's nebular theory, in
which planets condense from gaseous rings ejected from a spinning
protostar. In contrast, dualistic theories held that planets arise
from events that are completely distinct from star formation, and in
many such theories the events were extremely unlikely. An example was
the tidal theory in which planets condense from material stripped
away from a star during a chance encounter with another star. This was
predicted to occur with a probability of $\sim$$10^{-10}$ for a given star,
although \cite{Jeans1942} managed to increase the odds to $\sim$0.1 by
allowing the encounter to happen during pre--main-sequence
contraction.

Modern surveys using the Doppler, transit, and microlensing techniques
have shown that planets are prevalent. The probability that a random
star has a planet is of order unity for the stars that have been
searched most thoroughly: main-sequence dwarfs with masses
0.5-1.2~$M_\odot$. Monistic theories have prevailed, including the
currently favored theory of planet formation in which planets build up
from small particles within the gaseous disks that surround all young
stars.

The occurrence rate is the mean number of planets per star having
properties (such as mass and orbital distance) within a specified
range. The basic idea is to count the number of detected planets with
the stipulated properties and divide by the effective number of stars
in the survey for which such a planet could have been detected. The
word ``effective'' reminds us that it is not always clear-cut whether
or not a planet could have been detected; in such cases one must sum
the individual detection probabilities for each star.

\begin{figure}
\includegraphics[height=9cm]{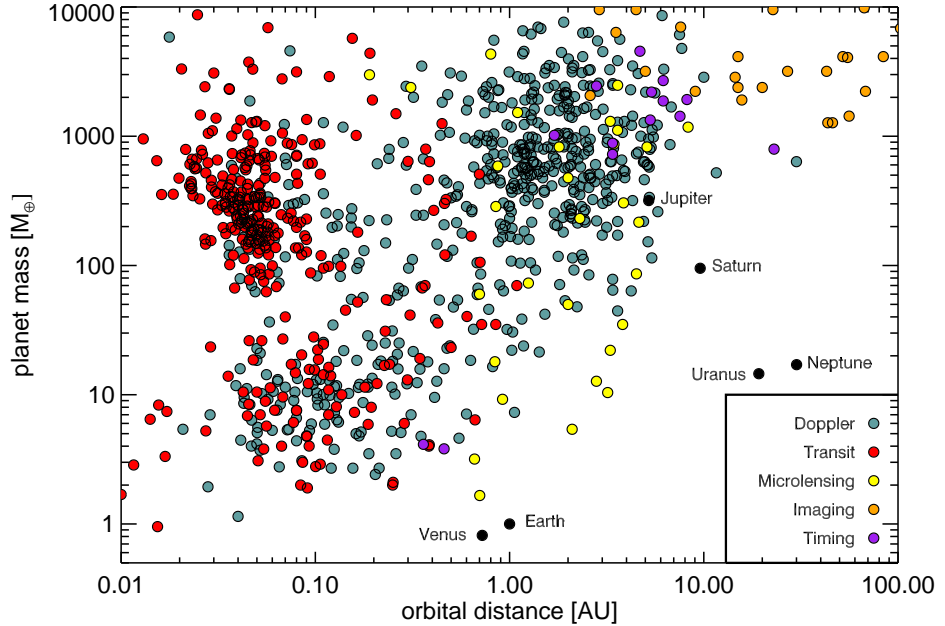}
\caption{
Approximate masses and orbital distances of known planets,
based on an October 2014 query of the {\tt exoplanets.eu} encyclopedia
\citep{Schneider+2011}.
This plot does not consider selection biases
and glosses over many important details.
For Doppler planets, the plotted mass is really $M_{\rm p}\sin i$.
For imaging planets, the plotted mass is based on
theoretical models relating the planet's age, luminosity, and mass.
With microlensing, the planet-to-star mass ratio is determined
more directly than the planet mass.
For microlensing and imaging planets, the plotted orbital distance is really the
sky-projected orbital distance.
For transiting planets, thousands of candidates
identified by the {\it Kepler} mission are missing; these
have unknown masses, but many of them are likely to be planets.
For timing planets, many are dubious cases of circumbinary
planets around evolved stars (see \S~\ref{sec:cbps}).
}
\label{fig:allplanets}
\end{figure}

\begin{figure}
\includegraphics[height=9cm]{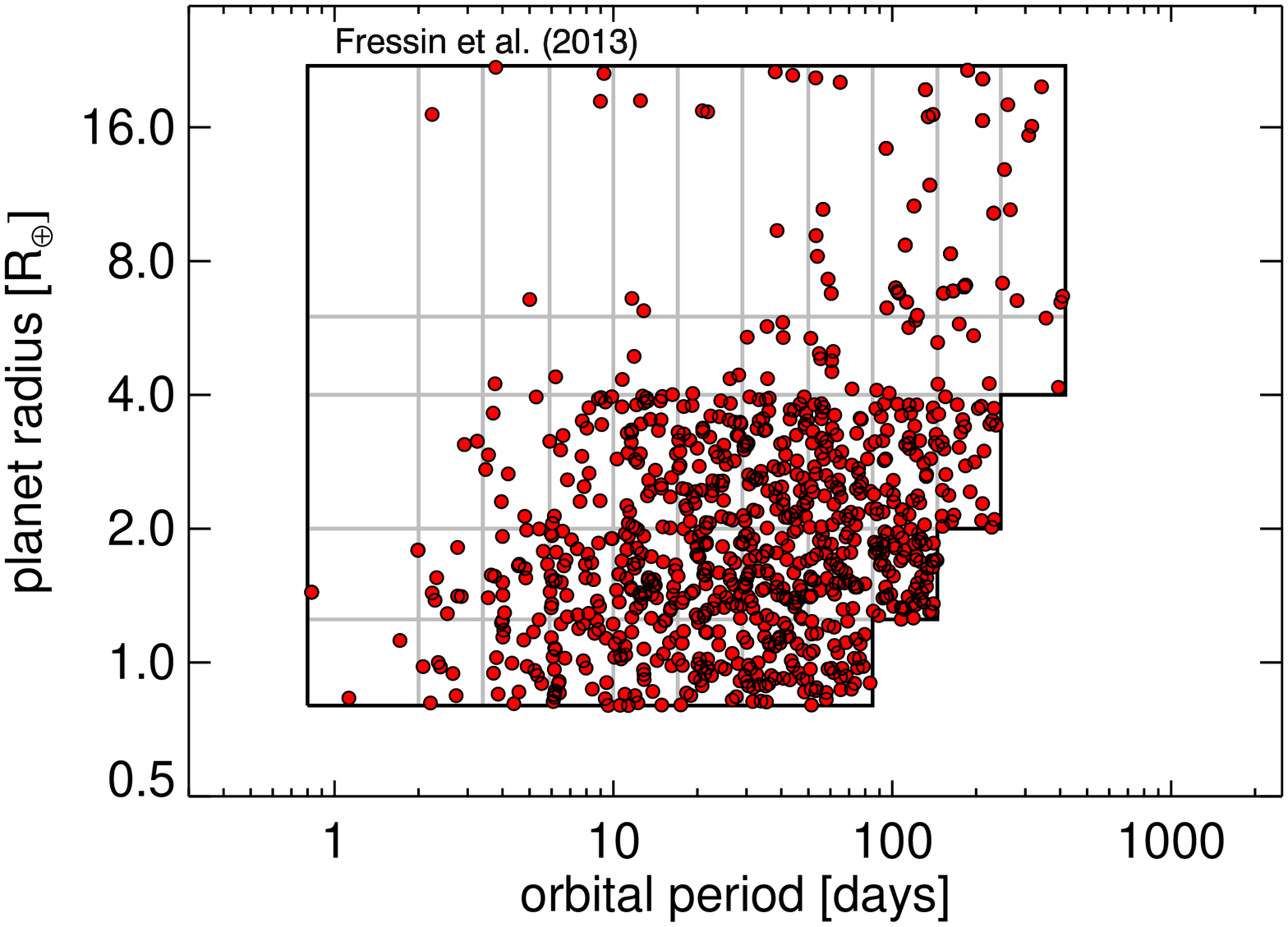}
\includegraphics[height=9cm]{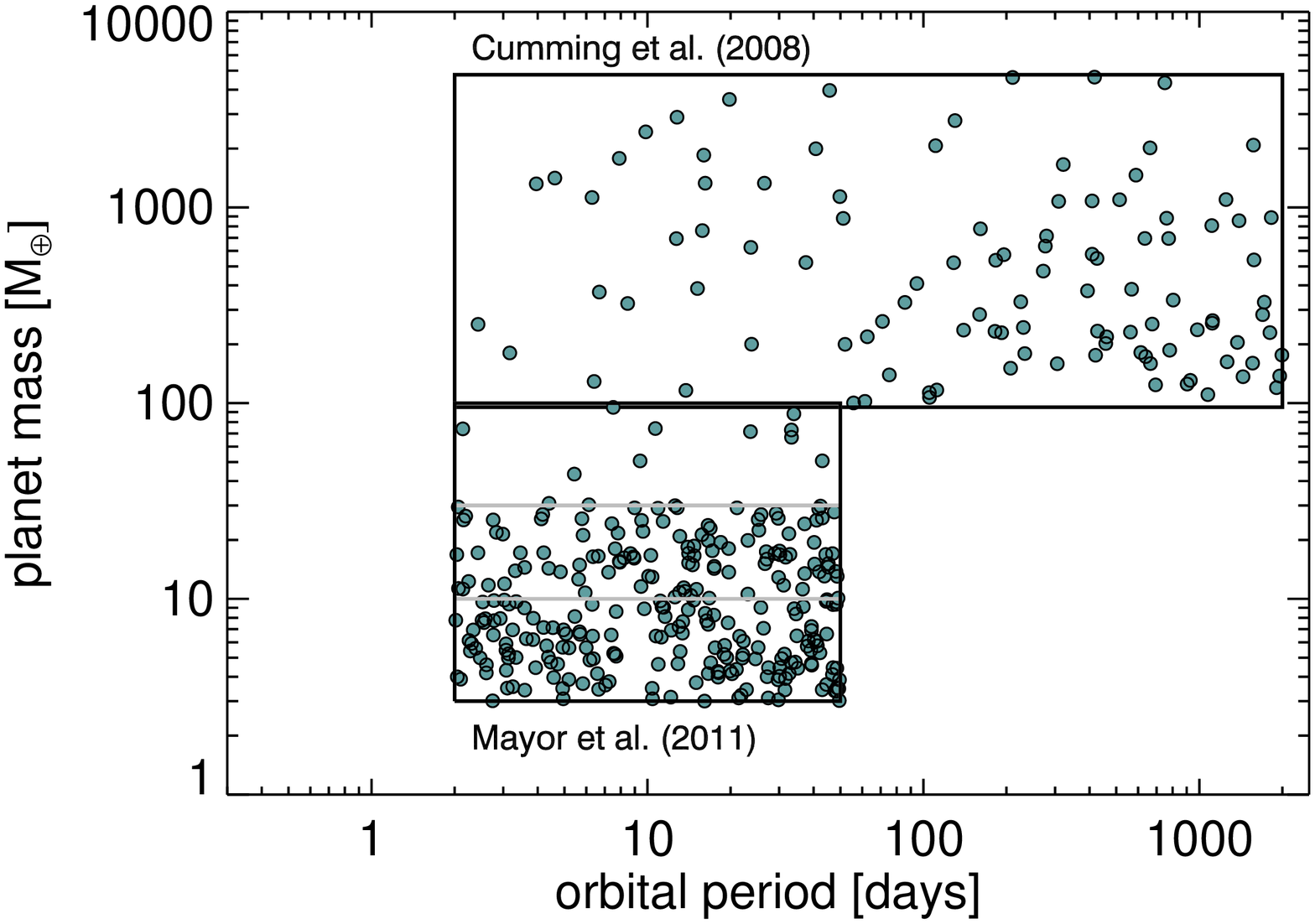}
\caption{Properties of a hypothetical sample of exoplanets around the
  nearest thousand FGK dwarfs, based on measured planet occurrence rates.
  {\it Top.}---Sizes and periods, based on the analysis of {\it
    Kepler} data by \cite{Fressin+2013}. The grid indicates
  the size and period range corresponding to each element in the
  matrix of planet occurrence rates provided by those authors.  {\it
    Bottom.}---Masses and periods, based on the Doppler
  results of \cite{Cumming+2008} for larger planets and
  \cite{Mayor+2011} for smaller planets.}
\label{fig:debiasplanets}
\end{figure}

Accurate measurement of occurrence rates requires a large sample of
stars that have been searched for planets and a good understanding of
the selection effects that favor the discovery of certain types of
planets. Because of these effects there can be major differences
between the observed sample and a truly representative
sample of planets. This is illustrated in Figs.~\ref{fig:allplanets}
and \ref{fig:debiasplanets}. Fig.~\ref{fig:allplanets} shows the
estimated masses and orbital distances of most of the known
exoplanets, labeled according to detection technique. Although this
provides a useful overview, it is misleading because it takes no
account of selection effects. Fig.~\ref{fig:debiasplanets} shows a
simulated volume-limited sample of planets around the nearest thousand
FGK dwarfs; the simulation is based on measurements of planet
occurrence rates that are described below.

Giant planets dominate the observed distribution, but smaller planets
are far more numerous in the simulated volume-limited sample. What
appears in the observed distribution to be a pile-up of hot Jupiters
(a few hundred Earth masses at a few hundredths of an AU) has vanished
in the volume-limited sample. In reality it is not clear that hot
Jupiters form a separate population from the giant planets at somewhat
larger orbital distances.

Table~\ref{tbl:occurrence} gives some key results of selected
studies. Before discussing these individual studies, we summarize the
basic picture:
\begin{enumerate}

\item Nature seems to distinguish between planets and brown
  dwarfs. For orbital periods shorter than a few years, companions
  with masses between 10 and 100~$M_{\rm Jup}$ are an order-of-magnitude
  rarer than less massive objects
  \citep{MarcyButler2000,GretherLineweaver2006,Sahlmann+2011}. This is
  known as the brown dwarf desert.
  
\item Nature also seems to distinguish between giant planets and
  smaller planets, with a dividing line at a radius of approximately
  4~$R_\oplus$ (0.4~$R_{\rm Jup}$) or a mass of approximately 30~$M_\oplus$
  (0.1~$M_{\rm Jup}$). Giants are less abundant than smaller planets
  per unit $\log R_{\rm p}$ or $\log M_{\rm p}$, within the period range that has
  been best investigated ($P\lsim 1$~year). Giants are also associated
  with a higher heavy-element abundance in the host star's photosphere
  and a broader eccentricity distribution than smaller planets.

\item Giant planets with periods shorter than a few years are found
  around $\approx$10\% of Sun-like stars, with a probability density
  nearly constant in $\log P$ between 2--2000 days. Giants are rarely
  found with $P<2$~days.

\item Smaller planets (1-4~$R_\oplus$) with $P\lsim 1$~year are found
  around approximately half of Sun-like stars, often in closely spaced
  multiplanet systems. The probability density is nearly constant in
  $\log P$ between about 10 and 300~days. For $P\lsim 10$~days the
  occurrence rate declines sharply with decreasing period.

\end{enumerate}

\begin{deluxetable}{lcrrc}
\tabletypesize{\small}
\tablewidth{0pt}
\tablecaption{Planet occurrence rates around FGK stars\label{tbl:occurrence}}

\tablehead{
\colhead{Study} &
\colhead{Technique} &
\colhead{Period range} &
\colhead{Size range} &
\colhead{Occurrence [\%]}
}

\startdata
\cite{Wright+2012}  & Doppler & $<10$~d & $>30~M_\oplus$ & $1.20\pm 0.38$ \\
\cite{Mayor+2011}  & Doppler & $<11$~d & $>50~M_\oplus$ & $0.89\pm 0.36$ \\
\cite{Cumming+2008} & Doppler & $<5.2$~yr & $>$100~$M_\oplus$ & $8.5\pm 1.3$ \\
  & & $<$100~d & $>$100~$M_\oplus$ & $2.4\pm 0.7$ \\
\cite{Howard+2010} & Doppler & $<$50~d & 3--10~$M_\oplus$ & $11.8_{-3.5}^{+4.3}$\\
  & & $<$50~d & 10--30~$M_\oplus$ & $6.5_{-2.3}^{+3.0}$ \\ 
\cite{Mayor+2011}  & Doppler & $<$50~d & 3--10~$M_\oplus$ & $16.6\pm 4.4$ \\
  & & $<$50~d & 10--30~$M_\oplus$ & $11.1\pm 2.4$ \\
  & & $<$10~yr & $>$50~$M_\oplus$ & $13.9\pm 1.7$ \\ 
\cite{Fressin+2013} & Transit & $<$10~d & 6--22~$R_\oplus$ & $0.43\pm 0.05$ \\
  & & $<$85~d & 0.8--1.25~$R_\oplus$ & $16.6\pm 3.6$ \\
  & & $<$85~d & 1.25--2~$R_\oplus$ & $20.3 \pm 2.0$ \\
  & & $<$85~d & 2--4~$R_\oplus$ & $19.9\pm 1.2$ \\
  & & $<$85~d & 1.25--22~$R_\oplus$ & $52.3\pm 4.2$ \\
Petigura et al.\ (2013) & Transit & 5--100~d & 1--2~$R_\oplus$&$26\pm 3$ \\
  & & 5--100~d & 8--16~$R_\oplus$ & $1.6\pm 0.4$
\enddata
\end{deluxetable}

\subsection{Doppler planets}
\label{sec:rv}

\cite{Cumming+2008} analyzed the results of a Doppler
campaign in which nearly 600 FGKM stars were monitored for 8
years. They fitted a simple function to the inferred probability
density for planets with mass $>100~M_\oplus$ and $P<5.5$~yr,
\be
\label{eq:cumming}
\frac{{\rm d}N}{{\rm d}\ln M_{\rm p}~d\ln P} \propto M_{\rm p}^{\alpha}P^{\beta},
\ee
and found $\alpha = -0.31\pm 0.20$, $\beta=0.26\pm 0.10$, and a
normalization such that the occurrence rate for $P<5.5$~yr and
$M_{\rm p}=0.3$-10~$M_{\rm Jup}$ is 10.5\%. This
function became a benchmark for many subsequent studies.
Udry et al.\ (2003) and \cite{Cumming+2008} found evidence for a few
features in the period distribution that are not captured in
Equation~(\ref{eq:cumming}). Specifically, there appears to be a
$\approx$2$\sigma$ excess of hot Jupiters, followed by a ``period
valley'' of lower probability. There is also a sharper rise in
probability as the period exceeds $\approx$1~year ($a\gsim 1$~AU). This
can be glimpsed in Figure~\ref{fig:allplanets}, but is not modeled in
Figure~\ref{fig:debiasplanets}.

\cite{Howard+2010} and \cite{Mayor+2011} explored the domain of
smaller planets with periods up to about 50~days, finding that such
planets are more common than giants. \cite{Mayor+2011} also
revealed that small short-period planets frequently come in compact
multiplanet systems, as later confirmed by the NASA {\it Kepler}
mission (see \S~\ref{sec:transits}).

Doppler surveys also revealed that giant-planet occurrence is strongly
associated with a high heavy-element abundance in the host star's
photosphere (Gonzalez 1997, Santos et al.\ 2004, Valenti \& Fischer 2005).
In contrast, small-planet occurrence is
not associated with high metallicity in FGK dwarfs \citep{Sousa+2008,
  Buchhave+2012}. Likewise, the occurrence of stellar companions
shows little or no correlation with metallicity \citep{Carney+2005,
  Raghavan+2010, GretherLineweaver2007}, supporting the notion that
giant planet formation is fundamentally different than binary-star
formation.

The M dwarfs are more plentiful per cubic parsec than any other type
of star, but they are more difficult to observe owing to their low
luminosities and relatively complex spectra. One secure result is that
they have fewer giant planets with $P\lsim 1$~year than FGK dwarfs
\citep{Butler+2004, Endl+2006, Johnson+2007, Bonfils+2013}. In
particular, \cite{Cumming+2008} showed that if the scaling law in
planetary mass and period of Equation~(\ref{eq:cumming}) holds for M
dwarfs, then the overall occurrence rate must be lower by a factor of
3--10 than that for FGK dwarfs.

Subgiant stars---those stars that have exhausted their core hydrogen
and swollen in size by a factor of a few---present an interesting
puzzle. Compared with main-sequence stars, subgiants are deficient in
short-period giant planets ($P\lsim 0.3$~year) and over-endowed with
longer-period giant planets ($\gsim 3$~year), and their planets tend
to have lower eccentricities \citep{Johnson+2010}. These differences
may be connected to the larger masses of the subgiants that were
surveyed; indeed, the surveys were designed to investigate
1.2-2.5~$M_\odot$ stars, which are not amenable to precise Doppler
observations when they are on the main sequence
\citep{Johnson+2006}. However, there is now evidence, based on
analyses of selection effects \citep{Lloyd2011} and kinematics
\citep{SchlaufmanWinn2013}, that the subgiants are not as massive as
previously thought. Possibly, the differences in planet populations
are attributable to stellar age or size rather than mass, although
none of the proposed scenarios can account for all the
observations. For example, the deficit of close-in planets may be
caused by the subgiants' enhanced rate of tidal dissipation, which
makes close-in planets vulnerable to orbit shrinkage and engulfment
\citep{VillaverLivio2009, SchlaufmanWinn2013}, but this would not
explain the over-abundance of giant planets in wider orbits.

\subsection{Transiting planets}
\label{sec:transits}

Selection effects are severe for transit surveys. In addition to the
obvious requirement that the planetary orbit be oriented nearly
perpendicular to the sky plane, there are strong biases favoring large
planets in tight orbits. In an idealized wide-field imaging survey,
the effective number of stars that can be searched for transits varies
as the orbital distance to the $5/2$ power and the planet radius to
the sixth power (Pepper et al.\ 2003). It is a struggle to bring
the occurrence rate to light when buried beneath such heavy biases.

The best opportunity to do so was provided by the {\it Kepler} space
telescope, which monitored $\approx$150,000~FGKM dwarfs for four
years and was uniquely sensitive to planets as small as Earth with
periods approaching one year. Ideally, occurrence rates would be based
on an analysis accounting for ($a$) the possibility that many
transit-like signals are actually eclipsing binaries, ($b$) the
efficiency of the transit-searching algorithm, ($c$) the uncertainties
and selection effects in the stellar parameters, and ($d$) the fact
that many target stars are actually multiple-star systems. No
published study has all these qualities. Significant steps toward this
ideal were made by \cite{Fressin+2013}, who estimated the number of
eclipsing binaries that are intermingled with the planetary signals,
and Petigura et al.\ (2013), who quantified the sensitivity of their
searching algorithm.

The flat distribution of planet occurrence in $\log P$, and the
relative deficit of planets inward of $P\lsim 10$~days, were seen
initially by \cite{Howard+2012} and supported by other studies,
including those cited above as well as by \cite{Youdin2011},
\cite{DongZhu2013}, and Silburt et al.\ (2014).  At the very shortest
periods ($P<1$~day) is a population of ``lava worlds'' smaller than
2~$R_\oplus$, which are approximately as common as hot Jupiters
\citep{SanchisOjeda+2014}.

Several groups have investigated how occurrence rates depend on the
host star's spectral type. \cite{Howard+2012} and Mulders et al.\ (2014) found
small planets to be more abundant around smaller stars (unlike giant
planets; see \S~\ref{sec:rv}). This finding was also supported by
\cite{DressingCharbonneau2013}, who focused exclusively on M dwarfs
because they offer the potential to detect the smallest planets.

One seemingly straightforward test is to compare the transit and Doppler
results for the occurrence rate of hot Jupiters, because those planets
are readily detected by both methods. The transit surveys give a
significantly lower rate (see Table~\ref{tbl:occurrence}), testifying
to the difficulty of accounting for selection effects. The discrepancy
might be attributable to differences in age, metallicity, or the
frequency of multiple-star systems between the two samples
\citep{Gould+2006, BaylissSackett2011, Wright+2012}.

Another interesting property of hot Jupiter hosts, seen in both
Doppler and transit data, is their tendency to have fewer
planets with periods between 10-100~days than stars without hot
Jupiters. \cite{Wright+2009} used Doppler data to show that systems
with more than one detected giant planet rarely include a hot Jupiter,
and \cite{Steffen+2012} used {\it Kepler} data to show that hot
Jupiters are less likely than longer-period giants to show evidence
for somewhat more distant planetary companions.

\subsection{Microlensing planets}

Microlensing surveys are more sensitive to distant planets than
Doppler and transit surveys. The sweet spot for planetary
microlensing is an orbital distance of a few astronomical units, a scale set by the
Einstein ring radius of a typical lensing star. Microlensing is
especially useful for probing M dwarfs, which are numerous enough to
provide most of the optical depth for lensing. Microlensing is also
capable of detecting planets that roam the galaxy untethered to any
star \citep{Sumi+2011}. Against these strengths must be weighed the
relatively small number of detected planetary events, the poor
knowledge of the host star's properties, and the difficulty of follow-up
observations to confirm or refine the planetary interpretation.

\cite{Gould+2010} analyzed a sample of 13 microlensing events that
led to six planet discoveries, allowing a measurement of the frequency
of planets with planet-to-star mass ratios $q\sim 10^{-4}$ (probably
Neptune-mass planets around M dwarfs) and sky-projected orbital
separations $s\sim 3$~AU. They expressed the result as
\be
\frac{{\rm d}N}{{\rm d}\log q~{\rm d}\log s} = 0.36\pm 0.15
\ee
This study, as well as those by \cite{Sumi+2011} and
\cite{Cassan+2012}, found that Neptune-mass planets are common around
M dwarfs; indeed, they are just as common as one would predict by extrapolating
Equation~(\ref{eq:cumming}). This sounds like good agreement until it is
recalled that Equation~(\ref{eq:cumming}) was based on Doppler surveys of
FGK dwarfs and that those same surveys found giant planets
($\gsim$0.3~$M_{\rm Jup}$) to be rarer around M dwarfs.
The combination of Doppler and microlensing surveys thereby suggests that
M dwarfs have few Jovian planets but a wealth of sub-Jovian planets at
orbital distances $\gsim 1$~AU \citep{ClantonGaudi2014}.

\subsection{Directly imaged planets}

The technological challenges associated with direct imaging have
limited the results to the outer regions of relatively young and
nearby stars. Young stars are preferred because young planets are
expected to be more luminous than older planets. In addition, direct
imaging is based on detection of planet luminosity, which must be
related to planet mass or size through uncertain theoretical
models. Some stunning individual systems have been reported
\citep{Marois+2010,Lagrange+2010}, but the surveys indicate that fewer
planets are found than would be predicted by extrapolating the
power-law of Equation~(\ref{eq:cumming}) out to 10-100~AU
\citep{Lafreniere+2007, NielsenClose2010, Biller+2013, Brandt+2014}.

\subsection{Earth-like planets}

How common are planets that are similar to the Earth? Until a few years ago the
main obstacle to an answer was the absence of data. Now that the
Doppler surveys and the {\it Kepler} mission have taken us to the
threshold of detecting Earth-like planets, the main obstacle is
turning the question into a well-posed calculation.

The criteria for being Earth-like are usually taken to be a range of
sizes or masses bracketing the Earth, and a range of values for the
incident flux of starlight $S = L_\star/4\pi a^2$, in which $L_\star$ is
the stellar luminosity and $a$ is the orbital distance. The range of
$S$ is chosen so that the planet is placed in the habitable zone, in which
liquid water would be stable on the surface of an Earth-like planet
(Kasting et al.\ 1993, Seager 2013, Guedel et al.\ 2014). The
$S$-range may depend on the type of star; this conclusion is based on models that account
for the stellar spectrum and its interaction with the presumed
constituents of the planet's atmosphere. Table~\ref{tbl:hz} summarizes
the recent estimates.

\begin{deluxetable}{ccccc}
\tabletypesize{\small}
\tablewidth{0pt}
\tablecaption{Occurrence rates of ``Earth-like planets''\label{tbl:hz}}

\tablehead{
\colhead{Type of} &
\colhead{Type of} &
\colhead{Approx.\ HZ} &
\colhead{Occurrence} &
\colhead{Reference} \\
\colhead{star} &
\colhead{planet} &
\colhead{boundaries$^\star$ [$S/S_\oplus$]} &
\colhead{rate [\%]} &
\colhead{}
}

\startdata
M & 1-10~$M_\oplus$ &
0.75-2.0 & $41_{-13}^{+54}$ & 1 \\
FGK & 0.8-2.0~$R_\oplus$ &
0.3-1.8 & $2.8^{+1.9}_{-0.9}$ & 2\\
FGK & 0.5-2.0~$R_\oplus$ &
0.8-1.8 & $34 \pm 14$ & 3\\
M & 0.5-1.4~$R_\oplus$ &
0.46-1.0 & $15^{+13}_{-6}$ & 4\\
M & 0.5-1.4~$R_\oplus$ &
0.22-0.80 & $48^{+12}_{-24}$ & 5\\
GK & 1-2~$R_\oplus$ &
0.25-4 & $22\pm 4$ & 6\\
FGK & 1-2~$R_\oplus$ &
$\sim$0.9-2.2$^\dagger$ & $\sim$0.01$^\dagger$ & 7 \\
FGK & 1-4~$R_\oplus$ &
0.35-1.0 & $6.4^{+3.4}_{-1.1}$ & 8 \\
G & 0.6-1.7~$R_\oplus$ & 0.51--1.95 & $1.7^{+1.8}_{-0.9}$ & 9 \\
\enddata
\tablecomments{
References:
(1) \cite{Bonfils+2013},
(2) \cite{CatanzariteShao2011},
(3) \cite{Traub2012},
(4) \cite{DressingCharbonneau2013},
(5) \cite{Kopparapu2013},
(6) Petigura et al.\ (2013),
(7) \cite{Schlaufman2014},
(8) Silburt et al.\ (2014),
(9) Foreman-Mackey et al.\ (2014).
In column 3, $S$ refers to the incident flux of starlight on the
planet, and $S_\oplus$ to the Earth's insolation.  All these works are
based on {\it Kepler} data except (1) which is based on the HARPS
Doppler survey, and (7) which is based on both {\it Kepler} and the
Keck Doppler survey.  $^\star$In many cases the actual HZ definitions
used by the authors were more complex; please refer to the original
papers for details.  $^\dagger$The result is much lower than the
others
because the author also required the Earth-sized planet to
have a long-period giant-planet companion.}
\end{deluxetable}

Much could be written about the virtues and defects of these studies
and why they disagree. It is probably more useful to make a broader
point. The occurrence rate is an integral of a probability density
over a chosen range of parameters, and in this case the uncertainty is
limited for the foreseeable future by our ignorance of the appropriate
limits of integration. For example, Petigura et al.\ (2013) reported
results differing by a factor of nine depending on the definition of
the habitable zone. Likewise, \cite{DressingCharbonneau2013} and
\cite{Kopparapu2013} used the same probability density to derive
occurrence rates differing by a factor of three.

In addition there may be other crucial aspects of habitability beyond
size and insolation. Must a planet's atmosphere be similar to that of
the Earth \citep{Zsom+2013}? Must the planet enjoy the white light of
a G star, as opposed to the infrared glow of an M dwarf
\citep{Tarter+2007}? For now, the best approach is probably to report
the probability density near Earth-like parameters
(Foreman-Mackey et al.\ 2014) and allow astrobiologists the
freedom of interpretation.

\section{ORBITAL DISTANCE AND ECCENTRICITY}
\label{sec:pe}

Planets in the Solar System have orbital distances between 0.3 and
30~AU, a much narrower range than is allowed from basic physical
considerations. For the minimum distance, a planet must be outside the
Roche limit ($a \gsim 0.01$~AU), the distance within which the star's
tidal force prevents a planet from maintaining hydrostatic
equilibrium. The Roche limit can be stated conveniently in terms of
the orbital period and planetary mean density \citep{Rappaport+2013}
and corresponds to $P\gsim 12$~hours for a gas giant and $\gsim$5~hours for
a rocky planet. As for the maximum distance, stable orbits are limited
to $a \lsim$$10^5$~AU by perturbations from random encounters with
other stars as well as the overall Galactic tidal field
(Heisler \& Tremaine 1986, Veras et al.\ 2009).

Regarding orbital shapes, in Newtonian gravity the eccentricity $e$
may range from 0~to~1, and the orbital energy is independent of
eccentricity for a given semimajor axis. This is why the nearly circular orbits in the Solar
System seem to require a special explanation, involving either the
initial conditions of planet formation or processes that tend to
circularize orbits over time.

\subsection{Doppler planets}

Most of our knowledge of eccentricities comes from the Doppler
technique. Many orbits are consistent with being circular; for some of
the closest-in planets the eccentricities have upper bounds on the order
of $\sim$$10^{-3}$. On the other side of the distribution are giant
planets with eccentricities of 0.8--0.9. Among the systems with very
secure measurements, the most eccentric orbit belongs to HD~80606b
($e=0.93$; Naef et al.\ 2001, H\'{e}brard et al.\ 2010).

\begin{figure}
\includegraphics[height=7cm]{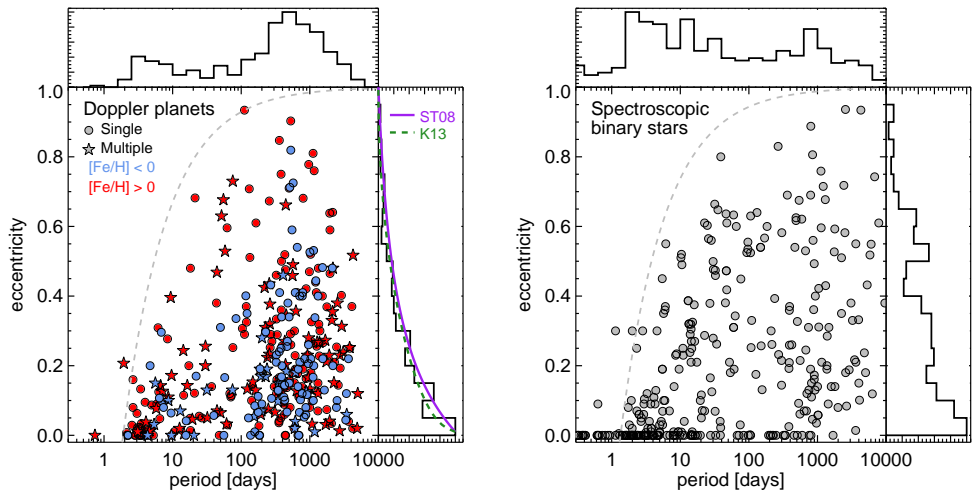}
\caption{{\it Left.}---Orbital eccentricity and period for Doppler planets detected
with a signal-to-noise ratio $>$10,
based on an October 2014 query of the {\tt exoplanets.org} database
\citep{exoplanetsorg}. The distribution has not been
corrected for selection effects. Red and blue points represent host star
metallicities above and below (respectively) [Fe/H]~$=0$.
Stars represent systems in which more than one
planet has been detected, whereas circles are single-planet systems.
The dashed line shows where the minimum orbital separation would be 0.03~AU
for a Sun-like star. The purple curve is
Equation~(\ref{eq:shen-turner}), taken from \cite{ShenTurner2008},
and the green dashed curve is Equation~(\ref{eq:beta}) from \cite{Kipping2013}.
{\it Right.}---Eccentricity and period for spectroscopic binary stars
from the SB9 catalog \citep{Pourbaix+2004}. A random subsample was
drawn, to match the number of exoplanet systems.
The dashed line shows where the minimum orbital separation would be 0.03~AU
for a pair of Sun-like stars. In the marginalized eccentricity
distribution, the very large number of systems consistent with $e=0$
is not shown.
}
\label{fig:PeriodEccentricity}
\end{figure}

The left panel of Figure~\ref{fig:PeriodEccentricity} shows the observed
eccentricity-period distribution. No correction was made for selection
effects; the apparent excess of planets with periods of a few days is
the result of the strong bias favoring the detection of short-period
giant planets. The data points have been given colors and shapes to
convey the metallicity of the host star, and whether or not additional
planets have been detected, for reasons to be described. For
comparison, the right panel of Figure~\ref{fig:PeriodEccentricity} shows
the $e$--$P$ distribution of the same number of eclipsing binary
stars, which have been drawn randomly from the SB9 catalog \citep{Pourbaix+2004}.

At a glance, the planetary and stellar distributions are similar: both
have preferences for circular orbits at the shortest periods, and
eccentricity distributions that broaden with increasing period. The
dashed lines are approximately where the closest approach between the
two bodies is 0.03~AU (6.5~$R_\odot$), which seems to set an upper
limit for the eccentricity as a function of period in both cases. The
likely explanation for this eccentricity envelope is tidal
dissipation. Time-varying tidal forces lead to periodic distortions
and fluid flows on both bodies. The inevitable friction associated
with those motions gradually dissipates energy, one consequence of
which is the circularization of the orbit \citep[see,
e.g.,][]{Ogilvie2014}.

For planets, the eccentricity distribution is often modeled with a
Rayleigh function,
\be
\frac{{\rm d}N}{{\rm d}e} = \frac{e}{\sigma^2} \exp\left(-\frac{e^2}{2\sigma^2} \right),
\ee
motivated in part by \cite{JuricTremaine2008},
who found this to be the expected long-term
outcome of planet-planet interactions in closely-spaced multiplanet
systems. However, a Rayleigh distribution only fits the
high-eccentricity portion of the observed distribution.

\cite{ShenTurner2008} proposed a different fitting function,
\be
\label{eq:shen-turner}
\frac{{\rm d}N}{{\rm d}e} \propto \frac{1}{(1+e)^a} - \frac{e}{2^a},
\ee
which gives a good single-parameter description of the data with $a=4$ (the purple curve in
Fig.~\ref{fig:PeriodEccentricity}). \cite{WangFord2011} considered
this model as well as more complex functions, such as a combination of
a Rayleigh function (which fits the high-$e$ end) and an exponential
function (for the more circular systems), finding that the sample
sizes were insufficient to make distinctions between these various models.
Hogg et al.\ (2010) and \cite{Kipping2013} advocated the two-parameter beta
distribution,
\be
\label{eq:beta}
\frac{{\rm d}N}{{\rm d}e} = \frac{\Gamma(a+b)}{\Gamma(a)\Gamma(b)} e^{a-1} (1-e)^{b-1},
\ee
which has a number of desirable mathematical properties
and provides a good fit for $a=0.867$, $b=3.03$ (the green
curve in Fig.~\ref{fig:PeriodEccentricity}).

\begin{figure}
\includegraphics[height=8cm]{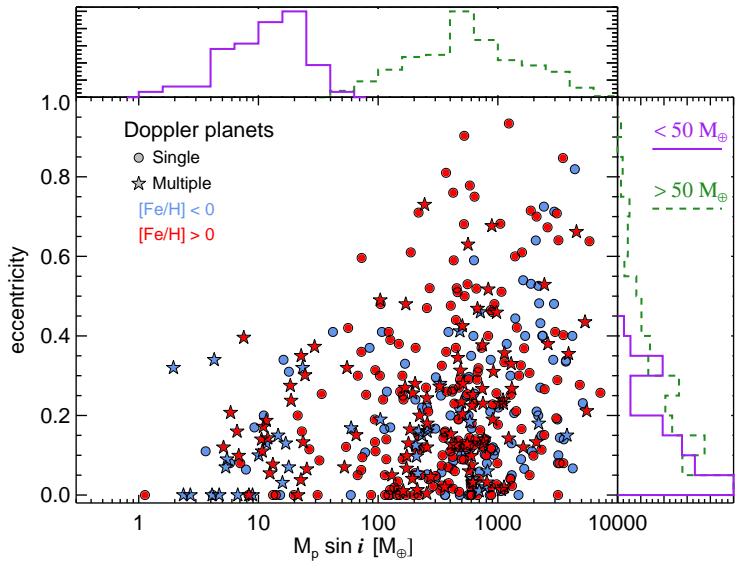}
\caption{Minimum mass ($M_{\rm p}\sin i$) and eccentricity of planets
  detected by the Doppler method with a signal-to-noise ratio $\geq$5,
  based on an October 2014 query of the
  {\tt exoplanets.org} database \citep{exoplanetsorg}. The distribution is {\it not}
  corrected for selection effects. Colors and
  symbol shapes follow the same conventions as in
  Fig.~\ref{fig:PeriodEccentricity}. In the marginalized histograms of
  period and $M_{\rm p}\sin i$, the blue curve is for planets with $M_{\rm p}\sin
  i < 50~M_\oplus$ and the orange curve is for planets with $M_{\rm p}\sin i
  > 50~M_\oplus$.}
\label{fig:MassEccentricity}
\end{figure}

Some authors have noted differences in the period--eccentricity
distributions among various subsamples of planets:
\begin{enumerate}

\item \cite{Wright+2009} noted that systems for which more than one
  planet has been detected tend to have lower eccentricities than
  those for which only a single planet is known. The effect is perhaps
  best articulated as an absence of the very highest eccentricities
  from the subsample of multiplanet systems, as can be seen by
  comparing the circles (singles) and stars (multiples) in
  Figure~\ref{fig:PeriodEccentricity}. It might be explained as a
  ``natural selection'' effect: In compact multiplanet systems, low
  eccentricities are more compatible with long-term dynamical
  stability.  \cite{LimbachTurner2014} went further to show that among
  the multiplanet systems, the median eccentricity declines with the
  number of known planets: $e_{\rm m} \propto N^{-1.2}$.

\item \cite{DawsonMurrayClay2013} found that giant planets around
  metal-rich stars have higher eccentricities than similar planets
  around metal-poor stars. This is manifested in
  Figure~\ref{fig:PeriodEccentricity} as a deficiency of blue
  (metal-poor) points with $P\approx10$-100~days and $e\gsim
  0.4$. \cite{DawsonMurrayClay2013} theorized that metal-rich stars
  had protoplanetary disks that were rich in solid material and
  consequently formed more planets that could engage in planet-planet
  interactions and excite eccentricities. \cite{Adibekyan+2013}
  reported another intriguing trend involving metallicity: The orbital
  periods tend to be longer around metal-poor stars than around
  metal-rich stars.

\item Smaller planets tend to have lower eccentricities
  \citep{Wright+2009,Mayor+2011}, as seen in
  Figure~\ref{fig:MassEccentricity}. Small planets are also more likely
  than giant planets to be found in multiplanet systems
  \citep{Latham+2011}, reinforcing the association between multiplanet
  systems and lower eccentricity. An important caveat is that
  eccentricity is especially difficult to measure when the
  signal-to-noise ratio is low, as is usually the case for small
  planets (see, e.g., Shen \& Turner 2008, Zakamska et al.\ 2011). For
  this and other reasons, Hogg et al.\ (2010) recommended modeling the
  eccentricity distribution on the basis of Bayesian analysis of the ensemble
  of Doppler measurements, rather than on a histogram of ``best
  values'' of each system's eccentricity; however, no such analysis
  has yet been published.

\end{enumerate}

\subsection{Transiting planets}

Most of the known planets smaller than Neptune were discovered by the
{\it Kepler} mission using the transit technique, which does not
directly reveal the orbital eccentricity. The usual way to measure the
eccentricity of a transiting planet is to undertake Doppler
observations, but for {\it Kepler} systems this has proven impractical
because the stars are too faint. Even in those few cases in which Doppler
observations have been performed, the signal-to-noise ratio is too low
for secure eccentricity measurement \citep{Marcy+2014}. Nevertheless,
enticed by the prospect of unveiling the eccentricity distribution for
small planets, investigators have pursued other approaches.

\subsubsection{Transit durations}
\label{sec:transit-durations} 
The transit duration is approximately
\be
\label{eq:transit-duration} 
T = \left( \frac{R_\star P}{\pi a} \sqrt{1-b^2} \right) \frac{\sqrt{1-e^2}}{1+e\sin\omega},
\ee
where $\omega$ is the argument of pericenter and $b$ is the impact
parameter, the minimum sky-plane distance between
the star and planet expressed in units of the stellar radius
$R_\star$. If the factor in parentheses is known, then measuring $T$
provides some eccentricity information. This type of analysis can be
performed for an individual transiting planet
\citep{DawsonJohnson2012}, a system of multiple transiting planets
\citep{Kipping+2012} or an ensemble of different stars with transiting
planets (Ford et al.\ 2008, Burke 2008).

\cite{Moorhead+2011} applied this method early in the {\it Kepler}
mission and found tentative evidence that smaller planets tend to have
lower eccentricities than giant planets, and that systems of multiple
transiting planets have lower eccentricities than those for which only
a single transiting planet was known. \cite{Kane+2012} and Plavchan et
al.\ (2014) attempted to improve upon this analysis by taking
advantage of several years of additional {\it Kepler} data, but the
results are limited by systematic uncertainties in the stellar
properties rather than data volume. \cite{SliskiKipping2014} examined
the small subset ($\approx$100) of planet-hosting stars that are
especially well characterized, due to the detection of asteroseismic
oscillations, and found that the eccentricity distribution could not be
distinguished from that of the Doppler sample.

\subsubsection{Dynamical modeling}
\label{subsec:dynamical-modeling}

In multiplanet systems, the non-Keplerian effects of planet-planet
interactions cause perturbations in the transit times and durations
that depend on the planets' masses and orbital elements
\citep{HolmanMurray2005, Agol+2005}. With enough data, those
parameters may be determined or bounded. This method works best when
all the relevant planets are transiting; otherwise the solutions are
not unique (see, e.g., Ballard et al.\ 2011, Veras et al.\ 2011).

Applications of this method have confirmed that the eccentricities are
low in compact multiplanet systems. For example, \cite{Lissauer+2013}
showed that at least five of the six known planets in the Kepler-11 system
have eccentricities $<$0.1, with the best-constrained planet having
$e<0.02$. Similar results were obtained for the Kepler-30 and
Kepler-36 systems \citep{SanchisOjeda+2012, Carter+2012}. It is not clear,
though, whether the systems that were selected for these intensive
studies are representative of small-planet systems.

Lithwick et al.\ (2012) made an important advance enabling the study
of a larger sample of small planets. They devised an analytic theory
for the transit-timing variations of pairs of planets near a
first-order mean-motion resonance (e.g., a period ratio of 2:1 or
3:2). In such a situation, seen in dozens of {\it Kepler} systems, the
sequence of timing anomalies is a nearly sinusoidal function of
time. The amplitude and phase of this function depend on a combination
of the planet masses and eccentricities. In general the mass and
eccentricity of a given planet cannot be determined uniquely, but the
method can be used to constrain the eccentricity distribution of a
population. \cite{WuLithwick2013} applied this method to 22 pairs of
small planets. They found that 16 pairs are consistent with very low
eccentricities; the best-fitting Rayleigh distribution had a mean
eccentricity of 0.009. The other 6 pairs have higher eccentricities,
probably in the range of 0.1--0.4.

\section{MULTIPLANET SYSTEMS}
\label{sec:multis}

\subsection{Multiplicity}

We do not know the average number of planets orbiting a given star;
this is due to the incompleteness of the detection
techniques. However, every technique that has successfully found
single planets has also provided at least one case of a multiplanet
system. The first known multiplanet system, PSR~1257+12, was shown to
have multiple planets on the basis of time delays in the arrival of
radio pulses from the central pulsar caused by the pulsar's orbital
motion \citep{WolszczanFrail1992}. The data were precise enough to
reveal non-Keplerian effects that were due to planet-planet
interactions, allowing \cite{KonackiWolszczan2003} to derive orbital
eccentricities, inclinations, and true masses (as opposed to $M_{\rm p}\sin
i$) for two of the three known planets. This gave a preview of the
type of architectural information that would only become available a
decade later for planets around normal stars.

Microlensing surveys have found two different two-planet systems
\citep{Gaudi+2008, Han+2013}, both of which feature giant planets with
orbital distances of a few astronomical units. Direct imaging has
delivered a spectacular system of four giant planets that have orbital
distances of $\approx$12-70~AU (HR~8799; Marois et al.\ 2010). Because
of the slow orbital motion, only small arcs of the four orbits have
been mapped out, preventing complete knowledge of the system's
architecture. The large planet masses and closely spaced orbits have
prompted theorists to suggest that long-term dynamical stability is
possible only if the planets are in particular resonances
\citep{FabryckyMurrayClay2010, GozdziewskiMigaszewski2014}.

Until recently, the Doppler technique provided the largest cache of
multiplanet systems \citep{Wright+2009, Wittenmyer+2009,
  Mayor+2011}. Then the {\it Kepler} mission vaulted the transit
technique to the top position, identifying hundreds of transiting
multiplanet systems \citep{Latham+2011, Burke+2014}.  Both the Doppler
and {\it Kepler} surveys have shown that compact multiplanet systems
with $P \lesssim 1$~year are usually composed of planets smaller than
Neptune and that Jovian planets are uncommon
\citep{Wright+2009,Latham+2011}. Also noteworthy is that many planets
in compact systems have very low densities ($\lsim$1~g~cm$^{-3}$; see,
e.g., Lissauer et al.\ 2013, Wu \& Lithwick 2013, Masuda 2014), which
have mainly been revealed through the transit-timing method. The
prevalence of small, low-density planets could be a clue about their
formation. It may also be related to the requirement for long-term
stability. When a closely packed system suffers from dynamical
instability, massive planets are more likely to eject one another,
whereas smaller planets are more likely to collide until the system
stabilizes \citep{FordRasio2008}.

\subsection{Orbital spacings}

In the Solar System, the orbital spacings of planets and their
satellites are organized around two themes. The first theme is the
nearly geometric progression of orbital distances, which is famously
encoded in the Titius-Bode ``law'':
\be
\label{eq:tb}
a_n~{\rm [AU]} = 0.4 + 0.3\times 2^n.
\ee
This charmingly simple formula loses much of its appeal after learning
that $n$ does not simply range over integers, but rather $n=-\infty,
0, 1, 2, \cdots$; $n=3$ corresponds to the asteroid belt rather than a
planet; and the formula fails for Neptune. Whether this formula has
any deep significance is a question that has bewitched investigators
since the 18th century. However, there is a sound physical reason to
expect a constant ratio of orbital spacings, rather than constant
differences or a random pattern: the scale-free nature of
gravitational dynamics \citep{HayesTremaine1998}.

The second theme is mean-motion resonance, characterized by period
ratios that are nearly equal to ratios of small integers. This is seen
most clearly in the satellites of the giant planets, most famously the
1:2:4 resonances of the inner Galilean satellites of
Jupiter. \cite{RoyOvenden1954} used the ensemble of satellite systems
known then to show that the period ratios are closer to
low-integer ratios than chance alone would produce, and
\cite{GoldreichSoter1966} summarized the arguments that at least some
of those configurations represent the outcome of long-term tidal
evolution.

\begin{figure}
\includegraphics[height=9cm]{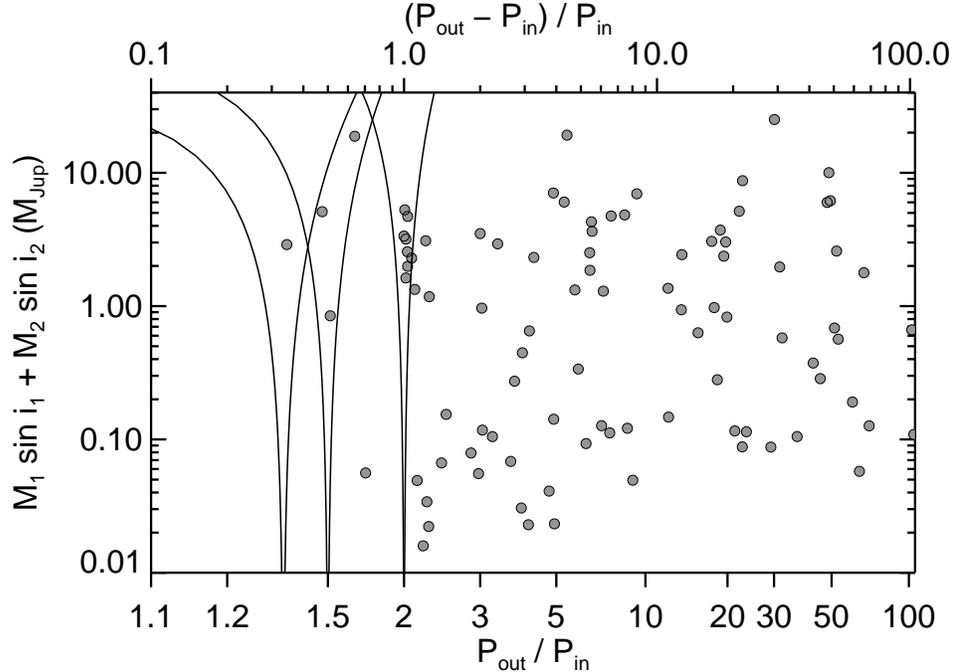}
\caption{ Period ratios of Doppler planet pairs.  The horn-shaped
  curves show the approximate resonance boundaries, $|P_{\rm out} - j
  P_{\rm in}|/P_{\rm in} = 0.05 ((M_1 + M_2)/M_{\rm Jup})^{1/2}$, for
  each $(j+1):j$ resonance. Pairs with larger planet masses are more
  often found in resonances. Based on an August 2014 query of {\tt
    exoplanets.org} \citep{exoplanetsorg}.  }
\label{fig:pratsrv}
\end{figure}

Geometric progressions and mean-motion resonances have also been seen
in exoplanetary systems. We begin this discussion by inspecting the
orbital spacings in the Doppler systems (which mainly involve giant
planets) and the {\it Kepler} systems (which mainly involve smaller
planets). Figure~\ref{fig:pratsrv} shows the period ratios of all
planet pairs discovered by the Doppler technique. For planet pairs
with a total mass exceeding 1~$M_{\rm Jup}$, there is a clustering of
points within the 2:1 resonance that is unlikely to be a statistical
fluke \citep{Wright+2011c}.

\begin{figure}
\includegraphics[height=9cm]{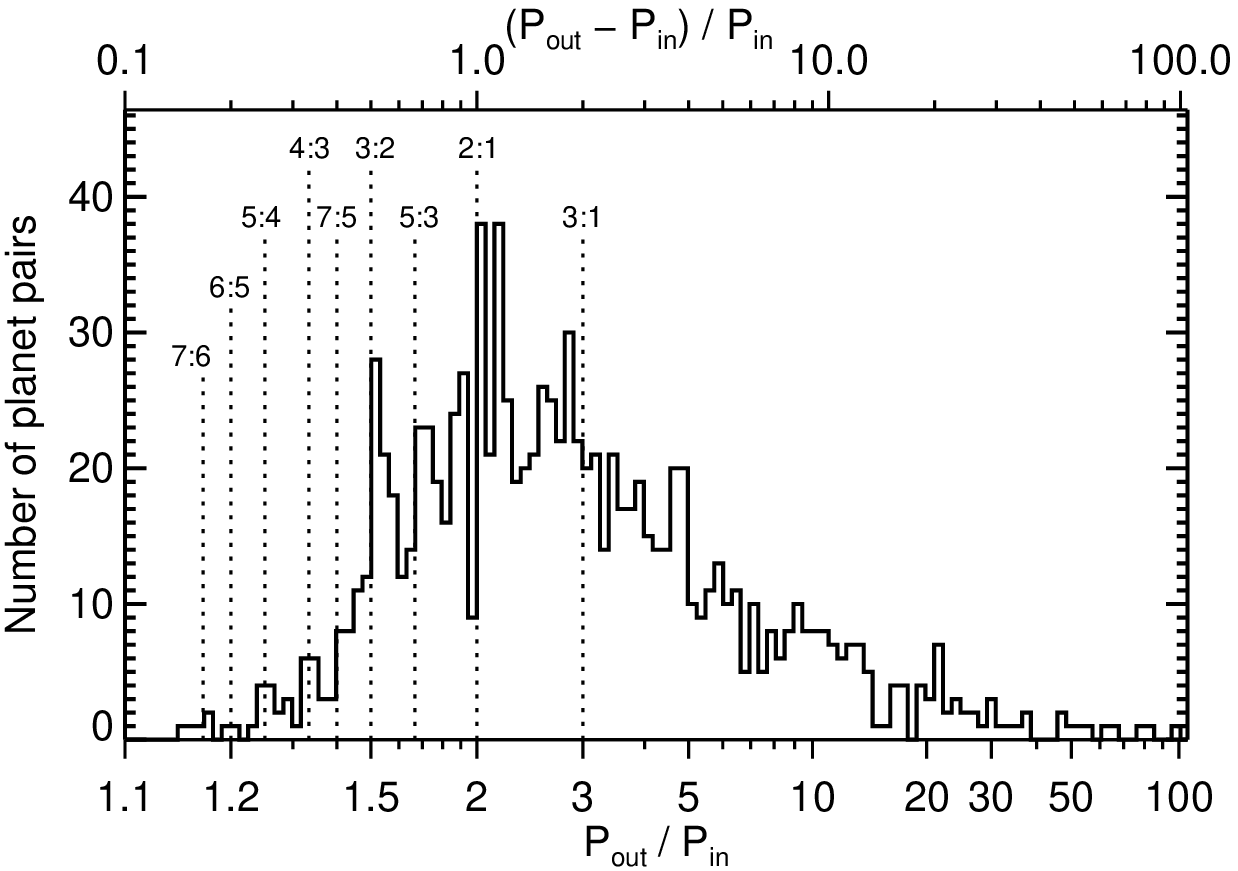}
\caption{ Period ratios of {\it Kepler} planet pairs, based on the
  sample by \cite{Burke+2014}.  The top axis indicates the fractional
  period difference, and the histogram bin widths represent 1/40 of
  the logarithm of this quantity.  The bottom axis indicates the
  period ratio.  }
\label{fig:prat}
\end{figure}

The situation is different for lower-mass
planets. Figure~\ref{fig:prat} shows the period ratios of {\it Kepler}
multiplanet systems \citep{Burke+2014}. Within this sample, dominated
by small planets with periods shorter than a few hundred days, there
is only a weak preference for period ratios near resonances
\citep{Lissauer+2011}. This finding was initially surprising because
the Doppler systems had already shown a stronger preference for
resonances, and because the theory of disk migration---by which
wide-orbiting planets are conveyed to smaller orbits through
gravitational interactions with the protoplanetary disk---predicted
that planet pairs would often be caught into resonances
\citep{GoldreichTremaine1980, LeePeale2002}. Retrospectively, it has
been possible to find ways to avoid this faulty conclusion
\citep{GoldreichSchlichting2014, ChatterjeeFord2015}.

Closer inspection of Figure~\ref{fig:prat} reveals subtle features:
there appears to be a tendency to avoid exact resonances
\citep{VerasFord2012}, and prefer period ratios slightly larger
than the resonant values \citep{Fabrycky+2014}. Specifically, there is
a deficit of pairs with period ratios of $1.99$--$2.00$ and an excess
of pairs with period ratios of $1.51$ and $2.02$. This finding was
anticipated by \cite{TerquemPapaloizou2007} as a consequence of tidal
dissipation within the inner planet that would cause the orbits to
spread apart, an idea that was picked up again by
\cite{BatyginMorbidelli2013} and \cite{Lithwick+2012} once the {\it
  Kepler} sample became available.

Despite this interest in the nearly-resonant systems, the main lesson
of Figure~\ref{fig:prat} is that resonances are uncommon among small
planets with periods shorter than a few years. The most commonly
observed period ratios are in the range of 1.5--3.0, a good match to
the period ratios of the Solar System (1.7--2.8). The smallest
confirmed period ratio of 1.17 belongs to Kepler-36. That system
features two unusually closely spaced planets with densities differing
by a factor of 8 and transit times exhibiting large and erratic
variations \citep{Carter+2012}. The fits to the data suggest the
planets' orbits are chaotic, with a Lyapunov time of approximately 20
years, and yet they manage to remain stable over long timescales
\citep{Deck+2012}.

\begin{figure}
\includegraphics[height=13cm]{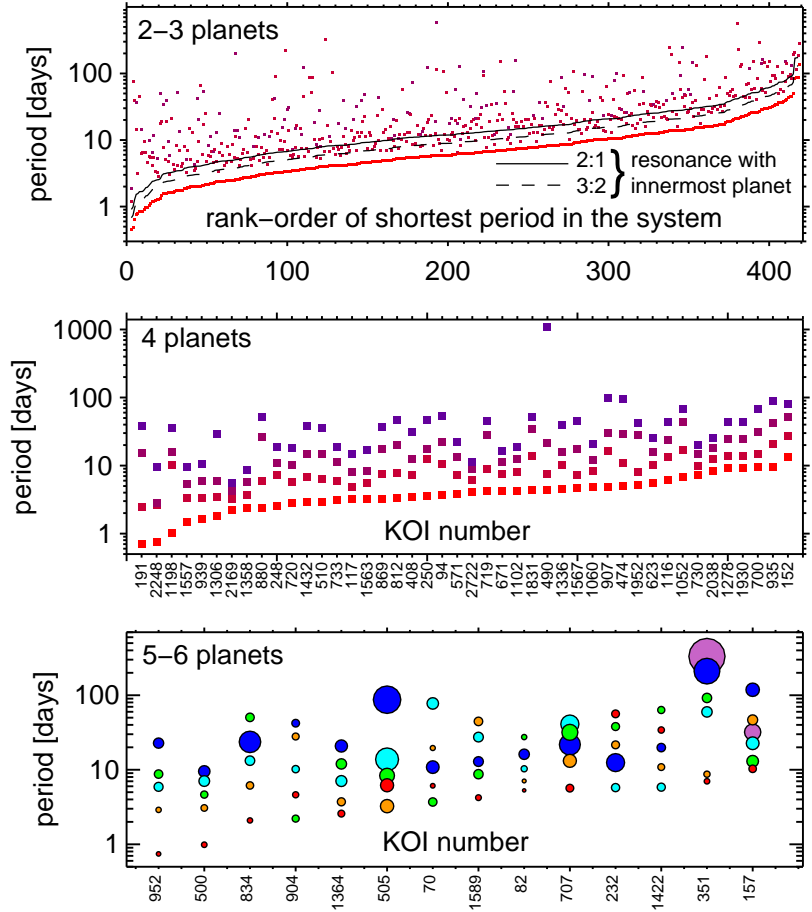}
\caption{Multitransiting planetary systems from \emph{Kepler}. Each
  symbol represents one planet. All the symbols along a given vertical
  line belong to the same star. The systems are ordered
  horizontally according to the innermost planet's orbital period.
  {\it Top.}---Systems with two or three detected transiting planets
  ($N_{\rm tra}= 2$-3).  {\it Middle.}---Systems with $N_{\rm tra}=4$.
  {\it Bottom.}---Systems with $N_{\rm tra}=5$-6 planets. The symbol
  size is proportional to planetary radius.}
\label{fig:gallery}
\end{figure}

One way in which the periods in the Solar System differ from a purely
geometric progression is that the period ratio tends to be larger for
the more distant planets; this is encoded in Equation~(\ref{eq:tb}) by the
constant term of $+0.4$. This progressive widening of period ratios
has not been seen in exoplanetary systems. In fact
\cite{SteffenFarr2013} found an opposite trend: The period ratios tend
to be larger for the very closest planets. This can be seen in
Figure~\ref{fig:gallery}, which displays the periods of the {\it
  Kepler} multiplanet systems. For those systems with an innermost
planet at $P<3$~days, the period ratio between the innermost two
planets is larger than that of other systems. Perhaps the innermost
planets are being pulled closer to the star by tidal dissipation
\citep{TeitlerKonigl2014, Mardling2007}.

\subsection{Mutual inclination}

The coplanarity of the Solar System is compelling evidence that the
planets originated within a flat rotating disk. The underlying logic
is that the laws of dynamics allow for higher mutual inclinations; the
current Solar System does not have any mechanism for damping
inclinations; therefore, the low inclinations must be attributed to
the initial conditions.

This is probably correct, although the argument is not watertight
because large mutual inclinations tend to make systems more vulnerable
to dynamical instability. For example, in the Kozai-Lidov instability,
mutual inclinations excite eccentricities and eventually cause the
orbits to intersect. This raises the possibility that highly inclined
systems are created in abundance but tend to destroy
themselves. \cite{VerasArmitage2004} derived inclination-dependent
formulas for the minimum separation between planetary orbits
compatible with long-term stability. They predict that even small
mutual inclinations are potentially lethal to closely packed systems
of small planets. For example, the minimum spacing between two
$3$~$M_\oplus$ planets around a Sun-like star is twice as large for
$\Delta i = 5^\circ$ as it is for coplanar orbits.

There are a few systems for which investigators have used the
requirement of long-term stability to place bounds on mutual
inclinations. For 47~Uma, \cite{Laughlin+2002} showed that the orbits
of the two giant planets are likely to be inclined by less than
40$^\circ$. Likewise, \cite{Nelson+2014} found the orbit of the
innermost planet of 55~Cnc to be inclined by $\lsim$40$^\circ$ from
the outer planets (which were assumed to be coplanar).

Directly measuring mutual inclinations is challenging. The astrometric
technique could reveal mutual inclinations, but it has proven
difficult to achieve the necessary precision. Likewise, direct imaging
can do the job, but the only directly imaged multiplanet system
(HR~8799) has not been observed for long enough. The Doppler technique
is sensitive to all the planets in a multiplanet system but is
generally blind to mutual inclinations. The transit technique would
likely miss many planets in a noncoplanar system, and even when
multiple transiting planets are seen, they could be mutually inclined
with nonparallel trajectories across the stellar disk. Nevertheless,
investigators have managed to learn about mutual inclinations by
exploiting loopholes in these statements, combining the results of
different techniques, and considering population statistics, as we
shall now discuss.

\subsubsection{Individual systems}

Although Doppler data do not ordinarily reveal mutual inclinations,
there are a few cases in which the planet-planet interactions produce
detectable perturbations in the host star's motion beyond the simple
superposition of Keplerian orbits. In one system, GJ~876, these
perturbations are large enough to have allowed the usual fitting
degeneracies to be broken and the mutual inclination to be
determined. In this system there are two giant planets in a 2:1
resonance along with a smaller inner planet and a Neptune-mass outer
planet. The data set is rich enough to have attracted many
investigators \citep{RiveraLissauer2001, Ji+2002, BeanSeifahrt2009,
  Correia+2010, Rivera+2010, Baluev2011} who all concluded that the
mutual inclination between the two giants is $\lsim$5$^\circ$.

Planet-planet perturbations have been detected in a few other systems,
but the Doppler data have too short a timespan to pin down the mutual
inclination. One example is the two-planet system HD~82493, for which
the data do not specify the mutual inclination, although they do
specify the sky-plane inclination of both orbits if they are assumed
to be coplanar \citep{Tan+2013}. What makes this interesting is that
the morphology of the star's directly imaged debris disk suggests that
it has the same inclination as the planetary orbits to within
$4^\circ$ \citep{Kennedy+2013}. A second system in which the
planet-disk inclination was measured is $\beta$~Pic. Both the planetary
orbit and the disk orientation have been measured through direct
imaging, and the planet is aligned with the inner portion of the disk
to within a few degrees \citep{Lagrange+2012, Nielsen+2014,
  Macintosh+2014}. The planet-disk alignment seen in these two systems
suggests that the systems are relatively flat.

Another special case is the three-planet system $\Upsilon$~And, for
which \cite{McArthur+2010} used a combination of Doppler and
astrometric data to show that the orbits of the two most massive
planets are mutually inclined by $30\pm 1^\circ$. This remains the
only direct demonstration of a mutual inclination greater than a few
degrees and is therefore an important result that deserves to be
checked against additional data. Earlier, the same technique gave an
inclination near 90$^\circ$ for one of the GJ~876 planets
\citep{Benedict+2002}, but this result was cast into doubt after being
found inconsistent with the seemingly more reliable results of the
Doppler analysis \citep{BeanSeifahrt2009}.

The four-planet system Kepler-89 was kind enough to schedule a
planet-planet eclipse during the {\it Kepler} mission, when one
planet eclipsed another planet while both were transiting the
star. \cite{Hirano+2012} used the timing and duration of this
extraordinary event to show that the angle between the two orbits was
smaller than a few degrees. The next such eclipse in Kepler-89 is not
expected until 2026, and no planet-planet eclipses have been
identified for any other system, apart from one problematic candidate
\citep{Masuda2014}.

Finally, a potentially powerful method to determine mutual
inclinations is the analysis of variations in transit times and
durations, as described earlier in the context of orbital
eccentricities (\S~\ref{subsec:dynamical-modeling}). For instance, the
Kepler-30 system features three transiting giant planets, and the
observed absence of transit duration variations requires that the
mutual inclinations be smaller than a few degrees
\citep{SanchisOjeda+2012}.  Dynamical modeling has also been used to
limit mutual inclinations to $\lsim$10$^\circ$ in the multitransiting
systems Kepler-9 \citep{Holman+2010}, Kepler-11 \citep{Lissauer+2011},
Kepler-36 \citep{Carter+2012} and Kepler-56 \citep{Huber+2013b}. These
results are all consistent with coplanarity, but one is left wondering
whether this is a selection effect because of the requirement that all
the planets transit the host star. Addressing this concern,
\cite{Nesvorny+2012} analyzed the Kepler-46 system, in which there is
only one transiting planet but it exhibits unusually large
transit-timing variations, allowing the inference of a nontransiting
outer planet with an orbit that is aligned with the transiting planet
to within about 5$^\circ$. Similar results, although with coarser
accuracy, have been obtained for several other systems
(Nesvorny et al.\ 2013, 2014; Dawson et al.\ 2014).

\subsubsection{Population analysis}
\label{sec:mutincl-pop}

Apart from the individual cases described above, statistical arguments
have been used to investigate the typical mutual
inclinations of multiplanet systems, thanks to the large number of
such systems that the Doppler and {\it Kepler} surveys have provided.
\cite{DawsonChiang2014} provided indirect evidence for large mutual
inclinations in five giant-planet systems. Their argument was based on
the inference that the major axes of the two orbits are more nearly
perpendicular than would have occurred by chance. This preference for
perpendicularity is predicted in a model for the origin of giant
planets with eccentric orbits and $a=0.1$--1~AU. In this model, the
planets' orbits are inclined by 35-65$^\circ$ and perturb each other
gravitationally, ultimately causing the inner planet's orbit to shrink
as a result of eccentricity cycling and intermittent tidal dissipation (see
also Dong et al.\ 2014).

All the other statistical investigations of mutual inclination have
concentrated on the compact systems of multiple planets rather than
on giant planets:

\begin{enumerate}

\item One approach is to analyze the transit multiplicity
  distribution, defined as the relative occurrence of systems with
  differing numbers of transiting planets. All other things being
  equal, systems of higher transit multiplicity are more likely to be
  detected if the mutual inclinations are low. \cite{TremaineDong2012}
  presented a general formalism for assessing the agreement between
  planet models with differing degrees of coplanarity and the observed
  number of transiting systems of different multiplicities. They
  concluded that, with only the transit data, mutual inclinations are
  not well-constrained because of a degeneracy with the typical
  number of planets per star: With more planets per star, one can
  reproduce the {\it Kepler} statistics with larger mutual
  inclinations. A similar conclusion was reached by
  \cite{Lissauer+2011}.

\item \cite{Lissauer+2011} noted a potentially interesting pattern in
  the multiplicity distribution. When modeling the mutual inclination
  distribution as a Rayleigh function, they had difficulty fitting the
  large observed ratio of single-transiting systems to
  multiple-transiting systems. Their interpretation was that the
  single-transiting systems are divided into two groups with different
  architectures. Some are flat systems for which only the inner planet
  happens to transit; the others are systems with fewer planets or
  higher mutual inclinations. Several other studies reached the same
  conclusion \citep{Johansen+2012, HansenMurray2013,
    BallardJohnson2014}. However, \cite{TremaineDong2012} found no
  such evidence for two separate populations. This may be because they
  used a more general model for the multiplicity distribution,
  allowing the fractions of systems with differing numbers of planets
  to be independent parameters, rather than requiring all stars to
  have a certain number of planets. Xie et al.\ (2014) found
  supporting evidence, based on the relative occurrence of detectable
  transit-timing variations, that the single-transiting and
  multitransiting systems are qualitatively different.

\item A different method, also based on {\it Kepler} data, employs the
  transit durations of different planets orbiting the same star. The
  transit duration varies as $\sqrt{1-b^2}$, in which $b$ is the impact
  parameter (defined in \S~\ref{sec:transit-durations}). In a
  perfectly flat system, more distant transiting planets should have
  larger impact parameters. This trend should not be present if the
  typical mutual inclinations are greater than $R_\star/a$. Therefore,
  by searching for this trend, one can check for
  near-coplanarity. \cite{Fabrycky+2014} found such a trend among the
  {\it Kepler} multiplanet systems and concluded the mutual
  inclinations are typically smaller than a few
  degrees. \cite{FangMargot2012} combined this technique with
  multiplicity statistics and came to a similar conclusion.

\item Another method is to combine the results of the Doppler and {\it
    Kepler} surveys. If mutual inclinations are typically larger than
  $R_\star/a$, then transit surveys would miss many planets in each
  system, but Doppler surveys would potentially detect them all. A
  constraint on the typical mutual inclination can therefore be
  obtained by requiring consistency among the occurrence rates of
  multiplanet systems seen in Doppler and transit surveys. Among the
  complications are the differences in the selection effects and
  stellar populations between the surveys and the
  need to measure or assume a planetary mass-radius relationship
  (because Doppler surveys measure mass while transit surveys measure
  radius). \cite{Figueira+2012} performed such a study and concluded
  the compact systems of small planets ($\lsim$2~$R_\oplus$,
  $\lsim$10~$M_\oplus$) have typical mutual inclinations smaller than
  a few degrees (see also \citealt{TremaineDong2012}).

\end{enumerate}

Taken together these studies paint a picture in which the systems of
small planets with periods less than approximately one year---which exist
around nearly half of all Sun-like stars---are essentially as flat as
the Solar System. The power of these statistical methods has put us in
the surprising situation of knowing more about the mutual inclinations
of small-planet systems than about giant-planet systems, because giant
planets (despite being easier to study in almost every other way) are
rarely found in compact multiplanet systems.

\section{STELLAR ROTATION}
\label{sec:rotation}

By the early 17th century it was known that the Sun rotates with a
period of approximately 25 days. This discovery invited comparisons
between the Sun's rotational angular momentum and the planets' orbital
angular momentum. The Sun's angular momentum is 3.4 times larger than
that of the four inner planets, and 185 times smaller than that of the
four outer planets. The solar obliquity---the angle between the Sun's
angular momentum vector and that of all the planets combined---is
only 6$^\circ$.

To our predecessors, the low obliquity did not seem remarkable because
it harmonized with the low mutual inclinations between the planetary
orbits and supported the idea that the Sun and planets inherited
their angular momentum from a rotating disk. In contrast, they
regarded the Sun's relatively slow rotation as a dissonance that
needed to be explained by the theory of planet formation. For example,
the demise of tidal theories of planet formation was partly on account
of the difficulty in drawing material from the Sun and endowing it
with enough angular momentum to become the outer planets.

In the modern exoplanet literature, the roles of rotation rate and
obliquity and have been reversed. Explaining the magnitude of a star's
angular momentum is now recognized as an issue for star-formation
theory and subsequent magnetic braking, as opposed to being directly
related to planet formation. However, stellar obliquity has become an
active subfield of exoplanetary science, with surprising observations
and creative theories.

\subsection{Stellar rotation rate}

Despite this role reversal, a few investigators have searched for
relationships between stellar rotation and planetary orbital motion.
Rotation rates can be estimated from the rotational contribution to
spectral line broadening ($v\sin i$, where $i$ is the inclination
angle between the stellar rotation axis and the line of sight) or from
the period of photometric modulations caused by rotating starspots.
The results show that any rotational differences between
planet-hosting stars and similar stars without known planets are
modest (Barnes 2001, Alves et al.\ 2010, Gonzalez 2011, Brown 2014).

The only clear differences pertain to systems with the shortest-period
planets, as one may have expected, because such planets are most
sensitive to spin-orbit interactions mediated by tides, magnetic
fields, or the innermost portion of the protoplanetary disk.  For
orbital periods shorter than the stellar rotation period, tidal
dissipation gradually shrinks the planet's orbit and spins up the host
star. Eventually this process synchronizes the rotational and orbital
periods unless the synchronous state would require the rotational
angular momentum to be more than 25\% of the total angular momentum,
in which case the planet spirals into the star \citep{Counselman1973,
  Hut1980}. A few cases are known in which spin-orbit synchronization
seems to have been achieved ($\tau$~Boo, Butler et al.\ 1997;
HD~162020, Udry et al.\ 2002; Corot-4b, Aigrain et al.\ 2008). In
addition, \cite{Pont2009}, \cite{Brown+2011}, \cite{Husnoo+2012}, and
\cite{PoppenhaegerWolk2014} highlighted particular cases in which a
planet-hosting star is rotating more rapidly than expected, suggesting
that tidal spin-up is underway.  McQuillan et al.\ (2013) used {\it
  Kepler} data to show that faster-rotating stars ($P_{\rm rot}\lsim
10$~days) do not host short-period planets ($P_{\rm orb} \lsim 3$~days)
as often as do slower-rotating stars. This has been interpreted as a
sign that the rapid rotators ingested their close-in planets
\citep{TeitlerKonigl2014}.

\subsection{Stellar obliquity}

Measurements of stellar obliquity have revealed a wide range of
configurations, including stars with lower obliquities than the Sun
\citep[e.g., HD~189733;][]{Winn+2006}, stars with moderate tilts
\citep[e.g., XO-3;][]{Hebrard+2008, Winn+2009, Hirano+2011}, stars
that are apparently spinning perpendicular to their orbits
\citep[e.g., WASP-7;][]{Albrecht+2012b} and retrograde systems in which
the star revolves in the opposite direction as the planet's rotation
\citep[e.g., WASP-17;][]{Triaud+2010}.

Much of this knowledge has been obtained by observing the
Rossiter-McLaughlin effect, a time-variable distortion in stellar
spectral lines caused by a transiting planet \citep{Queloz+2000}. Two
limitations of these results should be borne in mind. First, only the
sky projection of the obliquity can be measured. A small obliquity
must have a small sky projection, but a high obliquity may also have
a small sky projection. Second, almost all the existing data are for
hot Jupiters, for the practical reason that the effect is most easily
measured when transits are deep and frequent.

\begin{figure}
\includegraphics[height=8cm]{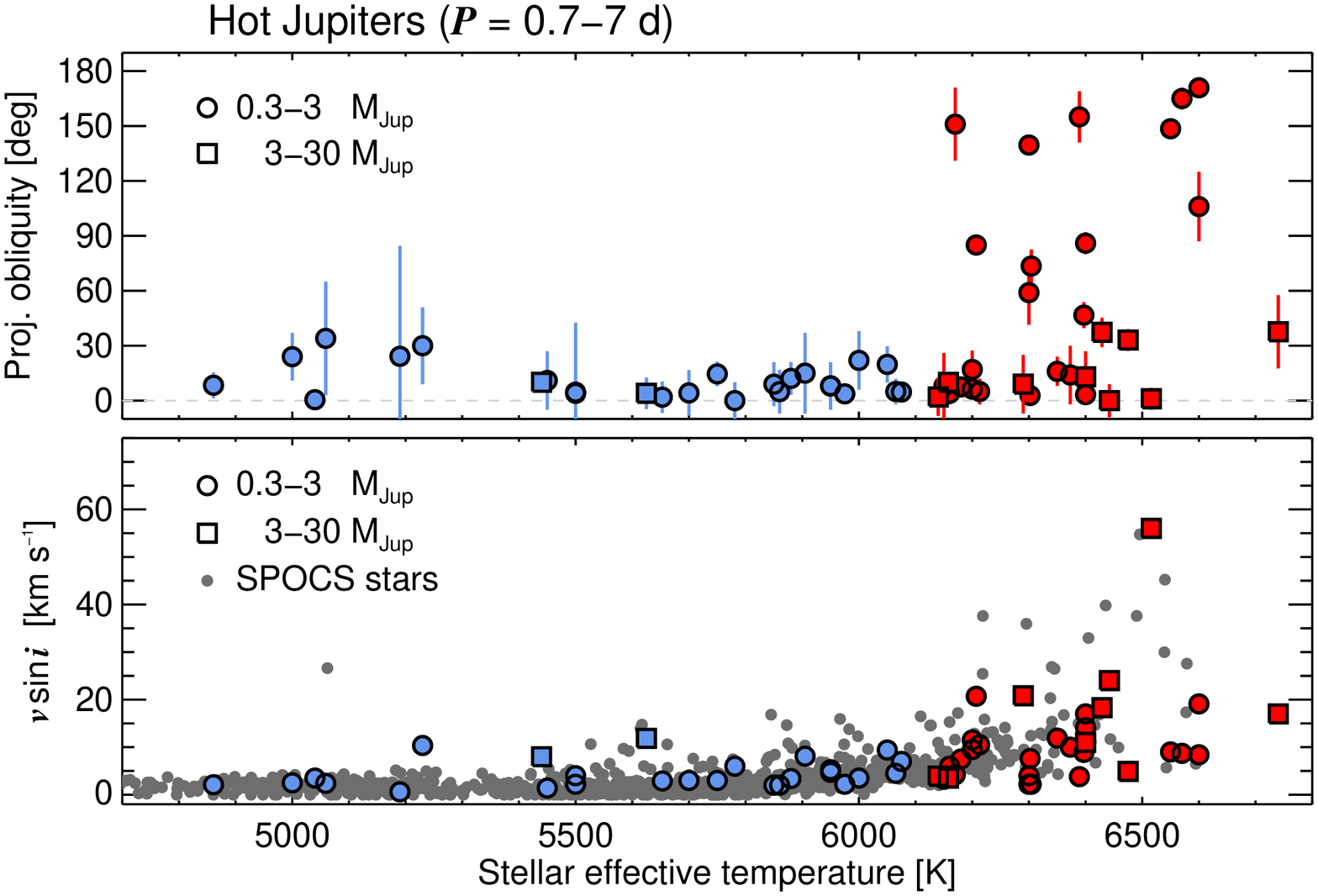}
\includegraphics[height=8cm]{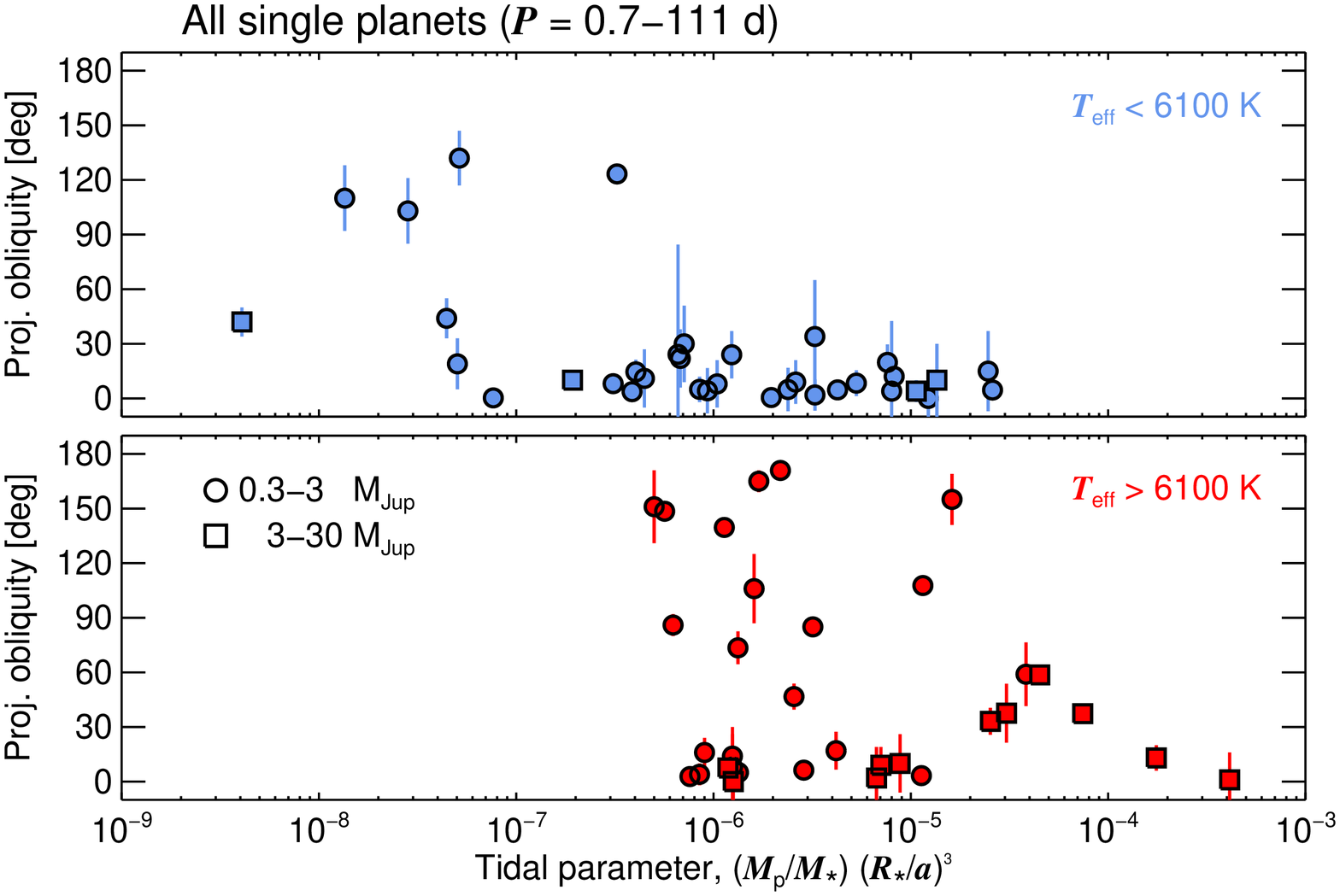}
\caption{
  {\it Top.}---Sky-projected stellar obliquity and rotation velocity
  as a function of effective temperature, for ``hot Jupiters'' ($M_{\rm p} >
  0.3~M_{\rm Jup}$, $P<7$~days) with secure measurements. Planet mass is
  encoded by the symbol shape.  Color is used to distinguish
  temperatures above and below 6100~K.  Gray dots are projected
  rotation rates of stars in the SPOCS catalog
  \citep{ValentiFischer2005}.
  {\it Bottom.}---Sky-projected stellar
  obliquity as a function of the relative strength of star-planet
  tidal forces, for all single-planet systems with secure
  measurements. These systems span a wider range of masses and orbital
  distances than the systems in the top panel. }
\label{fig:obliquity}
\end{figure}

The top panel of Figure~\ref{fig:obliquity} shows the hot Jupiter
data along with the two clearest patterns that have emerged. First,
stars with relatively cool photospheres ($\lsim 6100$~K) have low
obliquities, whereas hotter stars show a wider range of obliquities
\citep{Schlaufman2010, Winn+2010, Albrecht+2012a, Dawson2014}. Second,
the highest-mass planets ($\gsim 3$~$M_{\rm Jup}$) are associated with
lower obliquities (H\'ebrard et al.\ 2011).

Suggestively, the boundary of 6100~K coincides with the long-known
``rotational discontinuity'' above which stars are observed to rotate
significantly faster (Kraft 1967). This is demonstrated in
Figure~\ref{fig:obliquity} in the plot of $v\sin i$ versus $T_{\rm
  eff}$. The rotational discontinuity is thought to arise from the
differing internal structures of the stars on either side of the
boundary. Cool stars have thick convective envelopes and radiative
cores, whereas hot stars generally have radiative interiors with only
thin convective envelopes and a small convective core. These internal
differences cause cool stars to have stronger magnetic fields and
magnetic braking, explaining the rotational discontinuity. In
addition, cool stars are thought to be capable of dissipating tidal
oscillations more rapidly.

It is therefore natural to try and explain the obliquity patterns as
consequences of differing rates of rotation, magnetic braking, and
tidal dissipation. Regarding rotation, there seems to be no
correlation between obliquity and $v\sin i$ after controlling for
effective temperature. Magnetic braking and tidal dissipation are
complex and poorly understood processes, with no simple and
generally-accepted metrics that can be computed for all the systems
[although Albrecht et al.\ (2012) and Dawson et al.\ (2014) made
efforts in that direction]. A simple dimensionless ratio that
characterizes the ability of the planet to tidally deform the star is
$\epsilon_{\rm tide} \equiv (M_{\rm p}/M_\star)(R_\star/a)^3$, and indeed,
there is a suggestive pattern involving this tidal parameter. The
lower panel of Figure~\ref{fig:obliquity} shows projected obliquity as
a function of $\epsilon_{\rm tide}$ for all the single-planet systems
(not only hot Jupiters). The cool stars have a broad range of
obliquities for the weakest tides, and low obliquities for
$\epsilon_{\rm tide}\gsim 4\times 10^{-7}$. The hot stars have a broad
range of obliquities over essentially the entire range and a possible
trend toward low obliquity for $\epsilon_{\rm tide}\gsim 4\times
10^{-5}$. These trends suggest that many of the low obliquities are
the consequences of tides, which tend to align systems.

No doubt this story is oversimplified. Among the theorists who have
pursued more detailed descriptions are \cite{ValsecchiRasio2014}, who
found that misalignment is correlated with the depth of the outer
convective zone as calculated in stellar-evolutionary models; and
\cite{Dawson2014}, who modeled both tidal dissipation and magnetic
braking and concluded that the braking rate was the more important
difference between hot and cool stars.  Winds rob the star of most of
its angular momentum, allowing the planet to realign the rest of
it. Furthermore, even if the story is correct, it leaves many
questions unanswered. In many tidal theories, realignment is
accompanied by withdrawal of angular momentum from the planetary
orbit, leading to orbital decay. How, then, can the obliquity be
lowered without destroying the planet? \cite{Lai2012} provided one
possible solution by giving an example of a more complex tidal theory
in which realignment can be much faster than orbital decay.

Another unanswered question is the origin of the high obliquities.
Are hot Jupiters formed with a wide range of orbital orientations? An
affirmative answer would support theories for hot Jupiter production
involving planet-planet scattering \citep{RasioFord1996,
  WeidenschillingMarzari1996} or Kozai-Lidov oscillations (Mazeh et
al.\ 1997, Fabrycky \& Tremaine 2007, Naoz et al.\
2011). Alternatively, high stellar obliquities may simply be a
common outcome of star formation (Bate et al.\ 2010, Thies et al.\
2011, Fielding et al.\ 2014). Additional possibilities are magnetic
star-disk interactions (Lai et al.\ 2011), or torques from distant
stellar companions (Tremaine 1991, Batygin et al.\ 2011, Storch et
al.\ 2014). Rogers et al.\ (2012) proposed that high obliquities
originate from stochastic rearrangements of angular momentum within
hot stars, mediated by gravity waves; in this scenario, the
photosphere is rotating in a different direction from the interior.

Answering these questions would be easier if the domain of the
obliquity measurements could be expanded to include other types of
planets besides close-in giants. Because the Rossiter-McLaughlin
effect is more difficult to observe for smaller and longer-period
planets, new methods have been developed to gauge the stellar
obliquity, some of which are more widely applicable or at least are
suited to different types of systems:

\begin{enumerate}

\item In the $v\sin i$ method, one divides estimates of $v\sin
  i$ and $v$ to obtain $\sin i$. If this is significantly smaller than
  unity for a star with a transiting planet, then the star has a high
  obliquity \citep{Schlaufman2010}. Using this method, some candidate
  misaligned multitransiting systems have been identified
  \citep{WalkowiczBasri2013, Hirano+2014} but the results are
  uncertain because of the difficulty in measuring $v\sin i$ for cool
  stars, the most common type of stellar host. Morton \& Winn (2014)
  found a weak statistical tendency for multitransiting systems to
  have lower obliquities than the systems with only one known
  transiting planet.

\item The starspot-tracking method is based on events in which a
  transiting planet occults a starspot, temporarily reducing the loss
  of light and producing a glitch in the transit light curve. The
  obliquity can sometimes be decoded from a sequence of such anomalies
  \citep{SanchisOjeda+2011, Desert+2011}. A related method relies on
  the correlations between the timing of a starspot-crossing event,
  and the phase of the corresponding flux modulation produced by the
  rotation of the starspot (Nutzman et al.\ 2011, Mazeh et al.\
  2014). Both methods were used to show that the stellar obliquity in
  the three-planet Kepler-30 system is $\lsim 10^\circ$
  \citep{SanchisOjeda+2012}.

\item The starspot-variability method is based on the fact that
  the observed photometric variability due to rotating starspots
  should be larger for stars with $\sin i\approx 1$ than for stars
  that are viewed more nearly pole-on. \cite{Mazeh+2014} invented
  this technique and compared the photometric variability of {\it
    Kepler} stars with and without transiting planets. The relatively
  cool planet-hosting stars ($\lsim$5700~K) showed greater variability
  than non-planet hosts, suggesting a tendency toward spin-orbit
  alignment.

\item The gravity-darkening method is based on transits across a
  rapidly-rotating star, which is darker near its equator because of
  the lifting and cooling effect of the centrifugal force. When the
  obliquity is high, this effect causes an asymmetry in the light
  curve \citep{Barnes2009}, as seen during transits of Kepler-13
  (Barnes et al.\ 2011, Szabo et al.\ 2011).

\item The asteroseismic method is based on the observed frequencies of
  stellar photospheric oscillations, which are classified by the usual
  ``quantum numbers'' $nlm$. For a non-rotating star, modes with the
  same $nl$ and differing $m$ are degenerate. Rotation splits the
  modes into multiplets. Crucially, the relative visibilities of the
  modes within each multiplet depend on viewing angle; the observed
  amplitudes thereby divulge the stellar obliquity
  \citep{GizonSolanki2003}. The demanding observational
  requirements---years of ultraprecise photometry with $\sim$1~minute
  sampling---have only been met for a few {\it Kepler}
  stars. \cite{Chaplin+2013} found low obliquities for two stars with
  multiple transiting planets (Kepler-50 and 65), and
  \cite{VanEylen+2014} found a low obliquity for the host of a
  Neptune-sized planet. On the other hand, \cite{Huber+2013b} found a
  45$^\circ$ obliquity for the multiplanet system Kepler-56, and
  \cite{Benomar+2014} found (with less confidence) a moderate
  obliquity for Kepler-25. These are the first sightings of spin-orbit
  misalignment for planets that seem distinct from hot Jupiters.

\end{enumerate}

Obliquity, and its relation to planet formation, is a developing
story. We await another round of clarification from a larger sample of
more diverse systems, or the discovery of individually revealing
cases. Regarding the latter, one of the most intriguing planetary
candidates is PTFO~8-8695 \citep{vanEyken+2012}. This may be the first
known case of a giant planet transiting a very young star ($\lsim
5$~Myr). In addition, the light curve changes shape on a timescale of
months, a peculiarity which \cite{Barnes+2013} attributed to nodal
precession of a strongly misaligned orbit around the rapidly rotating,
gravity-darkened star. Thus, this object may provide clues to the
origin of high obliquities. However, the system has other strange and
unsettling properties: the transit and stellar-rotation periods are
both 16~hours (suggesting that starspots or accretion features may be
causing the transit-like dimmings); the hypothetical planet would be
at or within its Roche radius; giant planets on such short-period
orbits are known to be rare around mature stars; and the planet's
Doppler signal cannot be detected in the face of the spurious
variations produced by stellar activity. These factors cast doubt on
the planetary interpretation, although the scientific stakes are high
enough to warrant further investigation.

\subsection{Star-disk alignment}

We have focused on the alignment between a star's equator and its
planets' orbital planes, but work has also been done on alignment with
other planes, particularly the plane of a debris
disk. \cite{Watson+2011} used resolved images of debris disks and the
$v\sin i$ method to search for misalignments in 8 systems and found
none greater than 20-30$^\circ$. \cite{Greaves+2014} added 10 systems
observed with the {\it Herschel Space Observatory} and found alignment
to within $\lsim$10$^\circ$. In contrast, there is mild evidence for a
star-disk misalignment in the directly imaged system HR~8799, which
has at least three giant planets at orbital distances of
20-100~AU. The debris disk has an inclination of $26\pm 3^\circ$
\citep{Matthews+2014}, whereas the star has an inclination of $\gsim
40^\circ$ according to the asteroseismic analysis by
\cite{Wright+2011}. Finally, in a technical {\it tour de force},
\cite{LeBouquin+2009} used optical interferometry to measure the
position angle of the stellar equator of Fomalhaut~b, showing it to be
aligned within a few degrees of the debris disk.

\section{BINARY STAR SYSTEMS}
\label{sec:binaries}

Although the Sun has no stellar companion, a substantial fraction of
Sun-like stars are part of multiple-star systems
\citep{Raghavan+2010}. Naturally this makes one curious about the
architecture of planetary systems with more than one star. In
addition, investigating planets around stars with close stellar
companions may help to clarify some aspects of planet-formation
theory. The gravitational perturbations from a close companion would
stir up the protoplanetary disk, complicating the process of building
up large bodies from smaller ones. Whether or not planets manage to
form despite these perturbations may provide a test of our
understanding of this process \citep{ThebaultHaghighipour2014}.

Almost all the literature on this topic deals with the effect of
binary stars rather than systems of higher multiplicity. In that
case, we may distinguish systems with planets orbiting one member of
the binary from those with planets for which the orbit surrounds both
members. In the nomenclature of \cite{Dvorak1982} the former type of
system is $S$-type (for satellite) and the latter is $P$-type (for
planetary, although in this context a better term is
circumbinary).

The three-body problem famously allows for chaos and instability,
giving stringent restrictions on where we may find planetary orbits
in binaries, regardless of how they formed. \cite{HolmanWiegert1999}
used numerical integrations to explore these fundamental limits, as a
function of the binary semimajor axis $a_b$, mass ratio, and orbital
eccentricity. They inserted a test particle (representing the planet)
on an initially circular and prograde orbit, with varying orbital
distances, and searched for the most extreme orbital distance $a_c$ at
which planets remained stable for $10^4$ binary orbital periods. They
provided useful fitting formulas for $a_c$, which can be summarized by
the rough rule-of-thumb that the planet's period should differ by at
least a factor of three from the binary's period, even for a mass ratio as
low as $0.1$. The zone of instability near the binary's orbit widens
for more massive secondaries or more eccentric binary orbits, as one
would expect. These results were interpreted within the theory of resonance
overlap by \cite{MudrykWu2006} and extended to noncoplanar orbits by
\cite{DoolinBlundell2011}.

\subsection{Orbits around a single star}

The Doppler surveys have been the main source of $S$-type planets.
Approximately 70 examples are known \citep{Roell+2012}, but only about
5 of these systems have separations $\lsim$50~AU. These systems were
typically recognized by detecting the planet through the Doppler
method and then searching for additional stars through direct imaging
or long-term Doppler monitoring (see, e.g., Eggenberger et al.\ 2007,
Mugrauer et al.\ 2014). The selection effects are difficult to model
because close binaries (with angular separations $\lsim$2$\arcsec$)
are avoided in the Doppler planet surveys, as it is difficult to
achieve the necessary precision when the light from two stars is
blended on the spectrograph.

Several investigators have tried to discern whether planet occurrence
rates depend on the presence of a companion star. The broad synthesis
of these studies is that for $a_b\gsim 20$~AU, any such differences
are small \citep{Eggenberger+2011}. No planets are known in systems
with smaller binary separations, which is at least partially due to
selection bias. In no case has a planet been found with an orbit that
is close to the stability limit $a_c$, or even within a factor of
two. There is also an indication that for $a_b\gsim 10^3$~AU, giant
planets have slightly higher eccentricities (Kaib et al.\ 2013). In
fact the four planets with the largest known eccentricities ($e >
0.85$) are all members of $S$-type binaries \citep{Tamuz+2008}.

Another approach is to search for stellar companions to {\it Kepler}
planet-hosting stars, by following up with Doppler and direct imaging
observations. This gives access to smaller planets than are usually
found in the Doppler surveys but pays a penalty in sensitivity (and
complexity of selection effects) as a result of the dilution of photometric
transit signals by the light of the second star. Using this approach,
\cite{Wang+2014} found that binaries with $a_b=10$-10$^3$~AU have
fewer planets than single stars by about a factor of two. This result
pertains mainly to planets smaller than Neptune with periods of
$\lsim$50~days. At face value, then, the formation of these small
close-in planets is somehow more sensitive to the presence of a
companion star than the formation of giant planets found in Doppler
surveys.

\subsection{Orbits around two stars} 
\label{sec:cbps}

$P$-type planets are more exotic than $S$-type planets. They often
appear in science fiction, with evocative scenes of alien double
sunsets. The first known circumbinary planet was even more exotic than
had been anticipated in science fiction: its host stars are a pulsar
and a white dwarf \citep{Thorsett+1999}. Based on the observed
variations in the travel time of the radio pulses, caused by the
pulsar's orbital motion, it was possible to deduce that the pulsar
B1620-26 has a white-dwarf companion in a 191~day orbit and that both
are surrounded by a 1-3~$M_{\rm Jup}$ companion with an orbital period
of several decades \citep{Sigurdsson+2008}. The system is in a
globular cluster, and may have arisen through a dynamical exchange. In
this scenario the neutron star originally had a different white dwarf
companion that was exchanged for a main-sequence star during a close
encounter. A planet that was formerly orbiting the main-sequence star
was cast into a circumbinary orbit, and eventually the main-sequence
star evolved into the white dwarf seen today \citep{Ford+2000}.

Timing the eclipses of other types of stellar binaries has led to the
detection of candidate circumbinary planets, but with a less secure
status than that of B1620-20. Within this category are about six cases of
post-common-envelope binaries in which the observed eclipse timing
variations are consistent with Keplerian motion induced by one or more
circumbinary planets. Their properties and problems are reviewed by
\cite{Horner+2014}. One serious problem is that in at least four
cases, the proposed planetary orbits are dynamically unstable, casting
doubt on the planetary hypothesis for those systems and leaving one
wondering whether all the signals are the result of some other
physical phenomenon. In one of the best studied systems, HU Aquarii,
additional timing data refuted the planetary hypothesis
\citep{Bours+2014}, although the true origin of the timing variations
is unknown.

\begin{figure}
\includegraphics[height=12cm]{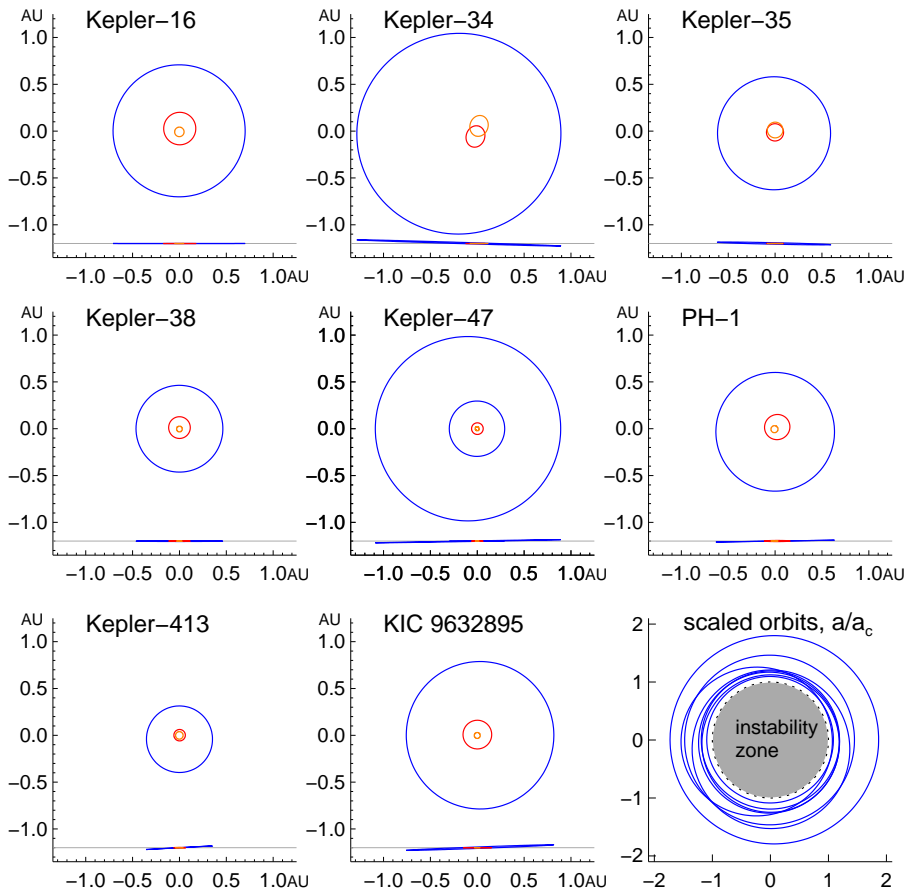}
\caption{ Gallery of the \emph{Kepler} circumbinary planetary systems.
  Each panel has the same scale.  Orange orbits are for the primary
  stars, red are for the secondary stars, and blue are for the
  planets. Near the bottom of each panel, a side view is shown, with
  the gray line indicating the stellar orbital plane. In the bottom
  right panel, the orbits are scaled to the size of the critical
  semi-major axis for stability ($a_c$) as estimated from the
  equations by \cite{HolmanWiegert1999}. The planets seem to bunch
  just outside the instability zone (although Kepler-47's exterior
  planet is too distant to fit in this plot).}
\label{fig:cbps}
\end{figure}

The {\it Kepler} survey has led to the discovery of nine (and counting)
transiting circumbinary planets around eclipsing binary stars. The
detection of transits and the satisfactory fits of simple dynamical
models to the observed transit and eclipse times leave very little
room for doubt about the interpretation. In fact, because of the
three-body effects, the dynamical models have in some cases provided
unusually precise measurements of the masses and sizes of all three
bodies \citep[see, e.g.,][]{Doyle+2011}. Table~\ref{tbl:cbps} gives
the basic properties of these systems, and Figure~\ref{fig:cbps}
depicts their orbital configurations. Kepler-47 is the only binary
with more than one known planet \citep{Orosz+2012a, Kostov+2013}; in
all the other cases, only one circumbinary planet is known in the
system. A few trends are worth noting, though they are difficult to
interpret at this stage:

\begin{enumerate}

\item The {\it Kepler} circumbinary planets all have sizes
  $>$3~$R_\oplus$. Smaller planets have not been excluded; they may
  simply be more difficult to detect. Unlike the case of isolated
  planets, it is not possible to build up the signal-to-noise ratio
  through simple period folding because the orbital motion of the
  stars causes substantial variations in the transit times and
  durations.

\item The planets seem to cluster just outside of the zone of
  instability (see the lower right panel of Fig.~\ref{fig:cbps}). Such
  a pile-up had been predicted by theoreticians as a consequence of
  migration of giant planets within a circumbinary disk
  \citep{PierensNelson2008}. Here as always, though, lurks the specter
  of selection effects: the transit method favors the detection of
  planets with the shortest possible periods.

\item The binary and planetary orbits are aligned to within a few
  degrees, as seen in Fig.~\ref{fig:cbps}. This too may be a selection
  effect: it is easier to recognize planets when multiple transits are
  detected, as is often the case in coplanar systems but not in
  inclined systems. We await the unambiguous discovery of circumbinary
  planets through methods that are less biased with respect to orbital
  coplanarity, such as eclipse timing variations
  \citep{Borkovits+2011} or transits of non-eclipsing binaries
  \citep{MartinTriaud2014}.

\end{enumerate}

\begin{deluxetable}{lrrrrrrrrrr}
\tabletypesize{\footnotesize}
\tablewidth{0pt}
\tablecaption{Key properties of {\it Kepler} circumbinary planets\label{tbl:cbps}}

\tablehead{
\colhead{Name}&
\colhead{$M_A$}&
\colhead{$M_B$}&
\colhead{$P_{\rm bin}$}&
\colhead{$e_{\rm bin}$}&
\colhead{$M_{\rm p}$}&
\colhead{$P_{\rm p}$}&
\colhead{$e_{\rm p}$}&
\colhead{$i_{\rm mut}$} &
\colhead{$P_{\rm p}/P_{\rm c}$ } &
\colhead{Ref.} \\
\colhead{ } &
\colhead{[$M_\odot$]} &
\colhead{[$M_\odot$]} &
\colhead{[days]} &
\colhead{}&
\colhead{[$M_{\rm Jup}$]}&
\colhead{[days]}&
\colhead{}&
\colhead{($^\circ$)}&
\colhead{}
}

\startdata
Kepler-16b & 0.690& 0.203& 41.07922 & 0.1594 & 0.333 & 228.776 & 0.0069 & 0.3 &1.14 & 1 \\
Kepler-34b & 1.0479& 1.0208& 27.79581 &0.5209 & 0.220 & 288.822 & 0.182 & 1.8 &1.21 & 2 \\
Kepler-35b & 0.8877 &0.8094 & 20.73367& 0.1421& 0.127 &131.458 &0.042 & 1.3 &1.24 & 2 \\
Kepler-38b & 0.941 &0.248 &18.79537 & 0.1042& $<0.2$ & 105.599 & 0.0 & 0.2 & 1.42 & 3 \\
Kepler-47b & 1.043 & 0.362 &7.44838 &0.0234 & $<0.1$ & 49.514& 0.01& 0.3 & 1.77 & 4 \\
Kepler-47c & 1.043 & 0.362 &7.44838 &0.0234 & $<0.1$ & 303.158& 0.1 & 1.1 & 10.8 & 4 \\
PH-1b& 1.384&0.386 & 20.00021& 0.2117& $<0.1$ &138.506 &0.052 &2.8 & 1.29 & 5 \\
Kepler-413b &0.820 & 0.542& 10.11615& 0.037& $<0.1$& 66.262& 0.117& 4.1 & 1.60& 6 \\
KIC~9632895b & 0.934 & 0.194& 27.32204 & 0.0510& $<0.1$ & 240.503 & 0.038 & 2.3 &2.41& 7\\
\enddata
\tablecomments{In column 10, $P_c$ refers to the critical orbital
  period, the minimum period compatible with long-term stability
  according to the equations of \cite{HolmanWiegert1999}. References:
(1) \cite{Doyle+2011},
(2) \cite{Welsh+2012},
(3) \cite{Orosz+2012b},
(4) \cite{Orosz+2012a},
(5) \cite{Schwamb+2013},
(6) \cite{Kostov+2013},
(7) \cite{Welsh+2014}.
}
\end{deluxetable}

As with $S$-type planets, it is interesting to compare the occurrence
rate for circumbinary planets with that of single-star planets. This
is not straightforward because the geometric transit probability
depends on the binary properties as well as the inclination between
the two orbital planes. Furthermore, precession of noncoplanar systems
causes the existence of transits to be ephemeral. A particular system
may display transits for a few months and then cease for years before
displaying them again \citep{Schneider1994,Kostov+2013}.

Despite these difficulties, \cite{Armstrong+2014} attempted to
calculate the occurrence rate of circumbinary planets. They evaluated
the observability of circumbinary planets within the sample of {\it
  Kepler} eclipsing binaries and tested for consistency between the
currently-detected systems and various proposed planet
distributions. They favored a model in which around $10\%$ of binaries
host a $6-10$~$R_\oplus$ planet on a nearly coplanar orbit ($\Delta
i\lsim 5^\circ$) somewhere between the stability limit and an outer
period of $300$~days. This fraction of $\approx$10\% is consistent
with the rates seen around single stars
(Table~\ref{tbl:occurrence}). The implication is that $\gsim
6$~$R_\oplus$ planets populate the space $\lesssim 1$~AU just as
readily around binaries as single stars. However, when they allowed
for the possibility of noncoplanar orbits, no firm conclusion could be
reached owing to a degeneracy between coplanarity and multiplicity.

An interesting trend is that when the {\it Kepler} eclipsing binaries
are sorted in order of orbital period, all eight circumbinary-planet hosts
are found among the longer-period half. This does not seem to be
purely a selection effect \citep{Armstrong+2014,
  MartinTriaud2014}. This observation, if confirmed by additional
analysis, would support the theory that very close binary stars
($\lsim 5$~days) form through a different mechanism than wider
binaries. Specifically, they may be the product of orbital shrinkage
due to gravitational perturbations from a third wide-orbiting star and
the long-term effect of tidal dissipation \citep{MazehShaham1979,
  FabryckyTremaine2007}. The perturbations and the shrinkage process
would likely be hostile to circumbinary planets, preventing their
formation or putting them at risk of ejection or accretion onto the
stars. This would be a delightful example of an advance in
exoplanetary science leading to progress in star-formation theory.

\section{SUMMARY AND DISCUSSION}
\label{sec:discussion}

Over the past few decades, astronomers have gradually become aware of the
following properties of planets around other stars:
\begin{itemize}
\item A Sun-like star has a $\approx$10\% chance of
  having a giant planet with a period shorter than a few years, and a
  $\approx$50\% chance of having a compact system of smaller planets with
  periods shorter than a year.
\item The giant planets have a broad eccentricity distribution,
  ranging from around 0--0.9 with a mean of $\approx$0.2. The
  smaller planets have lower eccentricities ($\lsim 0.1$),
  particularly those in multiplanet systems.
\item In compact systems of small planets, the orbits are typically
  aligned to within a few degrees. Giant planets
  may occasionally have larger
  mutual inclinations.
\item The ratios of orbital periods are often in the range of
  2--3 but are occasionally closer to unity, flirting with instability.
\item Giant planets are more often found in mean-motion resonances
  than smaller planets, which show only a slight preference for
  being near resonances.
\item The stellar rotation axis can be grossly misaligned with the
  planetary orbital axis, particularly for close-in giant planets orbiting
  relatively hot stars ($T_{\rm eff}\gsim 6100$~K);
\item Close binary stars host circumbinary giant planets just outside
  of the zone of instability, with an occurrence rate comparable with
  that of giant planets at similar orbital distances around single
  stars.
\end{itemize}

We are now in a position to compare this exoplanetary list with that
given for the Solar System (\S~\ref{sec:intro}). The hope has always
been that such a comparison would help answer one of the big questions
of exoplanetary science: How do planets form? We discuss this question
in \S~\ref{sec:formation}, but limit ourselves to broad-brush remarks.
One important difference to keep in mind is that the Solar System
properties are indisputable and provide a fairly complete picture. In
contrast the exoplanetary list is provisional and based on the myopic
views of planetary systems that are inherent to existing astronomical
techniques. In \S~\ref{sec:future} we review the prospects for
obtaining a clearer picture.

\subsection{Planet-formation theory}
\label{sec:formation}

By the mid-1990s the prevailing theoretical paradigm for planet
formation was that planets originate from the coagulation of very
small solid bodies. Although this notion cannot be traced back to
Laplace's theory, in which planets formed from gaseous rings, it is
nevertheless an old idea, dating back at least to Chamberlin~(1916)
who called it the ``planetesimal hypothesis.'' Between the 1960s and
1990s many theorists turned this hypothesis into a detailed
mechanistic theory, which included the further realization that a
large enough solid body could undergo runaway accretion of gas and
become a giant planet, a phenomenon that has become known as ``core
accretion'' \citep{Mizuno1980, Pollack+1996}.

The main predictions of this theory for the geometrical properties of
exoplanetary systems were that the eccentricities and inclinations
should be low. The inclinations are low because the processes take
place within a flat disk. The eccentricities are low because the
streamlines within the disk are nearly circular; this is a consequence
of viscous dissipation. Furthermore, even if an object managed to
acquire a moderate eccentricity or inclination, gravitational and
hydrodynamical interactions with the disk would coplanarize and
circularize its orbit \citep{Cresswell+2007,
  XiangGruessPapaloizou2013}. The theory also predicted that a
planet's composition should be linked to the location of its formation
within the disk. Giant planets, in particular, would be found beyond
the ``snow line'' at a few astronomical units, where water exists as a
solid, enhancing the abundance of solid material and promoting rapid
accretion. These predictions were taken seriously enough that
HD~114762b, now understood as a short-period giant planet on an
eccentric orbit, was not recognized as an exoplanet at the time of its
discovery in 1989 \citep{Latham2012}.

Even the most grotesque violations of these expectations---high
eccentricities, retrograde orbits, extremely short-period giant
planets---have not shattered the paradigm of planet-formation theory.
The only other theory that is occasionally discussed is gravitational
instability, in which the gaseous disk collapses directly into giant
planets \citep{Boss1997}. However, this theory leads to essentially the
same predictions for low eccentricities and inclinations as does core
accretion. Criticism of gravitational instability has been mainly on
non-geometric grounds, such as whether the thermodynamic properties of
the disk allow for collapse, and whether the theory is compatible with
the correlation between metallicity and giant-planet occurrence
\citep[as reviewed by][]{Durisen+2007}.

The failure of planet-formation theory to anticipate the geometric
properties of exoplanetary systems is generally understood as a
failure to pay attention to what may happen after planets
form. Attention had been focused too narrowly on the conglomeration of
solid material and the accretion of gas. Now, more attention is paid
to the interactions of the cores and planets with each other and with
the protoplanetary disk while it still exists. Gravitational
interactions with the disk can cause a planet to spiral inward or
``migrate'' (Lin et al.\ 1996, Ward et al.\ 1997), possibly explaining
the close-in giant planets. High eccentricities and inclinations can
be generated by resonant interactions with the disk (Goldreich \& Sari
2003, D'Angelo et al.\ 2006, Bitsch et al.\ 2013), or between planets
\citep{ThommesLissauer2003}. They can also be generated independently
of the disk, through planet-planet scattering \citep[see,
e.g.,][]{Chatterjee+2008}, or torquing from a distant massive
companion \citep{FabryckyTremaine2007}.

The closest we have seen to an attempt at paradigm shift is the proposal
that the small low-density planets within compact systems formed
essentially at the locations where they are seen today, rather than
having migrated inward from the snow line or beyond (Raymond et al.\
2008, Hansen \& Murray~2012, Chiang \& Laughlin~2013). This would
require the protoplanetary disk to have surface density approximately
ten times higher than is commonly thought. However, disk migration
theory also seems capable of explaining the formation of these systems
\citep{TerquemPapaloizou2007}. Disks have fluid, thermodynamic, and
electromagnetic properties of potentially crucial importance that are
poorly understood. From an observer's perspective there seem to be few
observations that cannot be accommodated by adjusting the properties
of the disk.

In short, the survival of the planetesimal hypothesis and core
accretion theory after the onslaught of exoplanetary data may be a
sign that the theory is correct, if lacking in some details. However,
it may also partially reflect the deep entrenchment of the theory
after having enjoyed a very long interval of time when the only known
planets were inside the Solar System. We wish we could travel back in
time to 1795, deliver this manuscript to Laplace, and follow the
subsequent development of planet-formation theory.

\subsection{Future observations}
\label{sec:future}

The Doppler and transit techniques have provided a fairly detailed
description of the inner regions of planetary systems. We may now
look forward to similar information about the outer regions of
exoplanetary systems as the astrometric, microlensing, and direct
imaging techniques reach greater maturity in the coming decade.

Astrometry has had little impact on exoplanetary science beyond a few
detections of previously-known planets \citep{Benedict+2002} and
spurious detections of new planets (see, e.g., Pravdo \& Shaklan 2009,
whose claim was refuted by Bean et al.\ 2010). The European {\em Gaia}
mission should improve this situation. \cite{Casertano+2008}
forecasted that the $\approx$5~$\mu$as precision of {\em Gaia}'s
positional measurements would lead to thousands of detections of giant
planets with orbits out to 3-4~AU. Astrometry also exposes the full
three-dimensional orbits of planets, apart from a twofold degeneracy
(an orbit cannot be distinguished from its sky-plane mirror
reflection). Thus, for some {\it Gaia} multiplanet systems, it should
be possible to measure mutual inclinations. In addition {\it Gaia}
will search for planets around massive stars, evolved stars, young
stars, and other types of stars that misbehave in Doppler and transit
surveys but are more cooperative in astrometric surveys.

Microlensing has already made substantial contributions and is poised
to offer more. The Korean Microlensing Telescope Network plans to use
three dedicated 1.6-m telescopes to perform a dedicated survey with
nearly continuous time coverage \citep{Park+2012}. The completed
network---scheduled to begin operation in 2015---is predicted to find
$\approx$60 planets~year$^{-1}$ ranging in mass from 0.1 to
10$^3$~$M_\oplus$ and in orbital distance from 0.4 to 16~AU
\citep{Henderson+2014}. An even more prodigious planet discovery rate
could be achieved with a space telescope \citep{BennettRhie2002}, such
as the 2.4-m wide-field infrared telescope currently being planned for
both cosmology and exoplanetary science. The mission's name
fluctuates; at the time of writing, it is {\it WFIRST-AFTA}.
\cite{Barry+2011} predicted that this survey will detect thousands of
planets with orbital distances of a few AU, thereby complementing {\it
  Kepler}'s view of sub-AU systems.  It will be a challenge to extract
any architectural information from these systems beyond the
instantaneous sky-projected orbital separations.

Direct imaging detections have been relatively few and limited to
massive planets with orbital distances $\gsim$20~AU, but the technique
is flourishing with the advent of new high-contrast imagers for large
ground-based telescopes \citep{Beuzit+2008, Close+2014,
  Macintosh+2014, Jovanovic+2014}. These are expected to provide
an order-of-magnitude improvement in contrast, enabling hundreds of young
stars to be searched for giant planets. In addition, the Atacama Large
Millimeter Array has enabled direct imaging of the inner regions of
gaseous protoplanetary disks, which may reveal structures associated
with planet formation. \cite{Biller+2014} and \cite{Reggiani+2014}
recently detected a ``blob'' that may represent such a
structure. Further ahead, {\it WFIRST-AFTA} may have a coronagraph
capable of detecting planets analogous to the Solar System's giant
planets.

Even though the Doppler and transit techniques have already reached a
high level of maturity, they show no signs of exhaustion. Numerous new
Doppler instruments are planned, with the goal of detecting
habitable-zone Earth-mass planets. They include traditional optical
spectrographs well suited for FGK dwarfs \citep{Pepe+2010,
  Szentgyorgyi+2012, Pasquini+2008} as well as new infrared
spectrographs more appropriate for M dwarfs \citep{Tamura+2012,
  Quirrenbach+2010, Mahadevan+2012, Delfosse+2013}. The main
attraction of M dwarfs is that the habitable zone occurs at shorter
orbital distances, expediting planet searches.

The sky will continue to be scoured for transiting planets. A new
round of ground-based surveys is underway, aiming to extend the hunt
beyond their traditional quarry of hot Jupiters into the realm of
Neptune-sized planets \citep{Bakos+2013, Wheatley+2013}, or even
smaller planets around M dwarfs \citep{NutzmanCharbonneau2008,
  Giacobbe+2012, Gillon+2013}. In space, the handicapped but still
potent {\it Kepler} telescope has begun monitoring several star fields
near the ecliptic plane, the only zone in which it can achieve stable
pointing \citep{Howell+2014}. The European {\it Characterising
  Exoplanet Satellite} ({\em CHEOPS}) mission
\citep{Broeg+2013} aims to achieve precise photometry for bright
stars, search for transits of Doppler planets, and improve upon the
light curves of previously detected transiting planets. NASA's
{\it Transiting Exoplanet Survey Satellite} ({\em TESS}) is scheduled to
perform an all-sky, bright-star survey for short-period transiting
planets in 2018-2019 \citep{Ricker+2015}. Further out, in the
mid-to-late 2020s, the European {\it Planetary Transits and
  Oscillations} ({\it PLATO}) mission intends to begin
longer-duration transit survey over half of the celestial sphere
\citep{Rauer+2014}.

The continued analysis of {\it Kepler} data may reveal the
architectures of additional systems through the measurement and
interpretation of transit-timing variations. Long-period companions in
those systems may be revealed through long-term Doppler observations
and {\it Gaia} astrometry. The current and forthcoming transit data
may also reveal planetary satellite systems, or planetary rotation
rates and obliquities, which are interesting features of the Solar
System but are essentially unknown for exoplanets. Current satellite
searches are limited to relatively large companions around
approximately a dozen of the most observationally favorable planets
\citep{Kipping2014}. A wildcard possibility is the discovery of the
third category of three-body system classified by \citep{Dvorak1982},
the so-called $L$-type (``librator''). This would be a Trojan
companion to a planet, one whose orbit surrounds the $L_4$ and $L_5$
Lagrange points of the star-planet system ($60^\circ$ ahead or behind
the planet's orbit). As for planetary spin, the only successful
measurement to date has been from direct imaging: \cite{Snellen+2014}
found the giant planet around $\beta$~Pic to be spinning more rapidly
than any Solar System planet ($v\sin i = 25\pm 3$~km~s$^{-1}$, as
compared to 13~km~s$^{-1}$ for Jupiter). This finding was based on the
Doppler broadening of the planet's spectrum.

Clearly there is vast scope for improving our understanding of the
occurrence and architecture of exoplanetary systems. Still, it seems
appropriate to admire the progress that has been made. Consider again
the thought experiment of traveling back in time to deliver this
information to Laplace. With so many startling results, it is
difficult to guess what would have impressed him the most: the high
eccentricities, the retrograde planets, the chaotic systems, the
circumbinary systems; the ceaseless technological developments that
have propelled the field; or the mere fact that we have learned so
much about faraway planetary systems on the basis of only the minuscule
changes in brightness and color of points of light?


\section{DISCLOSURE STATEMENT}

The authors are not aware of any affiliations, memberships, funding,
or financial holdings that might be perceived as affecting the
objectivity of this review.


\section{ACKNOWLEDGMENTS}

We are grateful to Eric Agol, Simon Albrecht, Eric Ford, Jonathan
Fortney, Andrew Howard, Jack Lissauer, Scott Tremaine, Amaury Triaud,
Bill Welsh, and Liang Yu for helpful comments on the
manuscript. J.N.W.\ thanks Avi Loeb, Matt Holman, and the Institute
for Theory and Computation at the Harvard-Smithsonian Center for
Astrophysics for their hospitality while this review was written.
D.C.F.\ thanks the Kepler TTV/Multis and Eclipsing Binary groups for
sharing an interesting scientific journey, as well as University of
Chicago students and postdocs for sharpening his thoughts on these
topics.  This work was supported by funding from the NASA Origins
program (NNX11AG85G) and Kepler Participating Scientist program
(NNX12AC76G, NNX14AB87G).


\bibliography{ms}{}

\begin{thebibliography}{}
\expandafter\ifx\csname natexlab\endcsname\relax\def\natexlab#1{#1}\fi

\bibitem[{{Adibekyan} et~al.(2013){Adibekyan}, {Figueira}, {Santos}, {Mortier},
  {Mordasini} et~al.}]{Adibekyan+2013}
{Adibekyan} VZ, {Figueira} P, {Santos} NC, {Mortier} A, {Mordasini} C, et~al.
  2013.
\newblock \textit{\aap} 560:A51

\bibitem[{{Agol} et~al.(2005){Agol}, {Steffen}, {Sari} \&
  {Clarkson}}]{Agol+2005}
{Agol} E, {Steffen} J, {Sari} R, {Clarkson} W. 2005.
\newblock \textit{\mnras} 359:567--579

\bibitem[{{Aigrain} et~al.(2008){Aigrain}, {Collier Cameron}, {Ollivier},
  {Pont}, {Jorda} et~al.}]{Aigrain+2008}
{Aigrain} S, {Collier Cameron} A, {Ollivier} M, {Pont} F, {Jorda} L, et~al.
  2008.
\newblock \textit{\aap} 488:L43--L46

\bibitem[{{Albrecht} et~al.(2012){Albrecht}, {Winn}, {Johnson}, {Howard},
  {Marcy} et~al.}]{Albrecht+2012}
{Albrecht} S, {Winn} JN, {Johnson} JA, {Howard} AW, {Marcy} GW, et~al. 2012.
\newblock \textit{\apj} 757:18

\bibitem[{{Alves}, {Do Nascimento} \& {de Medeiros}(2010)}]{Alves+2010}
{Alves} S, {Do Nascimento} Jr. JD, {de Medeiros} JR. 2010.
\newblock \textit{\mnras} 408:1770--1777

\bibitem[{{Armstrong} et~al.(2014){Armstrong}, {Osborn}, {Brown}, {Faedi},
  {G{\'o}mez Maqueo Chew} et~al.}]{Armstrong+2014}
{Armstrong} DJ, {Osborn} HP, {Brown} DJA, {Faedi} F, {G{\'o}mez Maqueo Chew} Y,
  et~al. 2014.
\newblock \textit{\mnras} 444:1873--1883

\bibitem[{{Bakos} et~al.(2013){Bakos}, {Csubry}, {Penev}, {Bayliss},
  {Jord{\'a}n} et~al.}]{Bakos+2013}
{Bakos} G{\'A}, {Csubry} Z, {Penev} K, {Bayliss} D, {Jord{\'a}n} A, et~al.
  2013.
\newblock \textit{\pasp} 125:154--182

\bibitem[{{Ballard} \& {Johnson}(2014)}]{BallardJohnson2014}
{Ballard} S, {Johnson} JA. 2014.
\newblock Submitted to {\it Ap.~J.}, {\tt arxiv:1410.4192}

\bibitem[{{Ballard} et~al.(2011){Ballard}, {Fabrycky}, {Fressin},
  {Charbonneau}, {Desert} et~al.}]{Ballard+2011}
{Ballard} S, {Fabrycky} D, {Fressin} F, {Charbonneau} D, {Desert} JM, et~al.
  2011.
\newblock \textit{\apj} 743:200

\bibitem[{{Baluev}(2011)}]{Baluev2011}
{Baluev} RV. 2011.
\newblock \textit{Celestial Mechanics and Dynamical Astronomy} 111:235--266

\bibitem[{{Barnes}(2009)}]{Barnes2009}
{Barnes} JW. 2009.
\newblock \textit{\apj} 705:683--692

\bibitem[{{Barnes}(2001)}]{Barnes2001}
{Barnes} SA. 2001.
\newblock \textit{\apj} 561:1095--1106

\bibitem[{{Barnes}, {Linscott} \& {Shporer}(2011)}]{Barnes+2011}
{Barnes} JW, {Linscott} E, {Shporer} A. 2011.
\newblock \textit{\apjs} 197:10

\bibitem[{{Barnes} et~al.(2013){Barnes}, {van Eyken}, {Jackson}, {Ciardi} \&
  {Fortney}}]{Barnes+2013}
{Barnes} JW, {van Eyken} JC, {Jackson} BK, {Ciardi} DR, {Fortney} JJ. 2013.
\newblock \textit{\apj} 774:53

\bibitem[{{Barry} et~al.(2011){Barry}, {Kruk}, {Anderson}, {Beaulieu},
    {Bennett} et~al.}]{Barry+2011} {Barry} R, {Kruk} J, {Anderson} J,
  {Beaulieu} JP, {Bennett} DP, et~al. 2011.  \newblock In
  \textit{Techniques and Instrumentation for Detection of Exoplanets
    V}, ed.\ S Shaklan. {\it Proc.\ SPIE Conf.\ Ser.} 8151:815110L. Bellingham,
  WA: SPIE

\bibitem[{{Bate}, {Lodato} \& {Pringle}(2010)}]{Bate+2010}
{Bate} MR, {Lodato} G, {Pringle} JE. 2010.
\newblock \textit{\mnras} 401:1505--1513

\bibitem[{{Batygin} \& {Morbidelli}(2013)}]{BatyginMorbidelli2013}
{Batygin} K, {Morbidelli} A. 2013.
\newblock \textit{\aj} 145:1

\bibitem[{{Batygin}, {Morbidelli} \& {Tsiganis}(2011)}]{Batygin+2011}
{Batygin} K, {Morbidelli} A, {Tsiganis} K. 2011.
\newblock \textit{\aap} 533:A7

\bibitem[{{Bayliss} \& {Sackett}(2011)}]{BaylissSackett2011}
{Bayliss} DDR, {Sackett} PD. 2011.
\newblock \textit{\apj} 743:103

\bibitem[{{Bean} \& {Seifahrt}(2009)}]{BeanSeifahrt2009}
{Bean} JL, {Seifahrt} A. 2009.
\newblock \textit{\aap} 496:249--257

\bibitem[{{Bean} et~al.(2010){Bean}, {Seifahrt}, {Hartman}, {Nilsson},
  {Reiners} et~al.}]{Bean+2010}
{Bean} JL, {Seifahrt} A, {Hartman} H, {Nilsson} H, {Reiners} A, et~al. 2010.
\newblock \textit{\apjl} 711:L19--L23

\bibitem[{{Benedict} et~al.(2002){Benedict}, {McArthur}, {Forveille},
  {Delfosse}, {Nelan} et~al.}]{Benedict+2002}
{Benedict} GF, {McArthur} BE, {Forveille} T, {Delfosse} X, {Nelan} E, et~al.
  2002.
\newblock \textit{\apjl} 581:L115--L118

\bibitem[{{Bennett} \& {Rhie}(2002)}]{BennettRhie2002}
{Bennett} DP, {Rhie} SH. 2002.
\newblock \textit{\apj} 574:985--1003

\bibitem[{{Benomar} et~al.(2014){Benomar}, {Masuda}, {Shibahashi} \&
  {Suto}}]{Benomar+2014}
{Benomar} O, {Masuda} K, {Shibahashi} H, {Suto} Y. 2014.
\newblock \textit{Publ.\ Astron.\ Soc.\ Japan}, {\tt 10.1093/pasj/psu069}

\bibitem[{{Beuzit} et~al.(2008){Beuzit}, {Feldt}, {Dohlen},
    {Mouillet}, {Puget} et~al.}]{Beuzit+2008}
{Beuzit} JL, {Feldt} M, {Dohlen} K, {Mouillet} D, {Puget} P, et~al.\ 2008.
\newblock In \textit{Ground-Based and Airborne Instrumentation for Astronomy II},
  ed.\ IS McLean, MM Casali. {\it Proc.\ SPIE Conf.\ Ser.} 7014:701418. Bellingham, WA: SPIE

\bibitem[{{Biller} et~al.(2013){Biller}, {Liu}, {Wahhaj}, {Nielsen}, {Hayward} et~al.}]{Biller+2013}
{Biller} BA, {Liu} MC, {Wahhaj} Z, {Nielsen} EL, {Hayward} TL, et~al. 2013.
\newblock \textit{\apj} 777:160

\bibitem[{{Biller} et~al.(2014){Biller}, {Males}, {Rodigas}, {Morzinski},
  {Close} et~al.}]{Biller+2014}
{Biller} BA, {Males} J, {Rodigas} T, {Morzinski} K, {Close} LM, et~al. 2014.
\newblock \textit{\apjl} 792:L22

\bibitem[{{Bitsch} et~al.(2013){Bitsch}, {Crida}, {Libert} \&
  {Lega}}]{Bitsch+2013}
{Bitsch} B, {Crida} A, {Libert} AS, {Lega} E. 2013.
\newblock \textit{\aap} 555:A124

\bibitem[{{Bonfils} et~al.(2013){Bonfils}, {Delfosse}, {Udry}, {Forveille},
  {Mayor} et~al.}]{Bonfils+2013}
{Bonfils} X, {Delfosse} X, {Udry} S, {Forveille} T, {Mayor} M, et~al. 2013.
\newblock \textit{\aap} 549:A109

\bibitem[{{Borkovits} et~al.(2011){Borkovits}, {Csizmadia}, {Forg{\'a}cs-Dajka}
  \& {Heged{\"u}s}}]{Borkovits+2011}
{Borkovits} T, {Csizmadia} S, {Forg{\'a}cs-Dajka} E, {Heged{\"u}s} T. 2011.
\newblock \textit{\aap} 528:A53

\bibitem[{{Borucki} et~al.(2013){Borucki}, {Agol}, {Fressin}, {Kaltenegger},
  {Rowe} et~al.}]{Borucki+2013}
{Borucki} WJ, {Agol} E, {Fressin} F, {Kaltenegger} L, {Rowe} J, et~al. 2013.
\newblock \textit{Science} 340:587--590

\bibitem[{{Boss}(1997)}]{Boss1997}
{Boss} AP. 1997.
\newblock \textit{Science} 276:1836--1839

\bibitem[{{Bours} et~al.(2014){Bours}, {Marsh}, {Breedt}, {Copperwheat},
  {Dhillon} et~al.}]{Bours+2014}
{Bours} M, {Marsh} T, {Breedt} E, {Copperwheat} C, {Dhillon} V, et~al. 2014.
\newblock \textit{MNRAS} 445:1924

\bibitem[{{Brandt} et~al.(2014){Brandt}, {Kuzuhara}, {McElwain}, {Schlieder},
  {Wisniewski} et~al.}]{Brandt+2014}
{Brandt} TD, {Kuzuhara} M, {McElwain} MW, {Schlieder} JE, {Wisniewski} JP,
  et~al. 2014.
\newblock \textit{\apj} 786:1

\bibitem[{{Broeg} et~al.(2013){Broeg}, {Fortier}, {Ehrenreich}, {Alibert},
  {Baumjohann} et~al.}]{Broeg+2013}
{Broeg} C, {Fortier} A, {Ehrenreich} D, {Alibert} Y, {Baumjohann} W, et~al.
  2013.
\newblock In \textit{European Physical Journal Web of Conferences}, 47:3005

\bibitem[{{Brown} et~al.(2011){Brown}, {Collier Cameron}, {Hall}, {Hebb} \&
  {Smalley}}]{Brown+2011}
{Brown} DJA, {Collier Cameron} A, {Hall} C, {Hebb} L, {Smalley} B. 2011.
\newblock \textit{\mnras} 415:605--618

\bibitem[{{Brown}(2014)}]{Brown2014}
{Brown} DJA. 2014.
\newblock \textit{\mnras} 442:1844--1862

\bibitem[{{Buchhave} et~al.(2012){Buchhave}, {Latham}, {Johansen}, {Bizzarro},
  {Torres} et~al.}]{Buchhave+2012}
{Buchhave} LA, {Latham} DW, {Johansen} A, {Bizzarro} M, {Torres} G, et~al.
  2012.
\newblock \textit{\nat} 486:375--377

\bibitem[{{Burke}(2008}]{Burke2008}
{Burke} C. 2008.
\newblock \textit{\apj} 679:1566--1573

\bibitem[{{Burke} et~al.(2014){Burke}, {Bryson}, {Mullally}, {Rowe},
  {Christiansen} et~al.}]{Burke+2014}
{Burke} CJ, {Bryson} ST, {Mullally} F, {Rowe} JF, {Christiansen} JL, et~al.
  2014.
\newblock \textit{\apjs} 210:19

\bibitem[{{Butler} et~al.(1997){Butler}, {Marcy}, {Williams}, {Hauser} \&
  {Shirts}}]{Butler+1997}
{Butler} RP, {Marcy} GW, {Williams} E, {Hauser} H, {Shirts} P. 1997.
\newblock \textit{\apjl} 474:L115--L118

\bibitem[{{Butler} et~al.(2004){Butler}, {Vogt}, {Marcy}, {Fischer}, {Wright}
  et~al.}]{Butler+2004}
{Butler} RP, {Vogt} SS, {Marcy} GW, {Fischer} DA, {Wright} JT, et~al. 2004.
\newblock \textit{\apj} 617:580--588

\bibitem[{{Carney} et~al.(2005){Carney}, {Aguilar}, {Latham} \&
  {Laird}}]{Carney+2005}
{Carney} BW, {Aguilar} LA, {Latham} DW, {Laird} JB. 2005.
\newblock \textit{\aj} 129:1886--1905

\bibitem[{{Carter} et~al.(2012){Carter}, {Agol}, {Chaplin}, {Basu}, {Bedding}
  et~al.}]{Carter+2012}
{Carter} JA, {Agol} E, {Chaplin} WJ, {Basu} S, {Bedding} TR, et~al. 2012.
\newblock \textit{Science} 337:556--

\bibitem[{{Casertano} et~al.(2008){Casertano}, {Lattanzi}, {Sozzetti},
  {Spagna}, {Jancart} et~al.}]{Casertano+2008}
{Casertano} S, {Lattanzi} MG, {Sozzetti} A, {Spagna} A, {Jancart} S, et~al.
  2008.
\newblock \textit{\aap} 482:699--729

\bibitem[{{Cassan} et~al.(2012){Cassan}, {Kubas}, {Beaulieu}, {Dominik},
  {Horne} et~al.}]{Cassan+2012}
{Cassan} A, {Kubas} D, {Beaulieu} JP, {Dominik} M, {Horne} K, et~al. 2012.
\newblock \textit{\nat} 481:167--169

\bibitem[{{Catanzarite} \& {Shao}(2011)}]{CatanzariteShao2011}
{Catanzarite} J, {Shao} M. 2011.
\newblock \textit{\apj} 738:151

\bibitem[{{Chamberlin}(1916)}]{Chamberlin1916}
Chamberlin TC. 1916.
\newblock \textit{J.\ R.\ Astron. Soc. Canada} 10:473

\bibitem[{{Chaplin} et~al.(2013){Chaplin}, {Sanchis-Ojeda}, {Campante},
  {Handberg}, {Stello} et~al.}]{Chaplin+2013}
{Chaplin} WJ, {Sanchis-Ojeda} R, {Campante} TL, {Handberg} R, {Stello} D,
  et~al. 2013.
\newblock \textit{\apj} 766:101

\bibitem[{{Chatterjee} \& {Ford}(2015)}]{ChatterjeeFord2015}
{Chatterjee} S, {Ford} EB. 2015.
\newblock \textit{\apj} 803:33

\bibitem[{{Chatterjee} et~al.(2008){Chatterjee}, {Ford}, {Matsumura} \&
  {Rasio}}]{Chatterjee+2008}
{Chatterjee} S, {Ford} EB, {Matsumura} S, {Rasio} FA. 2008.
\newblock \textit{\apj} 686:580--602

\bibitem[{{Chiang} \& {Laughlin}(2013)}]{ChiangLaughlin2013}
{Chiang} E, {Laughlin} G. 2013.
\newblock \textit{\mnras} 431:3444--3455

\bibitem[{{Clanton} \& {Gaudi}(2014)}]{ClantonGaudi2014}
{Clanton} C, {Gaudi} BS. 2014.
\newblock \textit{\apj} 791:91

\bibitem[{{Close} et~al.(2014){Close}, {Males}, {Follette}, {Hinz},
    {Morzinski} et~al.}]{Close+2014} {Close} LM, {Males} JR,
  {Follette} KB, {Hinz} P, {Morzinski} KM, et~al. 2014.  In {\it
    Adaptive Optics Systems IV}, ed.\ E Marchetti, LM Close, J-P
  V\'{e}ran.  {\it Proc.\ SPIE Conf.\ Ser.} 9148:91481M. Bellingham,
  WA: SPIE

\bibitem[{{Correia} et~al.(2010){Correia}, {Couetdic}, {Laskar}, {Bonfils},
  {Mayor} et~al.}]{Correia+2010}
{Correia} ACM, {Couetdic} J, {Laskar} J, {Bonfils} X, {Mayor} M, et~al. 2010.
\newblock \textit{\aap} 511:A21

\bibitem[{{Counselman}(1973)}]{Counselman1973}
{Counselman} III CC. 1973.
\newblock \textit{\apj} 180:307--316

\bibitem[{{Cresswell} et~al.(2007){Cresswell}, {Dirksen}, {Kley} \&
  {Nelson}}]{Cresswell+2007}
{Cresswell} P, {Dirksen} G, {Kley} W, {Nelson} RP. 2007.
\newblock \textit{\aap} 473:329--342

\bibitem[{{Cumming} et~al.(2008){Cumming}, {Butler}, {Marcy}, {Vogt}, {Wright}
  \& {Fischer}}]{Cumming+2008}
{Cumming} A, {Butler} RP, {Marcy} GW, {Vogt} SS, {Wright} JT, {Fischer} DA.
  2008.
\newblock \textit{\pasp} 120:531--554

\bibitem[{{D'Angelo}, {Lubow} \& {Bate}(2006)}]{DAngelo+2006}
{D'Angelo} G, {Lubow} SH, {Bate} MR. 2006.
\newblock \textit{\apj} 652:1698--1714

\bibitem[{{Dawson}(2014)}]{Dawson2014}
{Dawson} RI. 2014.
\newblock \textit{\apjl} 790:L31

\bibitem[{{Dawson} \& {Chiang}(2014)}]{DawsonChiang2014}
{Dawson} RI, {Chiang} E. 2014.
\newblock \textit{Science} 346:212--

\bibitem[{{Dawson} \& {Johnson}(2012)}]{DawsonJohnson2012}
{Dawson} RI, {Johnson} JA. 2012.
\newblock \textit{\apj} 756:122

\bibitem[{{Dawson} et~al.(2014){Dawson}, {Johnson}, {Fabrycky},
  {Foreman-Mackey}, {Murray-Clay} et~al.}]{Dawson+2014}
{Dawson} RI, {Johnson} JA, {Fabrycky} DC, {Foreman-Mackey} D, {Murray-Clay} RA,
  et~al. 2014.
\newblock \textit{\apj} 791:89

\bibitem[{{Dawson} \& {Murray-Clay}(2013)}]{DawsonMurrayClay2013}
{Dawson} RI, {Murray-Clay} RA. 2013.
\newblock \textit{\apjl} 767:L24

\bibitem[{{Deck} et~al.(2012){Deck}, {Holman}, {Agol}, {Carter}, {Lissauer}
  et~al.}]{Deck+2012}
{Deck} KM, {Holman} MJ, {Agol} E, {Carter} JA, {Lissauer} JJ, et~al. 2012.
\newblock \textit{\apjl} 755:L21

\bibitem[{{Delfosse} et~al.(2013){Delfosse}, {Donati}, {Kouach}, {H{\'e}brard},
  {Doyon} et~al.}]{Delfosse+2013}
{Delfosse} X, {Donati} JF, {Kouach} D, {H{\'e}brard} G, {Doyon} R, et~al. 2013.
\newblock In \textit{SF2A-2013: Proceedings of the Annual meeting of the French
  Society of Astronomy and Astrophysics}, eds. L~{Cambresy}, F~{Martins},
  E~{Nuss}, A~{Palacios}, June 4--7, pp.\ 497--508

\bibitem[{{D{\'e}sert} et~al.(2011){D{\'e}sert}, {Charbonneau}, {Demory},
  {Ballard}, {Carter} et~al.}]{Desert+2011}
{D{\'e}sert} JM, {Charbonneau} D, {Demory} BO, {Ballard} S, {Carter} JA, et~al.
  2011.
\newblock \textit{\apjs} 197:14

\bibitem[{{Dong} \& {Zhu}(2013)}]{DongZhu2013}
{Dong} S, {Zhu} Z. 2013.
\newblock \textit{\apj} 778:53

\bibitem[Dong et al.(2014)]{Dong+2014}
{Dong} S, {Katz} B., {Socrates} A. 2014.
\newblock \textit{\apjl} 781:L5 

\bibitem[{{Doolin} \& {Blundell}(2011)}]{DoolinBlundell2011}
{Doolin} S, {Blundell} KM. 2011.
\newblock \textit{\mnras} 418:2656--2668

\bibitem[{{Doyle} et~al.(2011){Doyle}, {Carter}, {Fabrycky}, {Slawson},
  {Howell} et~al.}]{Doyle+2011}
{Doyle} LR, {Carter} JA, {Fabrycky} DC, {Slawson} RW, {Howell} SB, et~al. 2011.
\newblock \textit{Science} 333:1602

\bibitem[{{Dressing} \& {Charbonneau}(2013)}]{DressingCharbonneau2013}
{Dressing} CD, {Charbonneau} D. 2013.
\newblock \textit{\apj} 767:95

\bibitem[{{Durisen} et~al.(2007){Durisen}, {Boss}, {Mayer}, {Nelson}, {Quinn}
  \& {Rice}}]{Durisen+2007}
{Durisen} RH, {Boss} AP, {Mayer} L, {Nelson} AF, {Quinn} T, {Rice} WKM. 2007.
\newblock In \textit{Protostars and Planets V}, eds.\ Reipurth B,
Jewitt D, Keil K, University of Arizona Press (Tucson), 607--622

\bibitem[{{Dvorak}(1982)}]{Dvorak1982}
{Dvorak} R. 1982.
\newblock \textit{Oesterreichische Akademie Wissenschaften Mathematisch
  naturwissenschaftliche Klasse Sitzungsberichte Abteilung} 191:423--437

\bibitem[{{Eggenberger} et~al.(2007){Eggenberger}, {Udry}, {Chauvin}, {Beuzit},
  {Lagrange} et~al.}]{Eggenberger+2007}
{Eggenberger} A, {Udry} S, {Chauvin} G, {Beuzit} JL, {Lagrange} AM, et~al.
  2007.
\newblock \textit{\aap} 474:273--291

\bibitem[{{Eggenberger} et~al.(2011){Eggenberger}, {Udry}, {Chauvin},
  {Forveille}, {Beuzit} et~al.}]{Eggenberger+2011}
{Eggenberger} A, {Udry} S, {Chauvin} G, {Forveille} T, {Beuzit} JL, et~al.
  2011.
\newblock In {\it Proc.\ IAU Symp.\ 276}, ed.\ A Sozzetti, MG
Lattanzi, AP Boss, pp.\ 409--10. Cambridge, UK: Cambridge Univ. Press 

\bibitem[{{Endl} et~al.(2006){Endl}, {Cochran}, {K{\"u}rster}, {Paulson},
  {Wittenmyer} et~al.}]{Endl+2006}
{Endl} M, {Cochran} WD, {K{\"u}rster} M, {Paulson} DB, {Wittenmyer} RA, et~al.
  2006.
\newblock \textit{\apj} 649:436--443

\bibitem[{{Fabrycky} \& {Tremaine}(2007)}]{FabryckyTremaine2007}
{Fabrycky} D, {Tremaine} S. 2007.
\newblock \textit{\apj} 669:1298--1315

\bibitem[{{Fabrycky} \& {Murray-Clay}(2010)}]{FabryckyMurrayClay2010}
{Fabrycky} DC, {Murray-Clay} RA. 2010.
\newblock \textit{\apj} 710:1408--1421

\bibitem[{{Fabrycky} et~al.(2014){Fabrycky}, {Lissauer}, {Ragozzine}, {Rowe},
  {Steffen} et~al.}]{Fabrycky+2014}
{Fabrycky} DC, {Lissauer} JJ, {Ragozzine} D, {Rowe} JF, {Steffen} JH, et~al.
  2014.
\newblock \textit{\apj} 790:146

\bibitem[{{Fang} \& {Margot}(2012)}]{FangMargot2012}
{Fang} J, {Margot} JL. 2012.
\newblock \textit{\apj} 761:92

\bibitem[{{Fielding} et~al.(2014){Fielding}, {McKee}, {Socrates}, {Cunningham}
  \& {Klein}}]{Fielding+2014}
{Fielding} DB, {McKee} CF, {Socrates} A, {Cunningham} AJ, {Klein} RI. 2014.
\newblock Submitted to \textit{MNRAS}, {\tt arxiv:1409.5148}

\bibitem[{{Figueira} et~al.(2012){Figueira}, {Marmier}, {Bou{\'e}}, {Lovis},
  {Santos} et~al.}]{Figueira+2012}
{Figueira} P, {Marmier} M, {Bou{\'e}} G, {Lovis} C, {Santos} NC, et~al. 2012.
\newblock \textit{\aap} 541:A139

\bibitem[Ford(2014)]{Ford2014} {Ford} EB.\ 2014, {\it Proc.\ Nat.\
Acad.\ Sci.} 111:12616

\bibitem[{{Ford} \& {Rasio}(2008)}]{FordRasio2008}
{Ford} EB, {Rasio} FA. 2008.
\newblock \textit{\apj} 686:621--636

\bibitem[{{Ford}, {Quinn} \& {Veras}(2008)}]{Ford+2008}
{Ford} EB, {Quinn} SN, {Veras} D. 2008.
\newblock \textit{\apj} 678:1407--1418

\bibitem[{{Ford} et~al.(2000){Ford}, {Joshi}, {Rasio} \& {Zbarsky}}]{Ford+2000}
{Ford} EB, {Joshi} KJ, {Rasio} FA, {Zbarsky} B. 2000.
\newblock \textit{\apj} 528:336--350

\bibitem[{{Foreman-Mackey}, {Hogg} \& {Morton}(2014)}]{ForemanMackey+2014}
{Foreman-Mackey} D, {Hogg} DW, {Morton} TD. 2014.
\newblock \textit{Ap.~J.}, 795:64

\bibitem[{{Fressin} et~al.(2013){Fressin}, {Torres}, {Charbonneau}, {Bryson},
  {Christiansen} et~al.}]{Fressin+2013}
{Fressin} F, {Torres} G, {Charbonneau} D, {Bryson} ST, {Christiansen} J, et~al.
  2013.
\newblock \textit{\apj} 766:81

\bibitem[{{Gaudi} et~al.(2008){Gaudi}, {Bennett}, {Udalski}, {Gould},
  {Christie} et~al.}]{Gaudi+2008}
{Gaudi} BS, {Bennett} DP, {Udalski} A, {Gould} A, {Christie} GW, et~al. 2008.
\newblock \textit{Science} 319:927

\bibitem[{{Giacobbe} et~al.(2012){Giacobbe}, {Damasso}, {Sozzetti}, {Toso},
  {Perdoncin} et~al.}]{Giacobbe+2012}
{Giacobbe} P, {Damasso} M, {Sozzetti} A, {Toso} G, {Perdoncin} M, et~al. 2012.
\newblock \textit{\mnras} 424:3101--3122

\bibitem[{{Gillon} et~al.(2013){Gillon}, {Jehin}, {Delrez}, {Magain}, {Opitom}
  \& {Sohy}}]{Gillon+2013}
{Gillon} M, {Jehin} E, {Delrez} L, {Magain} P, {Opitom} C, {Sohy} S. 2013.
\newblock In \textit{Protostars and Planets VI Posters},
{\tt http://www.mpia-hd.mpg.de/homes/ppvi/posters/2K066.html}

\bibitem[{{Gizon} \& {Solanki}(2003)}]{GizonSolanki2003}
{Gizon} L, {Solanki} SK. 2003.
\newblock \textit{\apj} 589:1009--1019

\bibitem[{{Goldreich} \& {Sari}(2003)}]{GoldreichSari2003}
{Goldreich} P, {Sari} R. 2003.
\newblock \textit{\apj} 585:1024--1037

\bibitem[{{Goldreich} \& {Schlichting}(2014)}]{GoldreichSchlichting2014}
{Goldreich} P, {Schlichting} HE. 2014.
\newblock \textit{\aj} 147:32

\bibitem[{{Goldreich} \& {Soter}(1966)}]{GoldreichSoter1966}
{Goldreich} P, {Soter} S. 1966.
\newblock \textit{Icarus} 5:375--389

\bibitem[{{Goldreich} \& {Tremaine}(1980)}]{GoldreichTremaine1980}
{Goldreich} P, {Tremaine} S. 1980.
\newblock \textit{\apj} 241:425--441

\bibitem[{{Gonzalez}(1997)}]{Gonzalez1997} {Gonzalez} G. 1997.
\newblock \textit{\mnras} 285:403--412

\bibitem[{{Gonzalez}(2011)}]{Gonzalez2011}
{Gonzalez} G. 2011.
\newblock \textit{\mnras} 416:L80--L83

\bibitem[{{Gould} et~al.(2010){Gould}, {Dong}, {Gaudi}, {Udalski}, {Bond}
  et~al.}]{Gould+2010}
{Gould} A, {Dong} S, {Gaudi} BS, {Udalski} A, {Bond} IA, et~al. 2010.
\newblock \textit{\apj} 720:1073--1089

\bibitem[{{Gould} et~al.(2006){Gould}, {Dorsher}, {Gaudi} \&
  {Udalski}}]{Gould+2006}
{Gould} A, {Dorsher} S, {Gaudi} BS, {Udalski} A. 2006.
\newblock \textit{Acta Astronomica} 56:1--50

\bibitem[{{Go{\'z}dziewski} \&
  {Migaszewski}(2014)}]{GozdziewskiMigaszewski2014}
{Go{\'z}dziewski} K, {Migaszewski} C. 2014.
\newblock \textit{\mnras} 440:3140--3171

\bibitem[{{Greaves} et~al.(2014){Greaves}, {Kennedy}, {Thureau}, {Eiroa},
  {Marshall} et~al.}]{Greaves+2014}
{Greaves} JS, {Kennedy} GM, {Thureau} N, {Eiroa} C, {Marshall} JP, et~al. 2014.
\newblock \textit{\mnras} 438:L31--L35

\bibitem[{{Grether} \& {Lineweaver}(2006)}]{GretherLineweaver2006}
{Grether} D, {Lineweaver} CH. 2006.
\newblock \textit{\apj} 640:1051--1062

\bibitem[{{Grether} \& {Lineweaver}(2007)}]{GretherLineweaver2007}
{Grether} D, {Lineweaver} CH. 2007.
\newblock \textit{\apj} 669:1220--1234

\bibitem[{{Guedel} et~al.(2014){Guedel}, {Dvorak}, {Erkaev}, {Kasting},
  {Khodachenko} et~al.}]{Guedel+2014}
{Guedel} M, {Dvorak} R, {Erkaev} N, {Kasting} J, {Khodachenko} M, et~al. 2014.
\newblock In {\it Protostars and Planets VI}, eds.\ H Beuther, RS
Klessen, CP Dullemond,
T Henning. Tucson: Univ. Ariz. Press, p.\ 883—906

\bibitem[{{Han} et~al.(2013){Han}, {Udalski}, {Choi}, {Yee}, {Gould}
  et~al.}]{Han+2013}
{Han} C, {Udalski} A, {Choi} JY, {Yee} JC, {Gould} A, et~al. 2013.
\newblock \textit{\apjl} 762:L28

\bibitem[{{Hansen} \& {Murray}(2012)}]{HansenMurray2012}
{Hansen} BMS, {Murray} N. 2012.
\newblock \textit{\apj} 751:158

\bibitem[{{Hansen} \& {Murray}(2013)}]{HansenMurray2013}
{Hansen} BMS, {Murray} N. 2013.
\newblock \textit{\apj} 775:53

\bibitem[{{Hayes} \& {Tremaine}(1998)}]{HayesTremaine1998}
{Hayes} W, {Tremaine} S. 1998.
\newblock \textit{Icarus} 135:549--557

\bibitem[{{H{\'e}brard} et~al.(2008){H{\'e}brard}, {Bouchy}, {Pont},
  {Loeillet}, {Rabus} et~al.}]{Hebrard+2008}
{H{\'e}brard} G, {Bouchy} F, {Pont} F, {Loeillet} B, {Rabus} M, et~al. 2008.
\newblock \textit{\aap} 488:763--770

\bibitem[{{H{\'e}brard} et~al.(2010){H{\'e}brard}, {D{\'e}sert}, {D{\'{\i}}az},
  {Boisse}, {Bouchy} et~al.}]{Hebrard+2010}
{H{\'e}brard} G, {D{\'e}sert} JM, {D{\'{\i}}az} RF, {Boisse} I, {Bouchy} F,
  et~al. 2010.
\newblock \textit{\aap} 516:A95

\bibitem[H{\'e}brard et al.(2011)]{Hebrard+2011}
{H{\'e}brard} G, {Ehrenreich} D, {Bouchy} F, et~al.\ 2011.
\newblock \textit{\aap} 527:L11 

\bibitem[{{Henderson} et~al.(2014){Henderson}, {Gaudi}, {Han}, {Skowron},
  {Penny} et~al.}]{Henderson+2014}
{Henderson} CB, {Gaudi} BS, {Han} C, {Skowron} J, {Penny} MT, et~al. 2014.
\newblock \textit{Ap.~J.} 794:52

\bibitem[{{Heisler} \& {Tremaine}(1986)}]{HeislerTremaine1986}
{Heisler} J, {Tremaine} S. 1986.
\newblock \textit{Icarus} 65:13--26

\bibitem[{{Hirano} et~al.(2011){Hirano}, {Narita}, {Sato}, {Winn}, {Aoki}
  et~al.}]{Hirano+2011}
{Hirano} T, {Narita} N, {Sato} B, {Winn} JN, {Aoki} W, et~al. 2011.
\newblock \textit{PASJ} 63:L57--L61

\bibitem[{{Hirano} et~al.(2012){Hirano}, {Narita}, {Sato}, {Takahashi},
  {Masuda} et~al.}]{Hirano+2012}
{Hirano} T, {Narita} N, {Sato} B, {Takahashi} YH, {Masuda} K, et~al. 2012.
\newblock \textit{\apjl} 759:L36

\bibitem[{{Hirano} et~al.(2014){Hirano}, {Sanchis-Ojeda}, {Takeda}, {Winn},
  {Narita} \& {Takahashi}}]{Hirano+2014}
{Hirano} T, {Sanchis-Ojeda} R, {Takeda} Y, {Winn} JN, {Narita} N, {Takahashi}
  YH. 2014.
\newblock \textit{\apj} 783:9

\bibitem[{{Hogg}, {Myers} \& {Bovy}(2010)}]{Hogg+2010}
{Hogg} DW, {Myers} AD, {Bovy} J. 2010.
\newblock \textit{\apj} 725:2166--2175

\bibitem[{{Holman} et~al.(2010){Holman}, {Fabrycky}, {Ragozzine}, {Ford},
  {Steffen} et~al.}]{Holman+2010}
{Holman} MJ, {Fabrycky} DC, {Ragozzine} D, {Ford} EB, {Steffen} JH, et~al.
  2010.
\newblock \textit{Science} 330:51

\bibitem[{{Holman} \& {Murray}(2005)}]{HolmanMurray2005}
{Holman} MJ, {Murray} NW. 2005.
\newblock \textit{Science} 307:1288--1291

\bibitem[{{Holman} \& {Wiegert}(1999)}]{HolmanWiegert1999}
{Holman} MJ, {Wiegert} PA. 1999.
\newblock \textit{\aj} 117:621--628

\bibitem[{{Horner} et~al.(2014){Horner}, {Wittenmyer}, {Hinse}, {Marshall} \&
  {Mustill}}]{Horner+2014}
{Horner} J, {Wittenmyer} R, {Hinse} T, {Marshall} J, {Mustill} A. 2014.
\newblock To appear in {\it Proc.\ Australian Space Science Conf.}, {\tt arxiv:1401:6742}

\bibitem[{{Howard} et~al.(2012){Howard}, {Marcy}, {Bryson}, {Jenkins}, {Rowe}
  et~al.}]{Howard+2012}
{Howard} AW, {Marcy} GW, {Bryson} ST, {Jenkins} JM, {Rowe} JF, et~al. 2012.
\newblock \textit{\apjs} 201:15

\bibitem[{{Howard} et~al.(2010){Howard}, {Marcy}, {Johnson}, {Fischer},
  {Wright} et~al.}]{Howard+2010}
{Howard} AW, {Marcy} GW, {Johnson} JA, {Fischer} DA, {Wright} JT, et~al. 2010.
\newblock \textit{Science} 330:653

\bibitem[{{Howell} et~al.(2014){Howell}, {Sobeck}, {Haas}, {Still}, {Barclay}
  et~al.}]{Howell+2014}
{Howell} SB, {Sobeck} C, {Haas} M, {Still} M, {Barclay} T, et~al. 2014.
\newblock \textit{\pasp} 126:398--408

\bibitem[{{Huber} et~al.(2013){Huber}, {Carter}, {Barbieri}, {Miglio}, {Deck}
  et~al.}]{Huber+2013b}
{Huber} D, {Carter} JA, {Barbieri} M, {Miglio} A, {Deck} KM, et~al. 2013.
\newblock \textit{Science} 342:331--334

\bibitem[{{Husnoo} et~al.(2012){Husnoo}, {Pont}, {Mazeh}, {Fabrycky},
  {H{\'e}brard} et~al.}]{Husnoo+2012}
{Husnoo} N, {Pont} F, {Mazeh} T, {Fabrycky} D, {H{\'e}brard} G, et~al. 2012.
\newblock \textit{\mnras} 422:3151--3177

\bibitem[{{Hut}(1980)}]{Hut1980}
{Hut} P. 1980.
\newblock \textit{\aap} 92:167--170

\bibitem[{{Jeans}(1942)}]{Jeans1942}
{Jeans} JH. 1942.
\newblock \textit{\nat} 149:695

\bibitem[{{Ji} et~al.(2002)}]{Ji+2002}
{Ji} J, {Li} G, {Liu} L. 2002.
\newblock \textit{\apj} 572:1041-1047

\bibitem[{{Johansen} et~al.(2012){Johansen}, {Davies}, {Church} \&
  {Holmelin}}]{Johansen+2012}
{Johansen} A, {Davies} MB, {Church} RP, {Holmelin} V. 2012.
\newblock \textit{\apj} 758:39

\bibitem[{{Johnson} et~al.(2010){Johnson}, {Aller}, {Howard} \&
  {Crepp}}]{Johnson+2010}
{Johnson} JA, {Aller} KM, {Howard} AW, {Crepp} JR. 2010.
\newblock \textit{\pasp} 122:905--915

\bibitem[{{Johnson} et~al.(2007){Johnson}, {Butler}, {Marcy}, {Fischer}, {Vogt}
  et~al.}]{Johnson+2007}
{Johnson} JA, {Butler} RP, {Marcy} GW, {Fischer} DA, {Vogt} SS, et~al. 2007.
\newblock \textit{\apj} 670:833--840

\bibitem[{{Johnson} et~al.(2006){Johnson}, {Marcy}, {Fischer}, {Henry},
  {Wright} et~al.}]{Johnson+2006}
{Johnson} JA, {Marcy} GW, {Fischer} DA, {Henry} GW, {Wright} JT, et~al. 2006.
\newblock \textit{\apj} 652:1724--1728

\bibitem[{{Jovanovic} et~al.(2014){Jovanovic}, {Guyon}, {Martinache}, {Clergeon}, {Singh} et~al.}]{Jovanovic+2014}
{Jovanovic} N, {Guyon} O, {Martinache} F, {Clergeon} C, {Singh} G, et~al.\ 2014.
\newblock In {\it Ground-Based and Airborne Instrumentation for Astronomy V},
ed.\ SK Ramsay, IS McLean, H Takami. {\it Proc. SPIE Conf.\ Ser.} 9147:91471Q

\bibitem[{{Juri{\'c}} \& {Tremaine}(2008)}]{JuricTremaine2008}
{Juri{\'c}} M, {Tremaine} S. 2008.
\newblock \textit{\apj} 686:603--620

\bibitem[{{Kaib}, {Raymond} \& {Duncan}(2013)}]{Kaib+2013}
{Kaib} NA, {Raymond} SN, {Duncan} M. 2013.
\newblock \textit{\nat} 493:381--384

\bibitem[{{Kane} et~al.(2012){Kane}, {Ciardi}, {Gelino} \& {von
  Braun}}]{Kane+2012}
{Kane} SR, {Ciardi} DR, {Gelino} DM, {von Braun} K. 2012.
\newblock \textit{\mnras} 425:757--762

\bibitem[{{Kasting}, {Whitmire} \& {Reynolds}(1993)}]{Kasting+1993}
{Kasting} JF, {Whitmire} DP, {Reynolds} RT. 1993.
\newblock \textit{Icarus} 101:108--128

\bibitem[{{Kennedy} et~al.(2013){Kennedy}, {Wyatt}, {Bryden}, {Wittenmyer} \&
  {Sibthorpe}}]{Kennedy+2013}
{Kennedy} GM, {Wyatt} MC, {Bryden} G, {Wittenmyer} R, {Sibthorpe} B. 2013.
\newblock \textit{\mnras} 436:898--903

\bibitem[{{Kipping}(2013)}]{Kipping2013}
{Kipping} DM. 2013.
\newblock \textit{\mnras} 434:L51--L55

\bibitem[{{Kipping}(2014)}]{Kipping2014}
{Kipping} DM. 2014.
\newblock {\it Proc.\ Frank N.\ Bash Symp.\ 2013: New Horizons in
  Astronomy},
Oct. 6-8, Austin, Tex. Austin: Univ. Tex.

\bibitem[{{Kipping} et~al.(2012){Kipping}, {Dunn}, {Jasinski} \&
  {Manthri}}]{Kipping+2012}
{Kipping} DM, {Dunn} WR, {Jasinski} JM, {Manthri} VP. 2012.
\newblock \textit{\mnras} 421:1166--1188

\bibitem[{{Konacki} \& {Wolszczan}(2003)}]{KonackiWolszczan2003}
{Konacki} M, {Wolszczan} A. 2003.
\newblock \textit{\apjl} 591:L147--L150

\bibitem[{{Kopparapu}(2013)}]{Kopparapu2013}
{Kopparapu} RK. 2013.
\newblock \textit{\apjl} 767:L8

\bibitem[{{Kostov} et~al.(2013){Kostov}, {McCullough}, {Hinse}, {Tsvetanov},
  {H{\'e}brard} et~al.}]{Kostov+2013}
{Kostov} VB, {McCullough} PR, {Hinse} TC, {Tsvetanov} ZI, {H{\'e}brard} G,
  et~al. 2013.
\newblock \textit{\apj} 770:52

\bibitem[{{Kraft}(1967)}]{Kraft1967}
{Kraft} RP. 1967.
\newblock \textit{\apj} 150:551 

\bibitem[{{Lafreni{\`e}re} et~al.(2007){Lafreni{\`e}re}, {Doyon}, {Marois},
  {Nadeau}, {Oppenheimer} et~al.}]{Lafreniere+2007}
{Lafreni{\`e}re} D, {Doyon} R, {Marois} C, {Nadeau} D, {Oppenheimer} BR, et~al.
  2007.
\newblock \textit{\apj} 670:1367--1390

\bibitem[{{Lagrange} et~al.(2012){Lagrange}, {Boccaletti}, {Milli}, {Chauvin},
  {Bonnefoy} et~al.}]{Lagrange+2012}
{Lagrange} AM, {Boccaletti} A, {Milli} J, {Chauvin} G, {Bonnefoy} M, et~al.
  2012.
\newblock \textit{\aap} 542:A40

\bibitem[{{Lagrange} et~al.(2010){Lagrange}, {Bonnefoy}, {Chauvin}, {Apai},
  {Ehrenreich} et~al.}]{Lagrange+2010}
{Lagrange} AM, {Bonnefoy} M, {Chauvin} G, {Apai} D, {Ehrenreich} D, et~al.
  2010.
\newblock \textit{Science} 329:57

\bibitem[{{Lai}, {Foucart} \& {Lin}(2011)}]{Lai+2011}
{Lai} D, {Foucart} F, {Lin} DNC. 2011.
\newblock \textit{\mnras} 412:2790--2798

\bibitem[{{Lai}(2012)}]{Lai2012}
{Lai} D. 2012.
\newblock \textit{\mnras} 423:486--492

\bibitem[{{Latham}(2012)}]{Latham2012}
{Latham} DW. 2012.
\newblock \textit{\nar} 56:16--18

\bibitem[{{Latham} et~al.(2011){Latham}, {Rowe}, {Quinn}, {Batalha}, {Borucki}
  et~al.}]{Latham+2011}
{Latham} DW, {Rowe} JF, {Quinn} SN, {Batalha} NM, {Borucki} WJ, et~al. 2011.
\newblock \textit{\apjl} 732:L24

\bibitem[{{Laughlin}, {Chambers} \& {Fischer}(2002)}]{Laughlin+2002}
{Laughlin} G, {Chambers} J, {Fischer} D. 2002.
\newblock \textit{\apj} 579:455--467

\bibitem[{{Le Bouquin} et~al.(2009){Le Bouquin}, {Absil}, {Benisty}, {Massi},
  {M{\'e}rand} \& {Stefl}}]{LeBouquin+2009}
{Le Bouquin} JB, {Absil} O, {Benisty} M, {Massi} F, {M{\'e}rand} A, {Stefl} S.
  2009.
\newblock \textit{\aap} 498:L41--L44

\bibitem[{{Lee} \& {Peale}(2002)}]{LeePeale2002}
{Lee} MH, {Peale} SJ. 2002.
\newblock \textit{\apj} 567:596--609

\bibitem[{{Limbach} \& {Turner}(2014)}]{LimbachTurner2014}
{Limbach} MA, {Turner} EL. 2014.
\newblock \textit{Proc.\ Nat.\ Acad.\ Sci.} 112:20-24

\bibitem[{{Lin}, {Bodenheimer} \& {Richardson}(1996)}]{Lin+1996}
{Lin} DNC, {Bodenheimer} P, {Richardson} DC. 1996.
\newblock \textit{\nat} 380:606--607

\bibitem[{{Lissauer}(2012)}]{Lissauer2012}
{Lissauer} JJ. 2012.
\newblock \textit{\nar} 56:1--1

\bibitem[{{Lissauer} et~al.(2013){Lissauer}, {Jontof-Hutter}, {Rowe},
  {Fabrycky}, {Lopez} et~al.}]{Lissauer+2013}
{Lissauer} JJ, {Jontof-Hutter} D, {Rowe} JF, {Fabrycky} DC, {Lopez} ED, et~al.
  2013.
\newblock \textit{\apj} 770:131

\bibitem[{{Lissauer} et~al.(2011){Lissauer}, {Ragozzine}, {Fabrycky},
  {Steffen}, {Ford} et~al.}]{Lissauer+2011}
{Lissauer} JJ, {Ragozzine} D, {Fabrycky} DC, {Steffen} JH, {Ford} EB, et~al.
  2011.
\newblock \textit{\apjs} 197:8

\bibitem[{{Lithwick}, {Xie} \& {Wu}(2012)}]{Lithwick+2012}
{Lithwick} Y, {Xie} J, {Wu} Y. 2012.
\newblock \textit{\apj} 761:122

\bibitem[{{Lloyd}(2011)}]{Lloyd2011}
{Lloyd} JP. 2011.
\newblock \textit{\apjl} 739:L49

\bibitem[{{Macintosh} et~al.(2014){Macintosh}, {Graham}, {Ingraham},
  {Konopacky}, {Marois} et~al.}]{Macintosh+2014}
{Macintosh} B, {Graham} JR, {Ingraham} P, {Konopacky} Q, {Marois} C, et~al.
  2014.
\newblock \textit{Proc.\ Nat.\ Acad.\ Sci.}, 111:12661

\bibitem[{{Mahadevan} et~al.(2012){Mahadevan}, {Ramsey}, {Bender}, {Terrien}, {Wright} et~al.}]{Mahadevan+2012}
{Mahadevan} S, {Ramsey} L, {Bender} C, {Terrien} R, {Wright} JT, et~al.\ 2012.
\newblock In {\it Ground-Based and Airborne Instrumentation for Astronomy IV},
ed.\ IS McLean, SK Ramsay, H Takami. {\it Proc.\ SPIE Conf.\ Ser.} 8446:84461S. Bellingham, WA: SPIE

\bibitem[{{Marcy} \& {Butler}(2000)}]{MarcyButler2000}
{Marcy} GW, {Butler} RP. 2000.
\newblock \textit{\pasp} 112:137--140

\bibitem[{{Marcy} et~al.(2014){Marcy}, {Isaacson}, {Howard}, {Rowe}, {Jenkins}
  et~al.}]{Marcy+2014}
{Marcy} GW, {Isaacson} H, {Howard} AW, {Rowe} JF, {Jenkins} JM, et~al. 2014.
\newblock \textit{\apjs} 210:20

\bibitem[{{Mardling}(2007)}]{Mardling2007}
{Mardling} RA. 2007.
\newblock \textit{\mnras} 382:1768--1790

\bibitem[{{Marois} et~al.(2010){Marois}, {Zuckerman}, {Konopacky}, {Macintosh}
  \& {Barman}}]{Marois+2010}
{Marois} C, {Zuckerman} B, {Konopacky} QM, {Macintosh} B, {Barman} T. 2010.
\newblock \textit{\nat} 468:1080--1083

\bibitem[{{Martin} \& {Triaud}(2014)}]{MartinTriaud2014}
{Martin} DV, {Triaud} AHMJ. 2014.
\newblock {\it Astron.\ \& Astroph.} 570:91

\bibitem[{{Masuda}(2014)}]{Masuda2014}
{Masuda} K. 2014.
\newblock \textit{\apj} 783:53

\bibitem[{{Matthews} et~al.(2014){Matthews}, {Kennedy}, {Sibthorpe}, {Booth},
  {Wyatt} et~al.}]{Matthews+2014}
{Matthews} B, {Kennedy} G, {Sibthorpe} B, {Booth} M, {Wyatt} M, et~al. 2014.
\newblock \textit{\apj} 780:97

\bibitem[{{Mayor} et~al.(2011){Mayor}, {Marmier}, {Lovis}, {Udry},
  {S{\'e}gransan} et~al.}]{Mayor+2011}
{Mayor} M, {Marmier} M, {Lovis} C, {Udry} S, {S{\'e}gransan} D, et~al. 2011.
\newblock Submitted to {\it Astron.\ \& Astroph.}, {\tt arxiv:1109.2497}

\bibitem[{{Mazeh}, {Holczer} \& {Shporer}(2014)}]{Mazeh+2014a}
{Mazeh} T, {Holczer} T, {Shporer} A. 2015.
\newblock \textit{Ap.~J.} 800:142

\bibitem[{{Mazeh}, {Krymolowski} \& {Rosenfeld}(1997)}]{Mazeh+1997}
{Mazeh} T, {Krymolowski} Y, {Rosenfeld} G. 1997.
\newblock \textit{\apjl} 477:L103--L106

\bibitem[{{Mazeh} \& {Shaham}(1979)}]{MazehShaham1979}
{Mazeh} T, {Shaham} J. 1979.
\newblock \textit{\aap} 77:145--151

\bibitem[{{Mazeh} et~al.(2014)}]{Mazeh+2014}
{Mazeh} T, {Perets} H, {McQuillan} A, {Goldstein} E. 2014.
\newblock \textit{\apj} 801:3

\bibitem[{{McArthur} et~al.(2010){McArthur}, {Benedict}, {Barnes}, {Martioli},
  {Korzennik} et~al.}]{McArthur+2010}
{McArthur} BE, {Benedict} GF, {Barnes} R, {Martioli} E, {Korzennik} S, et~al.
  2010.
\newblock \textit{\apj} 715:1203--1220

\bibitem[{{McQuillan}, {Mazeh} \& {Aigrain}(2013)}]{McQuillan+2013}
{McQuillan} A, {Mazeh} T, {Aigrain} S. 2013.
\newblock \textit{\apjl} 775:L11

\bibitem[{{Mizuno}(1980)}]{Mizuno1980}
{Mizuno} H. 1980.
\newblock \textit{Progress of Theoretical Physics} 64:544--557

\bibitem[{{Moorhead} et~al.(2011){Moorhead}, {Ford}, {Morehead}, {Rowe},
  {Borucki} et~al.}]{Moorhead+2011}
{Moorhead} AV, {Ford} EB, {Morehead} RC, {Rowe} J, {Borucki} WJ, et~al. 2011.
\newblock \textit{\apjs} 197:1

\bibitem[{{Morton} \& {Winn}(2014)}]{MortonWinn2014}
{Morton} TD, {Winn} JN. 2014.
\newblock \textit{\apj} 796:47

\bibitem[{{Mudryk} \& {Wu}(2006)}]{MudrykWu2006}
{Mudryk} LR, {Wu} Y. 2006.
\newblock \textit{\apj} 639:423--431

\bibitem[{{Mugrauer}, {Ginski} \& {Seeliger}(2014)}]{Mugrauer+2014}
{Mugrauer} M, {Ginski} C, {Seeliger} M. 2014.
\newblock \textit{\mnras} 439:1063--1070

\bibitem[{{Mulders}, {Pascucci} \& {Apai}(2014)}]{Mulders+2014}
{Mulders} GD, {Pascucci} I, {Apai} D. 2014.
\newblock \textit{\apj} 798:112

\bibitem[{{Naef} et~al.(2001){Naef}, {Latham}, {Mayor}, {Mazeh}, {Beuzit}
  et~al.}]{Naef+2001}
{Naef} D, {Latham} DW, {Mayor} M, {Mazeh} T, {Beuzit} JL, et~al. 2001.
\newblock \textit{\aap} 375:L27--L30

\bibitem[{{Naoz} et~al.(2011){Naoz}, {Farr}, {Lithwick}, {Rasio} \&
  {Teyssandier}}]{Naoz+2011}
{Naoz} S, {Farr} WM, {Lithwick} Y, {Rasio} FA, {Teyssandier} J. 2011.
\newblock \textit{\nat} 473:187--189

\bibitem[{{Nelson} et~al.(2014){Nelson}, {Ford}, {Wright}, {Fischer}, {von
  Braun} et~al.}]{Nelson+2014}
{Nelson} BE, {Ford} EB, {Wright} JT, {Fischer} DA, {von Braun} K, et~al. 2014.
\newblock \textit{\mnras} 441:442--451

\bibitem[{{Nesvorn{\'y}} et~al.(2014){Nesvorn{\'y}}, {Kipping}, {Terrell} \&
  {Feroz}}]{Nesvorny+2014}
{Nesvorn{\'y}} D, {Kipping} D, {Terrell} D, {Feroz} F. 2014.
\newblock \textit{\apj} 790:31

\bibitem[{{Nesvorn{\'y}} et~al.(2013){Nesvorn{\'y}}, {Kipping}, {Terrell},
  {Hartman}, {Bakos} \& {Buchhave}}]{Nesvorny+2013}
{Nesvorn{\'y}} D, {Kipping} D, {Terrell} D, {Hartman} J, {Bakos} G{\'A},
  {Buchhave} LA. 2013.
\newblock \textit{\apj} 777:3

\bibitem[{{Nesvorn{\'y}} et~al.(2012){Nesvorn{\'y}}, {Kipping}, {Buchhave},
  {Bakos}, {Hartman} \& {Schmitt}}]{Nesvorny+2012}
{Nesvorn{\'y}} D, {Kipping} DM, {Buchhave} LA, {Bakos} G{\'A}, {Hartman} J,
  {Schmitt} AR. 2012.
\newblock \textit{Science} 336:1133

\bibitem[{{Nielsen} \& {Close}(2010)}]{NielsenClose2010}
{Nielsen} EL, {Close} LM. 2010.
\newblock \textit{\apj} 717:878--896

\bibitem[{{Nielsen} et~al.(2014){Nielsen}, {Liu}, {Wahhaj}, {Biller}, {Hayward}
  et~al.}]{Nielsen+2014}
{Nielsen} EL, {Liu} MC, {Wahhaj} Z, {Biller} BA, {Hayward} TL, et~al. 2014.
\newblock \textit{Ap.\ J.}, 794:158

\bibitem[{{Nutzman} \& {Charbonneau}(2008)}]{NutzmanCharbonneau2008}
{Nutzman} P, {Charbonneau} D. 2008.
\newblock \textit{\pasp} 120:317--327

\bibitem[{{Nutzman}, {Fabrycky} \& {Fortney}(2011)}]{Nutzman+2011}
{Nutzman} PA, {Fabrycky} DC, {Fortney} JJ. 2011.
\newblock \textit{\apjl} 740:L10

\bibitem[{{Ogilvie}(2014)}]{Ogilvie2014}
{Ogilvie} GI. 2014.
\newblock \textit{\araa} 52:171--210

\bibitem[{{Orosz} et~al.(2012{\natexlab{a}}){Orosz}, {Welsh}, {Carter},
  {Brugamyer}, {Buchhave} et~al.}]{Orosz+2012b}
{Orosz} JA, {Welsh} WF, {Carter} JA, {Brugamyer} E, {Buchhave} LA, et~al.
  2012{\natexlab{a}}.
\newblock \textit{\apj} 758:87

\bibitem[{{Orosz} et~al.(2012{\natexlab{b}}){Orosz}, {Welsh}, {Carter},
  {Fabrycky}, {Cochran} et~al.}]{Orosz+2012a}
{Orosz} JA, {Welsh} WF, {Carter} JA, {Fabrycky} DC, {Cochran} WD, et~al.
  2012{\natexlab{b}}.
\newblock \textit{Science} 337:1511

\bibitem[{{Park} et~al.(2012){Park}, {Kim}, {Lee}, {Lee}, {Lee} et~al.}]{Park+2012}
{Park} BG, {Kim} SL, {Lee} JW, {Lee} BC, {Lee} CU, et~al. 2012.
\newblock In {\it Ground-Based and Airborne Telescopes I},
ed.\ LM Stepp, R Gilmozzi, HJ Hall. {\it Proc.\ SPIE Conf.\ Ser.}
8444:844447. Bellingham, WA: SPIE

\bibitem[{{Pasquini} et~al.(2008){Pasquini}, {Avila}, {Dekker}, {Delabre},
  {D'Odorico} et~al.}]{Pasquini+2008}
{Pasquini} L, {Avila} G, {Dekker} H, {Delabre} B, {D'Odorico} S, et~al. 2008.
\newblock In {\it Ground-Based and Airborne Instrumentation for
  Astronomy II},
ed.\ IS McLean, MM Casali. {\it Proc.\ SPIE Conf.\ Ser.}
7014:70141I. Bellingham, WA: SPIE

\bibitem[{{Pepe} et~al.(2010){Pepe}, {Cristiani}, {Rebolo Lopez}, {Santos},
  {Amorim} et~al.}]{Pepe+2010}
{Pepe} FA, {Cristiani} S, {Rebolo Lopez} R, {Santos} NC, {Amorim} A, et~al.
  2010.
\newblock In {\it Ground-Based and Airborne Instrumentation for
  Astronomy III},
ed.\ IS McLean, SK Ramsay, H Takami. {\it Proc.\ SPIE Conf.\ Ser.}
7735:77350F. Bellingham, WA: SPIE

\bibitem[{{Pepper}, {Gould} \& {Depoy}(2003)}]{Pepper+2003}
{Pepper} J, {Gould} A, {Depoy} DL. 2003.
\newblock \textit{Acta Astronomica} 53:213--228

\bibitem[Petigura et al.(2013)]{Petigura+2013}
{Petigura} EA, {Howard} AW, Marcy GW. 2013.
{\it Proc.\ Nat.\ Acad.\ Sci.} 110:19273

\bibitem[{{Pierens} \& {Nelson}(2008)}]{PierensNelson2008}
{Pierens} A, {Nelson} RP. 2008.
\newblock \textit{\aap} 483:633--642

\bibitem[{{Plavchan} et~al.(2014)}]{Plavchan2014}
{Plavchan} P, {Bilinski} C, {Currie} T. 2014.
{\it Publ.\ Astron.\ Soc.\ Pacific} 126:34--47

\bibitem[{{Pollack} et~al.(1996){Pollack}, {Hubickyj}, {Bodenheimer},
  {Lissauer}, {Podolak} \& {Greenzweig}}]{Pollack+1996}
{Pollack} JB, {Hubickyj} O, {Bodenheimer} P, {Lissauer} JJ, {Podolak} M,
  {Greenzweig} Y. 1996.
\newblock \textit{Icarus} 124:62--85

\bibitem[{{Pont}(2009)}]{Pont2009}
{Pont} F. 2009.
\newblock \textit{\mnras} 396:1789--1796

\bibitem[{{Poppenhaeger} \& {Wolk}(2014)}]{PoppenhaegerWolk2014}
{Poppenhaeger} K, {Wolk} SJ. 2014.
\newblock \textit{\aap} 565:L1

\bibitem[{{Pourbaix} et~al.(2004){Pourbaix}, {Tokovinin}, {Batten}, {Fekel},
  {Hartkopf} et~al.}]{Pourbaix+2004}
{Pourbaix} D, {Tokovinin} AA, {Batten} AH, {Fekel} FC, {Hartkopf} WI, et~al.
  2004.
\newblock \textit{\aap} 424:727--732

\bibitem[{{Pravdo} \& {Shaklan}(2009)}]{PravdoShaklan2009}
{Pravdo} SH, {Shaklan} SB. 2009.
\newblock \textit{\apj} 700:623--632

\bibitem[{{Queloz} et~al.(2000){Queloz}, {Eggenberger}, {Mayor}, {Perrier},
  {Beuzit} et~al.}]{Queloz+2000}
{Queloz} D, {Eggenberger} A, {Mayor} M, {Perrier} C, {Beuzit} JL, et~al. 2000.
\newblock \textit{\aap} 359:L13--L17

\bibitem[{{Quirrenbach} et~al.(2010){Quirrenbach}, {Amado}, {Mandel},
  {Caballero}, {Mundt} et~al.}]{Quirrenbach+2010}
{Quirrenbach} A, {Amado} PJ, {Mandel} H, {Caballero} JA, {Mundt} R, et~al.
  2010.
\newblock In {\it Ground-Based and Airborne Instrumentation for
  Astronomy III},
ed.\ IS McLean, SK Ramsay, H Takami. {\it Proc.\ SPIE Conf.\ Ser.}
7735:773513. Bellingham, WA: SPIE

\bibitem[{{Raghavan} et~al.(2010){Raghavan}, {McAlister}, {Henry}, {Latham},
  {Marcy} et~al.}]{Raghavan+2010}
{Raghavan} D, {McAlister} HA, {Henry} TJ, {Latham} DW, {Marcy} GW, et~al. 2010.
\newblock \textit{\apjs} 190:1--42

\bibitem[{{Rappaport} et~al.(2013){Rappaport}, {Sanchis-Ojeda}, {Rogers},
  {Levine} \& {Winn}}]{Rappaport+2013}
{Rappaport} S, {Sanchis-Ojeda} R, {Rogers} LA, {Levine} A, {Winn} JN. 2013.
\newblock \textit{\apjl} 773:L15

\bibitem[{{Rasio} \& {Ford}(1996)}]{RasioFord1996}
{Rasio} FA, {Ford} EB. 1996.
\newblock \textit{Science} 274:954--956

\bibitem[{{Rauer} et~al.(2014){Rauer}, {Catala}, {Aerts},
    {Appourchaux}, {Benz} et~al.}]{Rauer+2014} {Rauer} H, {Catala} C,
  {Aerts} C, {Appourchaux} T, {Benz} W, et~al. 2014. {\it
    Experimental Astronomy}, 41

\bibitem[{{Raymond}, {Barnes} \& {Mandell}(2008)}]{Raymond+2008}
{Raymond} SN, {Barnes} R, {Mandell} AM. 2008.
\newblock \textit{\mnras} 384:663--674

\bibitem[{{Reggiani} et~al.(2014){Reggiani}, {Quanz}, {Meyer}, {Pueyo}, {Absil}
  et~al.}]{Reggiani+2014}
{Reggiani} M, {Quanz} SP, {Meyer} MR, {Pueyo} L, {Absil} O, et~al. 2014.
\newblock \textit{\apjl} 792:L23

\bibitem[{{Ricker} et~al.(2015){Ricker}, {Winn}, {Vanderspek}, {Latham},
  {Bakos} et~al.}]{Ricker+2015}
{Ricker} GR, {Winn} JN, {Vanderspek} R, {Latham} DW, {Bakos} G{\'A}, et~al.
  2014.
\newblock {\it Journal of Astronomical Telescopes,
Instruments, and Systems} 1:014003

\bibitem[{{Rivera} et~al.(2010){Rivera}, {Laughlin}, {Butler}, {Vogt},
  {Haghighipour} \& {Meschiari}}]{Rivera+2010}
{Rivera} EJ, {Laughlin} G, {Butler} RP, {Vogt} SS, {Haghighipour} N,
  {Meschiari} S. 2010.
\newblock \textit{\apj} 719:890--899

\bibitem[{{Rivera} \& {Lissauer}(2001)}]{RiveraLissauer2001}
{Rivera} EJ, {Lissauer} JJ. 2001.
\newblock \textit{\apj} 558:392--402

\bibitem[{{Roell} et~al.(2012){Roell}, {Neuh{\"a}user}, {Seifahrt} \&
  {Mugrauer}}]{Roell+2012}
{Roell} T, {Neuh{\"a}user} R, {Seifahrt} A, {Mugrauer} M. 2012.
\newblock \textit{\aap} 542:A92

\bibitem[{{Rogers}, {Lin} \& {Lau}(2012)}]{Rogers+2012}
{Rogers} TM, {Lin} DNC, {Lau} HHB. 2012.
\newblock \textit{\apjl} 758:L6

\bibitem[{{Roy} \& {Ovenden}(1954)}]{RoyOvenden1954}
{Roy} AE, {Ovenden} MW. 1954.
\newblock \textit{\mnras} 114:232

\bibitem[{{Sahlmann} et~al.(2011){Sahlmann}, {S{\'e}gransan}, {Queloz}, {Udry},
  {Santos} et~al.}]{Sahlmann+2011}
{Sahlmann} J, {S{\'e}gransan} D, {Queloz} D, {Udry} S, {Santos} NC, et~al.
  2011.
\newblock \textit{\aap} 525:A95

\bibitem[{{Sanchis-Ojeda} et~al.(2012){Sanchis-Ojeda}, {Fabrycky}, {Winn},
  {Barclay}, {Clarke} et~al.}]{SanchisOjeda+2012}
{Sanchis-Ojeda} R, {Fabrycky} DC, {Winn} JN, {Barclay} T, {Clarke} BD, et~al.
  2012.
\newblock \textit{\nat} 487:449--453

\bibitem[{{Sanchis-Ojeda} et~al.(2014){Sanchis-Ojeda}, {Rappaport}, {Winn},
  {Kotson}, {Levine} \& {El Mellah}}]{SanchisOjeda+2014}
{Sanchis-Ojeda} R, {Rappaport} S, {Winn} JN, {Kotson} MC, {Levine} A, {El
  Mellah} I. 2014.
\newblock \textit{\apj} 787:47

\bibitem[{{Sanchis-Ojeda} et~al.(2011){Sanchis-Ojeda}, {Winn}, {Holman},
  {Carter}, {Osip} \& {Fuentes}}]{SanchisOjeda+2011}
{Sanchis-Ojeda} R, {Winn} JN, {Holman} MJ, {Carter} JA, {Osip} DJ, {Fuentes}
  CI. 2011.
\newblock \textit{\apj} 733:127

\bibitem[{{Santos}, {Israelian} \& {Mayor}(2004)}]{Santos+2004}
{Santos} NC, {Israelian} G, {Mayor} M. 2004.
\newblock \textit{\aap} 415:1153--1166

\bibitem[{{Schlaufman}(2010)}]{Schlaufman2010}
{Schlaufman} KC. 2010.
\newblock \textit{\apj} 719:602--611

\bibitem[{{Schlaufman}(2014)}]{Schlaufman2014}
{Schlaufman} KC. 2014.
\newblock \textit{\apj} 790:91

\bibitem[{{Schlaufman} \& {Winn}(2013)}]{SchlaufmanWinn2013}
{Schlaufman} KC, {Winn} JN. 2013.
\newblock \textit{\apj} 772:143

\bibitem[{{Schneider}(1994)}]{Schneider1994}
{Schneider} J. 1994.
\newblock \textit{Planetary and Space Science} 42:539--544

\bibitem[{{Schneider} et~al.(2011){Schneider}, {Dedieu}, {Le Sidaner},
  {Savalle} \& {Zolotukhin}}]{Schneider+2011}
{Schneider} J, {Dedieu} C, {Le Sidaner} P, {Savalle} R, {Zolotukhin} I. 2011.
\newblock \textit{\aap} 532:A79

\bibitem[{{Schwamb} et~al.(2013){Schwamb}, {Orosz}, {Carter}, {Welsh},
  {Fischer} et~al.}]{Schwamb+2013}
{Schwamb} ME, {Orosz} JA, {Carter} JA, {Welsh} WF, {Fischer} DA, et~al. 2013.
\newblock \textit{\apj} 768:127

\bibitem[{{Seager}(2011)}]{Seager2011}
{Seager} S, ed. 2011.
\newblock \textit{{Exoplanets}}.
\newblock University of Arizona Press (Tucson, AZ)

\bibitem[{{Seager}(2013)}]{Seager2013}
{Seager} S. 2013.
\newblock \textit{Science} 340:577--581

\bibitem[{{Shen} \& {Turner}(2008)}]{ShenTurner2008}
{Shen} Y, {Turner} EL. 2008.
\newblock \textit{\apj} 685:553--559

\bibitem[{{Sigurdsson} et~al.(2008){Sigurdsson}, {Stairs}, {Moody},
  {Arzoumanian} \& {Thorsett}}]{Sigurdsson+2008}
{Sigurdsson} S, {Stairs} IH, {Moody} K, {Arzoumanian} KMZ, {Thorsett} SE. 2008.
\newblock In \textit{Extreme Solar Systems}, ed.\ D~{Fischer}, FA~{Rasio},
  SE~{Thorsett}, A~{Wolszczan}. \textit{ASP Conf.\ Ser.} 398119. San
  Francisco: ASP

\bibitem[{{Silburt}, {Gaidos} \& {Wu}(2015)}]{Silburt+2015}
{Silburt} A, {Gaidos} E, {Wu} Y. 2015.
\newblock \textit{\apj} 799:180

\bibitem[{{Sliski} \& {Kipping}(2014)}]{SliskiKipping2014}
{Sliski} DH, {Kipping} DM. 2014.
\newblock \textit{\apj} 788:148

\bibitem[{{Snellen} et~al.(2014){Snellen}, {Brandl}, {de Kok}, {Brogi},
  {Birkby} \& {Schwarz}}]{Snellen+2014}
{Snellen} IAG, {Brandl} BR, {de Kok} RJ, {Brogi} M, {Birkby} J, {Schwarz} H.
  2014.
\newblock \textit{\nat} 509:63--65

\bibitem[{{Sousa} et~al.(2008){Sousa}, {Santos}, {Mayor}, {Udry}, {Casagrande}
  et~al.}]{Sousa+2008}
{Sousa} SG, {Santos} NC, {Mayor} M, {Udry} S, {Casagrande} L, et~al. 2008.
\newblock \textit{\aap} 487:373--381

\bibitem[{{Steffen} \& {Farr}(2013)}]{SteffenFarr2013}
{Steffen} JH, {Farr} WM. 2013.
\newblock \textit{\apjl} 774:L12

\bibitem[{{Steffen} et~al.(2012){Steffen}, {Ragozzine}, {Fabrycky}, {Carter},
  {Ford} et~al.}]{Steffen+2012}
{Steffen} JH, {Ragozzine} D, {Fabrycky} DC, {Carter} JA, {Ford} EB, et~al.
  2012.
\newblock \textit{Proceedings of the National Academy of Science}
  109:7982--7987

\bibitem[{{Storch}, {Anderson} \& {Lai}(2014)}]{Storch+2014}
{Storch} NI, {Anderson} KR, {Lai} D. 2014.
\newblock \textit{Science} 345:1317

\bibitem[{{Sumi} et~al.(2011){Sumi}, {Kamiya}, {Bennett}, {Bond}, {Abe}
  et~al.}]{Sumi+2011}
{Sumi} T, {Kamiya} K, {Bennett} DP, {Bond} IA, {Abe} F, et~al. 2011.
\newblock \textit{\nat} 473:349--352

\bibitem[{{Szab{\'o}} et~al.(2011){Szab{\'o}}, {Szab{\'o}}, {Benk{\H o}},
  {Lehmann}, {Mez{\H o}} et~al.}]{Szabo+2011}
{Szab{\'o}} GM, {Szab{\'o}} R, {Benk{\H o}} JM, {Lehmann} H, {Mez{\H o}} G,
  et~al. 2011.
\newblock \textit{\apjl} 736:L4

\bibitem[{{Szentgyorgyi} et~al.(2012){Szentgyorgyi}, {Frebel}, {Furesz},
  {Hertz}, {Norton} et~al.}]{Szentgyorgyi+2012}
{Szentgyorgyi} A, {Frebel} A, {Furesz} G, {Hertz} E, {Norton} T, et~al. 2012.
\newblock In {\it Ground-Based and Airborne Instrumentation for
  Astronomy IV},
ed.\ IS McLean, SK Ramsay, H Takami. {\it Proc.\ SPIE Conf.\ Ser.}
8446:84461H. Bellingham, WA: SPIE

\bibitem[{{Tamura} et~al.(2012){Tamura}, {Suto}, {Nishikawa}, {Kotani}, {Sato}
  et~al.}]{Tamura+2012}
{Tamura} M, {Suto} H, {Nishikawa} J, {Kotani} T, {Sato} B, et~al. 2012.
\newblock In {\it Ground-Based and Airborne Instrumentation for
  Astronomy IV},
ed.\ IS McLean, SK Ramsay, H Takami. {\it Proc.\ SPIE Conf.\ Ser.}
8446:84461T. Bellingham, WA: SPIE

\bibitem[{{Tamuz} et~al.(2008){Tamuz}, {S{\'e}gransan}, {Udry}, {Mayor},
  {Eggenberger} et~al.}]{Tamuz+2008}
{Tamuz} O, {S{\'e}gransan} D, {Udry} S, {Mayor} M, {Eggenberger} A, et~al.
  2008.
\newblock \textit{\aap} 480:L33--L36

\bibitem[{{Tan} et~al.(2013){Tan}, {Payne}, {Lee}, {Ford}, {Howard}
  et~al.}]{Tan+2013}
{Tan} X, {Payne} MJ, {Lee} MH, {Ford} EB, {Howard} AW, et~al. 2013.
\newblock \textit{\apj} 777:101

\bibitem[{{Tarter} et~al.(2007){Tarter}, {Backus}, {Mancinelli}, {Aurnou},
  {Backman} et~al.}]{Tarter+2007}
{Tarter} JC, {Backus} PR, {Mancinelli} RL, {Aurnou} JM, {Backman} DE, et~al.
  2007.
\newblock \textit{Astrobiology} 7:30--65

\bibitem[{{Teitler} \& {K{\"o}nigl}(2014)}]{TeitlerKonigl2014}
{Teitler} S, {K{\"o}nigl} A. 2014.
\newblock \textit{\apj} 786:139

\bibitem[{{Terquem} \& {Papaloizou}(2007)}]{TerquemPapaloizou2007}
{Terquem} C, {Papaloizou} JCB. 2007.
\newblock \textit{\apj} 654:1110--1120

\bibitem[{{Thebault} \& {Haghighipour}(2014)}]{ThebaultHaghighipour2014}
{Thebault} P, {Haghighipour} N. 2014.
\newblock In {\it Planetary Exploration and Science: Recent Advances
  and Applications},
ed.\ S Jin, N Haghighipour, WH Ip. Dordrect, Neth.: Springer, pp.\ 309-340 

\bibitem[{{Thies} et~al.(2011){Thies}, {Kroupa}, {Goodwin}, {Stamatellos} \&
  {Whitworth}}]{Thies+2011}
{Thies} I, {Kroupa} P, {Goodwin} SP, {Stamatellos} D, {Whitworth} AP. 2011.
\newblock \textit{\mnras} 417:1817--1822

\bibitem[{{Thommes} \& {Lissauer}(2003)}]{ThommesLissauer2003}
{Thommes} EW, {Lissauer} JJ. 2003.
\newblock \textit{\apj} 597:566--580

\bibitem[{{Thorsett} et~al.(1999){Thorsett}, {Arzoumanian}, {Camilo} \&
  {Lyne}}]{Thorsett+1999}
{Thorsett} SE, {Arzoumanian} Z, {Camilo} F, {Lyne} AG. 1999.
\newblock \textit{\apj} 523:763--770

\bibitem[{{Traub}(2012)}]{Traub2012}
{Traub} WA. 2012.
\newblock \textit{\apj} 745:20

\bibitem[{{Tremaine}(1991)}]{Tremaine1991}
{Tremaine} S 1991.
\newblock \textit{Icarus} 89:85--92

\bibitem[{{Tremaine} \& {Dong}(2012)}]{TremaineDong2012}
{Tremaine} S, {Dong} S. 2012.
\newblock \textit{\aj} 143:94

\bibitem[{{Triaud} et~al.(2010){Triaud}, {Collier Cameron}, {Queloz},
  {Anderson}, {Gillon} et~al.}]{Triaud+2010}
{Triaud} AHMJ, {Collier Cameron} A, {Queloz} D, {Anderson} DR, {Gillon} M,
  et~al. 2010.
\newblock \textit{\aap} 524:A25

\bibitem[{{Udry} et~al.(2002){Udry}, {Mayor}, {Naef}, {Pepe}, {Queloz}
  et~al.}]{Udry+2002}
{Udry} S, {Mayor} M, {Naef} D, {Pepe} F, {Queloz} D, et~al. 2002.
\newblock \textit{\aap} 390:267--279

\bibitem[{{Udry}, {Mayor} \& {Santos}(2003)}]{Udry+2003}
{Udry} S, {Mayor} M, {Santos} NC. 2003.
\newblock \textit{\aap} 407:369--376

\bibitem[{{Valenti} \& {Fischer}(2005)}]{ValentiFischer2005}
{Valenti} JA, {Fischer} DA. 2005.
\newblock \textit{\apjs} 159:141--166

\bibitem[{{Valsecchi} \& {Rasio}(2014)}]{ValsecchiRasio2014}
{Valsecchi} F, {Rasio} FA. 2014.
\newblock \textit{\apj} 786:102

\bibitem[{{van Eyken} et~al.(2012){van Eyken}, {Ciardi}, {von Braun}, {Kane},
  {Plavchan} et~al.}]{vanEyken+2012}
{van Eyken} JC, {Ciardi} DR, {von Braun} K, {Kane} SR, {Plavchan} P, et~al.
  2012.
\newblock \textit{\apj} 755:42

\bibitem[{{Van Eylen} et~al.(2014)}]{VanEylen+2014}
{Van Eylen} V, {Lund} MN, {Silva} V, et al. 2014.
\newblock \textit{\apj} 782:14

\bibitem[{{Veras} \& {Armitage}(2004)}]{VerasArmitage2004}
{Veras} D, {Armitage} PJ. 2004.
\newblock \textit{Icarus} 172:349--371

\bibitem[{{Veras}, {Crepp} \& {Ford}(2009)}]{Veras+2009}
{Veras} D, {Crepp} JR, {Ford} EB. 2009.
\newblock \textit{\apj} 696:1600--1611

\bibitem[{{Veras} \& {Ford}(2012)}]{VerasFord2012}
{Veras} D, {Ford} EB. 2012.
\newblock \textit{\mnras} 420:L23--L27

\bibitem[{{Veras}, {Ford} \& {Payne}(2011)}]{Veras+2011}
{Veras} D, {Ford} EB, {Payne} MJ. 2011.
\newblock \textit{\apj} 727:74

\bibitem[{{Villaver} \& {Livio}(2009)}]{VillaverLivio2009}
{Villaver} E, {Livio} M. 2009.
\newblock \textit{\apjl} 705:L81--L85

\bibitem[{{Walkowicz} \& {Basri}(2013)}]{WalkowiczBasri2013}
{Walkowicz} LM, {Basri} GS. 2013.
\newblock \textit{\mnras} 436:1883--1895

\bibitem[{{Wang} et~al.(2014){Wang}, {Fischer}, {Xie} \& {Ciardi}}]{Wang+2014}
{Wang} J, {Fischer} DA, {Xie} JW, {Ciardi} DR. 2014.
\newblock \textit{\apj} 791:111

\bibitem[{{Wang} \& {Ford}(2011)}]{WangFord2011}
{Wang} J, {Ford} EB. 2011.
\newblock \textit{\mnras} 418:1822--1833

\bibitem[{{Ward}(1997)}]{Ward1997}
{Ward} WR. 1997.
\newblock \textit{Icarus} 126:261--281

\bibitem[{{Watson} et~al.(2011){Watson}, {Littlefair}, {Diamond}, {Collier
  Cameron}, {Fitzsimmons} et~al.}]{Watson+2011}
{Watson} CA, {Littlefair} SP, {Diamond} C, {Collier Cameron} A, {Fitzsimmons}
  A, et~al. 2011.
\newblock \textit{\mnras} 413:L71--L75

\bibitem[{{Weidenschilling} \& {Marzari}(1996)}]{WeidenschillingMarzari1996}
{Weidenschilling} SJ, {Marzari} F. 1996.
\newblock \textit{\nat} 384:619--621

\bibitem[{{Welsh} et~al.(2012){Welsh}, {Orosz}, {Carter}, {Fabrycky}, {Ford}
  et~al.}]{Welsh+2012}
{Welsh} WF, {Orosz} JA, {Carter} JA, {Fabrycky} DC, {Ford} EB, et~al. 2012.
\newblock \textit{\nat} 481:475--479

\bibitem[{{Welsh} et~al.(2014){Welsh}, {Orosz}, {Short}, {Haghighipour},
  {Buchhave} et~al.}]{Welsh+2014}
{Welsh} WF, {Orosz} JA, {Short} DR, {Haghighipour} N, {Buchhave} LA, et~al.
  2014.
\newblock Submitted to \textit{Ap.~J.}, {\tt arxiv:1409.1605}

\bibitem[{{Wheatley} et~al.(2013){Wheatley}, {Pollacco}, {Queloz}, {Rauer}, {Watson} et~al.}]{Wheatley+2013}
{Wheatley} PJ, {Pollacco} DL, {Queloz} D, {Rauer} H, {Watson} CA, et~al.\ 2013.
\newblock In \textit{Eur.\ Phys.\ J.\ Web Conf.}, 47:13002

\bibitem[{{Williams} \& {Cremin}(1968)}]{WilliamsCremin1968}
{Williams} IP, {Cremin} AW. 1968.
\newblock \textit{Q.\ Jl.\ R.\ Ast.\ Soc.} 9:40, p.~41

\bibitem[{{Winn} et~al.(2010){Winn}, {Fabrycky}, {Albrecht} \&
  {Johnson}}]{Winn+2010}
{Winn} JN, {Fabrycky} D, {Albrecht} S, {Johnson} JA. 2010.
\newblock \textit{\apjl} 718:L145--L149

\bibitem[{{Winn} et~al.(2009){Winn}, {Johnson}, {Fabrycky}, {Howard}, {Marcy}
  et~al.}]{Winn+2009}
{Winn} JN, {Johnson} JA, {Fabrycky} D, {Howard} AW, {Marcy} GW, et~al. 2009.
\newblock \textit{\apj} 700:302--308

\bibitem[{{Winn} et~al.(2006){Winn}, {Johnson}, {Marcy}, {Butler}, {Vogt}
  et~al.}]{Winn+2006}
{Winn} JN, {Johnson} JA, {Marcy} GW, {Butler} RP, {Vogt} SS, et~al. 2006.
\newblock \textit{\apjl} 653:L69--L72

\bibitem[{{Wittenmyer} et~al.(2009){Wittenmyer}, {Endl}, {Cochran}, {Levison}
  \& {Henry}}]{Wittenmyer+2009}
{Wittenmyer} RA, {Endl} M, {Cochran} WD, {Levison} HF, {Henry} GW. 2009.
\newblock \textit{\apjs} 182:97--119

\bibitem[{{Wolszczan} \& {Frail}(1992)}]{WolszczanFrail1992}
{Wolszczan} A, {Frail} DA. 1992.
\newblock \textit{\nat} 355:145--147

\bibitem[{{Wright} et~al.(2011{\natexlab{a}}){Wright}, {Chen{\'e}}, {De Cat},
  {Marois}, {Mathias} et~al.}]{Wright+2011}
{Wright} DJ, {Chen{\'e}} AN, {De Cat} P, {Marois} C, {Mathias} P, et~al.
  2011{\natexlab{a}}.
\newblock \textit{\apjl} 728:L20

\bibitem[{{Wright} et~al.(2011{\natexlab{b}}){Wright}, {Fakhouri}, {Marcy},
  {Han}, {Feng} et~al.}]{exoplanetsorg}
{Wright} JT, {Fakhouri} O, {Marcy} GW, {Han} E, {Feng} Y, et~al.
  2011{\natexlab{b}}.
\newblock \textit{\pasp} 123:412--422

\bibitem[{{Wright} \& {Gaudi}(2013)}]{WrightGaudi2013}
{Wright} JT, {Gaudi} BS. 2013.
\newblock In \textit{{Exoplanet Detection Methods}}, eds.\ Oswalt TD,
French LM, Kalas P., p.~489. Dordrecht, Neth: Springer

\bibitem[{{Wright} et~al.(2012){Wright}, {Marcy}, {Howard}, {Johnson}, {Morton}
  \& {Fischer}}]{Wright+2012}
{Wright} JT, {Marcy} GW, {Howard} AW, {Johnson} JA, {Morton} TD, {Fischer} DA.
  2012.
\newblock \textit{\apj} 753:160

\bibitem[{{Wright} et~al.(2009){Wright}, {Upadhyay}, {Marcy}, {Fischer}, {Ford}
  \& {Johnson}}]{Wright+2009}
{Wright} JT, {Upadhyay} S, {Marcy} GW, {Fischer} DA, {Ford} EB, {Johnson} JA.
  2009.
\newblock \textit{\apj} 693:1084--1099

\bibitem[{{Wright} et~al.(2011{\natexlab{c}}){Wright}, {Veras}, {Ford},
  {Johnson}, {Marcy} et~al.}]{Wright+2011c}
{Wright} JT, {Veras} D, {Ford} EB, {Johnson} JA, {Marcy} GW, et~al.
  2011{\natexlab{c}}.
\newblock \textit{\apj} 730:93

\bibitem[{{Wu} \& {Lithwick}(2013)}]{WuLithwick2013}
{Wu} Y, {Lithwick} Y. 2013.
\newblock \textit{\apj} 772:74

\bibitem[{{Xiang-Gruess} \& {Papaloizou}(2013)}]{XiangGruessPapaloizou2013}
{Xiang-Gruess} M, {Papaloizou} JCB. 2013.
\newblock \textit{\mnras} 431:1320--1336

\bibitem[{{Xie}, {Wu} \& {Lithwick}(2014)}]{Xie+2014}
{Xie} JW, {Wu} Y, {Lithwick} Y. 2014.
\newblock \textit{\apj} 789:165

\bibitem[{{Youdin}(2011)}]{Youdin2011}
{Youdin} AN. 2011.
\newblock \textit{\apj} 742:38

\bibitem[{{Zakamska}, {Pan} \& {Ford}(2011)}]{Zakamska+2011}
{Zakamska} NL, {Pan} M, {Ford} EB. 2011.
\newblock \textit{\mnras} 410:1895--1910

\bibitem[{{Zsom} et~al.(2013){Zsom}, {Seager}, {de Wit} \& {Stamenkovi{\'c}}}]{Zsom+2013}
{Zsom} A, {Seager} S, {de Wit} J, {Stamenkovi{\'c}} V. 2013.
\newblock \textit{\apj} 778:109

\end{thebibliography}

\end{document}